\documentclass[structabstract]{aa}

\usepackage{natbib}
\bibpunct{(}{)}{;}{a}{}{,} 
\usepackage{graphicx,amsmath,amssymb}

\usepackage{txfonts}

\begin{document}

\title{
Non-axisymmetric instabilities of neutron star with 
toroidal magnetic fields
}
\author{
  Kenta Kiuchi\inst{\ref{inst1}}
  \and
  Shijun Yoshida\inst{\ref{inst2}}
  \and
  Masaru Shibata\inst{\ref{inst1}}
}

\institute{Yukawa Institute for Theoretical Physics, 
Kyoto University, Kyoto, 606-8502, Japan~\label{inst1}
\and
Astronomical Institute, Tohoku University, Sendai 980-8578, Japan~\label{inst2}
}

\date{Received day month year / Accepted day month year}

{\abstract 
{
Neutron stars with strong toroidal magnetic fields are often 
produced in nature. We show that isentropic neutron stars 
with purely toroidal magnetic fields are unstable against 
the interchange, Parker and/or Taylor instabilities irrespective of 
the toroidal magnetic field configurations. 
}
{
The aim of this paper is to clarify the stabilities of neutron
stars with strong toroidal magnetic fields against non-axisymmetric
perturbation. The motivation comes from the fact that super magnetized
neutron stars of $\sim 10^{15}$G, magnetars, and magnetized
proto-neutron stars born after the magnetically-driven supernovae
are likely to have such strong toroidal magnetic fields.
}
{
Long-term, three-dimensional general relativistic magneto-hydrodynamic
simulations are performed, preparing isentropic neutron stars with
toroidal magnetic fields in equilibrium as initial conditions.  To
explore the effects of rotations on the stability, simulations are
done for both non-rotating and rigidly rotating models.  }
{
We find the emergence of the Parker and/or Tayler
instabilities in both the non-rotating and rotating models. For both
non-rotating and rotating models, the Parker instability is the
primary instability as predicted by the local linear perturbation
analysis. The interchange instability also appears in the rotating models. 
It is found that rapid rotation is not enough to suppress
the Parker instability, and this finding does not agree with the
perturbation analysis.  The reason for this is that rigidly and
rapidly rotating stars are marginally stable, and hence, in the
presence of stellar pulsations by which the rotational profile is
deformed, unstable regions with negative gradient of angular momentum
profile is developed. After the onset of the instabilities, a
turbulence is excited.  Contrary to the axisymmetric case, the
magnetic fields never reach an equilibrium state after the development
of the turbulence.  }
{
Isentropic neutron stars with strong toroidal magnetic
fields are likely to be always unstable against the Parker
instability. A turbulence motion is induced and maintained for a long
time. This conclusion is different from that in axisymmetric
simulations and suggests that three-dimensional simulation is
indispensable for exploring the formation of magnetars or prominence
activities of magnetars such as giant flares.  } 

 }

\maketitle

\section{Introduction}\label{Sec:Intro}

There are a lot of observational evidences that suggest the presence
of neutron stars with strong magnetic fields. Observed spin periods
and their time derivatives in conjunction with the assumption of a
magnetic dipole radiation give us magnetic field strength as $B\propto
(P\dot{P})^{1/2}$ where $P$ and $\dot{P}$ are the spin period and its
derivative, respectively.  For radio pulsars, of which more than 1800
are known today~\citep{Manchester:2005}, the inferred value of the
magnetic field strength is in the range $10^{11}$--$10^{14}$ G. For a
smaller population of older, millisecond pulsars, the typical magnetic
field strength is $B \sim 10^{8}$--$10^{9}$ G. For anomalous X-ray
pulsars (AXPs) and soft gamma repeaters (SGRs), super strong magnetic
fields of $10^{14}-10^{15}$ G are again inferred from the measured
values of $P$ and $\dot{P}$~\citep{Woods:2004kb}.  Various observed
properties of AXPs and SGRs like the giant flares from the three SGRs
and bursts are often explained in connection with a super strong
magnetic field~\citep{Thompson:1995gw,Thompson:1996pe,Thompson:2001}
rather than with rotation because their spin down luminosities are
much smaller than the observed luminosity. In addition, temporary 
detections of spectral lines during SGR/AXP bursts have been reported
in several systems~\citep{Gavriil:2002mc,Ibrahim:2003,Rea:2003mx}.  If
we assume that they are associated with proton cyclotron lines, the
magnetic field strength is estimated to give $B \sim 10^{15}$~G. 

For about a dozen accreting X-ray pulsars in binary systems, electron
cyclotron line features have been detected, suggesting that $B\sim
10^{12}-10^{13}$ G according to the formula for the electron cyclotron
energy, $E_{\rm ce}=\hbar B/(m_e c)=11.58(B/10^{12}~{\rm G})$ keV
~\citep{Orlandini:2001ue}. For many other X-ray pulsars with no
detectable electron cyclotron line features, typical magnetic fields
are $B\sim 10^{12}$ G if one assumes that spin up due to accretion of
matter is balanced by magnetic braking~\citep{Bildsten:1997}.

From a theoretical point of view, these strongly magnetized objects
deserve a detailed study, because their magnetic fields sometimes
exceed the quantum critical field strength, $B_{\rm
QED}=m_e^2c^3/(e\hbar)=4.144\times 10^{13}$ G, at which gyration
radius of the electron $pc/(eB)$ is shorter than the de Broglie
wavelength $\hbar /p$.  Above this limit, magnetic fields affect the
properties of atoms, molecules, and condensed
matters~\citep{Lai:2001}, propagation of photons, radiative processes,
equation of states, and thermal conductivity in crusts (see
\cite{Harding:2006qn} and references therein). The origin of such 
extremely large magnetic fields has been a big issue since their
discovery. Specifically, there are two hypotheses for its origin; in
one scenario, the magnetic fields are assumed to be generated in a
rapidly rotating proto-neutron star formed after stellar core collapse
of a massive star~\citep{Thompson:1993hn} and in the other, it is
assumed to be descended from the main sequence stars, i.e., the strong
magnetic field is assumed to be a fossil of a strongly magnetized main
sequence star~\citep{Wickramasinghe:2005}.  For exploring the magnetar
formation, there are a plenty of magneto-hydrodynamic simulations for
supernova core collapse both in Newtonian
gravity~\citep{Yamada:2004,Kotake:2004,OAM2006,Scheidegger:2007nk,Takiwaki:2007sf,Burrows:2007yx},
and in general relativity~\citep{Shibata:2006hr,CerdaDuran:2007cr}.
For these works, the simulations were performed in axisymmetric
spacetime.  In most of these simulations, it was found that toroidal
magnetic fields are dominantly enhanced in the proto-neutron stars
after the core bounce by the magnetic winding mechanism, and
eventually, the proto-neutron star settles to a quasi-equilibrium
state.  However, it has been well known that purely toroidal fields in
equilibria are often unstable due to the interchange, Tayler, and
Parker
instabilities~\citep{Acheson:1978,Goossens:1980,Parker:1955,Parker:1966,Tayler:1973}.
The stability analyses have suggested that non-axisymmetric modes
would play an essential role for these instabilities and rapid
rotation could suppress these instabilities~\citep{Acheson:1978}.
\footnote{Besides the instabilities listed here, magneto-rotational instability
(MRI)~\citep{Balbus:1991} could play an important role for enhancing
the magnetic field strength and for modifying the magnetic field
profile~\citep{OAM2006,Shibata:2006hr}. However, any accurate MHD
simulation, in which the fastest growing mode of MRI is resolved, has
not been performed yet.}

Motivated by these facts, we studied the axisymmetric instability of
neutron stars with toroidal magnetic fields in the previous
work~\citep{Kiuchi:2008ss}. In that work, magnetized neutron stars in
equilibria are prepared as initial conditions with varying its profile
and strength and with changing the angular
velocity~\citep{Kiuchi:2008ch}. Performing the general relativistic
magneto-hydrodynamic (GRMHD) simulations, we found that slowly
rotating neutron stars with the toroidal fields, whose profile is
proportional to the power of cylindrical radius $\varpi$ as
$B_{(\varphi)}\propto \varpi^{2k-1}$ with $k \ge 2$, are unstable. The
growth time scale is of order the Alfv\'{e}n time scale, and the type
of the instability is the interchange instability in the absence of
stellar rotation.  Only for the case $k=1$, slowly rotating neutron
stars are stable against the axisymmetric perturbation, and thus, we
concluded that the configuration with $k=1$ will be the attractor for
the unstable neutron stars.  We also found that rapid rotation, with
which the rotational kinetic energy is much greater than the magnetic 
energy, suppresses the onset of the interchange instability, and
stabilizes the neutron stars. These results qualitatively and
semi-quantitatively agree with the local linear analysis.

In the magneto-rotational explosion scenarios, the magnetic energy is
composed primarily of the toroidal field and is at most as large as
the rotational kinetic energy. This indicates that the axisymmetric
interchange instability would not play an important role in the
proto-neutron star.  However, it is still possible that the neutron
star becomes unstable against the Parker and Tayler instabilities
which grow in a non-axisymmetric way. Stability of magnetars is also
the important issue along this
line~\citep{Braithwaite:2005xi,Braithwaite:2005ps}.  For the magnetar,
the rotational effect is negligible.  Thus, the toroidal field profile
should be similar to that of $k=1$ to avoid the onset of the
axisymmetric interchange instability.  However, such configuration may
be still unstable if the Parker and/or Tayler instabilities are 
taken into account.

Motivated by these facts, we extend our previous
work~\citep{Kiuchi:2008ss}.  The main aim of this article is to
explore the non-axisymmetric instabilities of neutron stars with
toroidal magnetic fields.  Following our previous work, we prepare
equilibrium neutron stars with purely toroidal magnetic fields as
initial conditions. This time, we perform three-dimensional GRMHD
simulations with varying the field strength and/or rotation velocity. 
As mentioned above, the three-dimensional simulation is inevitable 
for exploring the Parker and Tayler instabilities. 

This paper is organized as follows. In Section~\ref{Sec:Method}, we
briefly review the formulation and numerical methods employed in our
numerical-relativity simulation.  Set up of numerical simulation and
initial models for our GRMHD simulations are described in
Section~\ref{Sec:Numerics}.  In Section~\ref{Sec:Linear}, we present
the results of a local linear perturbation analysis as a forecast of
numerical-simulation results.  Section~\ref{Sec:Result} is devoted to
presenting the numerical results.  A summary and discussion are given
in Section~\ref{Sec:Summary}.

Throughout this paper, we adopt the geometrical units in which ${\rm
c}={\rm G}=1$ with c and G being the speed of light and gravitational
constant, respectively. Cartesian coordinates are denoted by
$x^k=(x,y,z)$. The coordinates are oriented so that the rotation axis
is along the $z$-direction.  We define the coordinate radius 
$r=\sqrt{x^2+y^2+z^2}$, cylindrical radius $\varpi=\sqrt{x^2+y^2}$,
and azimuthal angle $\varphi=\tan^{-1}(y/x)$.  Coordinate time is
denoted by $t$. Greek indices $\mu, \nu, \cdots$ denote spacetime
components, and small Latin indices $i, j, \cdots$ denote spatial
components.

\section{Formulation and Method}\label{Sec:Method}

The stability of magnetized neutron stars are studied by three
dimensional GRMHD simulation assuming that the ideal MHD condition
holds.  In this paper we focus on the Parker and/or Tayler
instabilities against non-axisymmetric perturbations.  The simulation
is performed upgrading our axisymmetric GRMHD numerical code
in~\citet{Shibata:2005gp,Kiuchi:2008ss} to that for three dimensions. 

Formulation and numerical scheme for solving Einstein's equation are
essentially the same as in~\citet{Shibata:1995we}.  For solving 
Einstein's evolution equation, we use the original version of the
Baumgarte-Shapiro-Shibata-Nakamura
formulation~\citep{Shibata:1995we,Baumgarte:1998te}: We evolve the
inverse square of the conformal factor $W\equiv\exp(-2\phi)$ with
$\phi=\ln(\gamma)/12$, the trace part of the extrinsic curvature, $K$,
the conformal three-metric, 
$\tilde{\gamma}_{ij}\equiv\gamma^{-1/3}\gamma_{ij}$, the tracefree
extrinsic curvature, $\tilde{A}_{ij}\equiv
\gamma^{-1/3}(K_{ij}-K\gamma_{ij}/3)$, and a three-auxiliary variable,
$F_i\equiv\delta^{jk}\partial_j\tilde{\gamma}_{ik}$. Here,
$\gamma_{ij}$ is the three-metric, $K_{ij}$ the extrinsic curvature,
$\gamma\equiv{\rm det}(\gamma_{ij})$, and $K \equiv
K_{ij}\gamma^{ij}$.  Note that we evolve $W$, not the conformal factor
as in~\citet{Kiuchi:2009jt}, because our code is designed to simulate
black hole spacetimes with the moving puncture
method~\citep{Baker:2006,Campanelli:2005dd,Brugmann:2008zz}.  For the
conditions of the lapse, $\alpha$, and the shift vector, $\beta^i$, we
adopt a dynamical gauge condition in the following forms~\citep{S03},
\begin{eqnarray}
&&
(\partial_t-\beta^i\partial_i) \ln\alpha = - 2 K,\label{eq:gaug1}\\
&&
\partial_t \beta^i = 0.75\tilde{\gamma}^{ij}(F_j + \Delta t
\partial_t F_j),\label{eq:gaug2}
\end{eqnarray}
where $\Delta t$ denotes the time step in the numerical simulations,
and the second term in the right-hand side of (\ref{eq:gaug2}) is
introduced for stabilizing the numerical computations.  The
finite-differencing schemes for solving Einstein's equation is
essentially the same as those in~\citet{Kiuchi:2009jt}. We use the
fourth-order finite-differencing scheme in the spatial direction and a
fourth-order Runge-Kutta scheme in the time integration, where the
advection terms such as $\beta^i\partial_i W$ are evaluated by a
fourth-order non-centered difference scheme, as proposed, e.g., 
in~\citet{Brugmann:2008zz}.

A conservative shock-capturing scheme is employed to integrate the
GRMHD equations. Specifically we use a high-resolution central
scheme~\citep{KT,lucas} with the third-order piece-wise parabolic
interpolation and with a steep min-mod limiter in which the limiter
parameter $b$ is set to be 2.5 (see appendix A of \cite{S03}).


Magnetized neutron stars in equilibrium, employed as the initial 
condition, are computed giving the polytropic equation of 
state~\citep{Kiuchi:2008ch}, 
\begin{eqnarray}
P=\kappa \rho^{\Gamma},\label{Eq:EOS}
\end{eqnarray}
where $P$, $\rho$, $\kappa$, and $\Gamma$ are the pressure, rest-mass
density, polytropic constant, and adiabatic constant, respectively.
In this work, we choose $\Gamma=2$. Because $\kappa$ is arbitrarily
chosen or else completely scaled out of the problem, we adopt the
units of $\kappa=1$ in the following (i.e., the polytropic units of
c$=$G=$\kappa=1$ are employed). In the numerical simulation, we adopt
the $\Gamma$-law equation of state as
\begin{eqnarray}
P=(\Gamma-1)\rho \varepsilon,
\end{eqnarray}
where $\varepsilon$ is the specific internal thermal energy.

We monitor the total baryon rest mass $M_b$, Arnowitt-Deser-Misner
(ADM) mass $M$, internal energy $U_{\rm int}$, rotational kinetic
energy $T_{\rm rot}$, total kinetic energy $T_{\rm kin}$, and magnetic
energy $H_{\rm mag}$, defined by
\begin{eqnarray}
&&M_b = \int \rho w \sqrt{\gamma} d^3x\\
&&M   = \int 
e^{-\phi}\Big[
\rho h w^2 - P +\frac{1}{16\pi}\{
K_{ij}K^{ij}- K^2 - \tilde{R} e^{-4\phi}
\}\Big]\sqrt{\gamma}d^3x\\
&&U_{\rm int}=\int \rho w \varepsilon \sqrt{\gamma} d^3x \label{eq:int}\\
&&T_{\rm rot}=\frac{1}{2}\int \rho h w u_\varphi \Omega \sqrt{\gamma} d^3x\\
&&T_{\rm kin}=\frac{1}{2}\int \rho h w u_iv^i \sqrt{\gamma}d^3x\\
&&H_{\rm mag}=\frac{1}{8\pi}\int b^2 w \sqrt{\gamma}d^3x 
\end{eqnarray}
where $\tilde{R}$ is the Ricci scalar with respect to 
$\tilde{\gamma}_{ij}$, $u^\mu$ is the four velocity of the fluid, $h$
is the specific enthalpy defined by $1+\varepsilon+P/\rho$, $w\equiv
\alpha u^t$, $v^i=u^i/u^t$, $\Omega=v^\varphi$, and $b^2=b^\mu
b_\mu$. $b^\mu$ is the magnetic field observed in the frame co-moving
with the fluid element. $M_0$ denotes the initial value of the ADM
mass. Once each energy component is obtained, we define the
gravitational potential energy by
\begin{eqnarray}
W=M-(M_b+U_{\rm int}+T_{\rm kin}+H_{\rm mag}).
\end{eqnarray}
Following \citet{Kiuchi:2008ss}, we define an averaged Alfv\'{e}n
time scale as
\begin{eqnarray}
\bar{v}_A \equiv \sqrt{
\frac{2 H_{\rm mag}}{M_b+\Gamma U_{\rm int}+2H_{\rm mag}}, 
}
\end{eqnarray}
where we use the relation $h=1+\Gamma \varepsilon$, which holds in the
$\Gamma$-law equation of state.  Note that the Alfv\'{e}n velocity in
relativity is given by $\sqrt{b^2/(4\pi\rho h+ b^2)}$.  Then, we 
define the averaged Alfv\'{e}n time scale as 
\begin{eqnarray}
\bar{\tau}_A \equiv \frac{R}{\bar{v}_A}, 
\end{eqnarray}
where $R$ is the equatorial stellar radius. Because the magnetic
field instability grows on an order of the Alfv\'{e}n time scale
(cf. Section~\ref{Sec:Linear}), $\bar{\tau}_A$ is useful to judge 
whether or not the instability found in numerical simulation is 
associated with a magnetic field effect.

\section{Model and Numerical setup}\label{Sec:Numerics}

\subsection{Initial condition}

Neutron stars with toroidal magnetic fields in equilibrium, employed
as initial conditions, are computed by the code described
in~\citet{Kiuchi:2008ch}. Several key quantities characterizing these
magnetized neutron stars are listed in Table~\ref{Tab:model}.  The
instability associated with the presence of toroidal magnetic fields
depends on the profile of the magnetic field as shown
by~\citet{Acheson:1978,Goossens:1980,Spruit:1999,Tayler:1973}. 
We assume the toroidal magnetic field profile confined inside the neutron star 
to be given by
\begin{eqnarray}
 b_{(\varphi)}= B_0 u^t (\rho h \alpha^2 \gamma_{\varphi\varphi} )^k 
\gamma_{\varphi\varphi}^{-1/2}, \label{Eq:MAG}
\end{eqnarray}
where $k$ and $B_0$ are constants which determine the field profile
and strength, respectively. The regularity condition of magnetic
fields near the axis of $\varpi=0$ requires $k \ge 1$. Because of
$\gamma_{\varphi\varphi} \propto \varpi^2$ for $\varpi \rightarrow 0$,
the toroidal magnetic field is proportional to $\varpi^{2k-1}$ near
the axis.  Previous works predict that the profile with $k \ge 2$ is
unstable against axisymmetric
perturbations~\citep{Acheson:1978,Goossens:1980,Spruit:1999,Tayler:1973}
and we confirmed this prediction in the previous
paper~\citep{Kiuchi:2008ss}.  On the other hand, the profile of $k=1$
is not unstable against axisymmetric perturbations but may be unstable
against non-axisymmetric ones (see
also~\citet{Lander:2009ws,Lander:2010br}).  Hence in this paper, we
focus on the profile of $k=1$.

Magnetic field strength, $B_0$, is chosen so as to satisfy $8\times
10^{-3} \le H/|W| \le 5\times 10^{-2}$.  These values imply the field
strength of $10^{16}$--$10^{17}~{\rm G}$ for a typical neutron star of
mass $1.4M_\odot$, radius $10~{\rm km}$, and $|W|\sim 6\times
10^{53}~{\rm erg}$.  These magnetic field strengths are extremely
large even for models of magnetar and might be less realistic.  We
here give such strong magnetic fields simply to save the computational
costs; note that the growth time scales of the instabilities by the
presence of the toroidal magnetic field are of order Alfv\'{e}n time
scale, which is still much longer than the dynamical time scale of the
system in the present set up. Thus, a scaling relation should hold for
a weaker magnetic field strength.  We may apply the scaling relation
to derive a generic physical essence from the results obtained in the
present set up.  Namely, if the magnetic field strength becomes half,
the growth time scale of the instabilities becomes approximately twice
longer, although qualitative properties of the evolution of the
unstable neutron star are essentially the same.

The initial conditions for the non-rotating model is specified if the
central density $\rho_c$ (in the polytropic units) is determined. We
choose it to be $\rho_c \approx 0.22$. With this choice, the neutron
star has a realistic compactness; e.g, if we assume $M\approx
1.35M_\odot$, circumferential radius is $R\approx 11$km
(cf. Table~\ref{Tab:model}).  Note that the maximum rest mass and
gravitational mass of the spherical neutron star for $\Gamma=2$ are,
respectively, $0.1799$ and $0.1637$, and the corresponding central
density is $\approx 0.318$ in our unit. 

Simulations are also performed for rotating neutron star models. In
this paper, we focus only on rigidly rotating neutron stars with a
moderate compactness; $\rho_c$ is again chosen to be $0.22$. For the
neutron stars with the $\Gamma=2$ polytrope, the maximum values of
$T_{\rm rot}/|W|$ is $\sim 0.09$ for $M_0/R_{\rm cir}\sim
0.1$~\citep{Cook:1994}.  Local linear stability analysis predicts that
rapid rotations will suppress the growth of some of the instabilities
associated with the presence of toroidal magnetic
fields~(\citet{Acheson:1978}, see also Section 4.2). To study whether this would be indeed the
case, we prepare rapidly rotating models with $T_{\rm rot}/|W| \approx
0.08$, and vary the magnetic field strength. 

\subsection{Grid settings}

The simulations are performed on the cell-centered Cartesian,
$(x,y,z)$, grid. Equatorial plane symmetry with respect to $z=0$ plane
is assumed. The computational domain of $-L\le x \le L$, $-L\le y \le
L$, and $0\le z \le L$ is covered by the grid size $(2N,2N,N)$ for
$(x, y, z)$, where $L$ and $N$ are
constants. Following~\citet{Kiuchi:2008ss}, we adopt a non-uniform
grid as follows; an inner domain is covered with an uniform grid of
spacing $\Delta x$ and with the grid size, $(2N_0,2N_0,N_0)$. Outside
this inner domain, the grid spacing is increased according to the
relation, $\xi\tanh[(i-N_0)/\Delta i]\Delta x$, where $i$ denotes the
$i$-th grid point in each positive direction, and $N_0$, $\Delta i$,
and $\xi$ are constants.  Then, the location of $i$-th grid, $x^k(i)$ 
($i \geq 0$), for each direction is
\begin{eqnarray}
x^k(i)=\left\{
\begin{array}{ll}
(i+1/2)\Delta x & i \leq N_0 \\ 
(i+1/2)\Delta x + \xi \Delta i
\Delta x \log [ \cosh \{(i-N_0)/\Delta i \}] & i > N_0\\
\end{array}
\right.
\end{eqnarray}
and $x^k(-i-1)=-x^k(i)$, where $i=0,1,\cdots N-1$ for $x^k=x,y,$ and
$z$. The chosen parameters of the grid structure for each simulation
are listed in Table~\ref{tab:numset}.  To check the validity of our
numerical results, the simulations for models N22H5 and R22H2T8 are
performed with three different grid resolutions.  With the highest
(lowest) resolution, the coordinate equatorial radii of neutron stars,
$L_{\rm NS}$, are covered by 130 (80) grid points while the outer
boundary location is chosen to be $\simeq 8 L_{\rm NS}$ in all the simulations. We
confirmed the modest resolution in which $L_{\rm NS}$ is covered by
100 grid points is high enough to draw a resolution-invariant 
conclusion.

\begin{table*}
\centering
\begin{minipage}{140mm}
\caption{
List of characteristic quantities for neutron stars with toroidal
magnetic fields.
\label{Tab:model}}
\begin{tabular}{lcccccccc}
\hline\hline
Model             &
$\rho_c$          &
$U_{\rm int}/|W|$ &
$H_{\rm mag}/|W|$ &
$T_{\rm rot}/|W|$ &
$M_0$             &
$M_{\rm b}$       &
$M_0/R_{\rm cir}$ &
$\bar{\tau}_A/M_0$ \\
\hline\hline
N22H5     & 2.2E-1 (1.24)& 5.12E-1 & 5.01E-2  & 0       & 1.59E-1 (1.67)& 1.73E-1 (1.81) & 1.78E-1 &37.4 (0.31)\\
N22H1     & 2.2E-1 (1.24)& 5.32E-1 & 1.01E-2  & 0       & 1.60E-1 (1.68)& 1.75E-1 (1.96) & 1.87E-1 &76.7 (0.63)\\
\hline
R22H2T8  & 2.2E-1 (1.24)& 4.71E-1 & 1.50E-2  & 7.97E-2 & 1.84E-1 (1.93)& 2.01E-1 (1.63) & 1.55E-1 &63.5 (0.63)\\
R22H08T8 & 2.2E-1 (1.24)& 4.74E-1 & 7.86E-3  & 8.10E-2 & 1.85E-1 (1.94)& 2.02E-1 (1.73) & 1.65E-1 &97.3 (0.93)\\
\hline\hline
\end{tabular}
 \tablefoot{
Central density ($\rho_c$), ratio of the internal
energy to the gravitational potential energy $W$ ($U_{\rm int}/|W|$),
ratio of the magnetic energy to $W$ ($H_{\rm mag}/|W|$),
ratio of the rotational kinetic energy to $W$ ($T_{\rm rot}/|W|$), ADM
mass ($M_0$), baryon rest mass ($M_{\rm b}$), and compactness
($M_0/R_{\rm cir}$) with $R_{\rm cir}$ being equatorial
circumferential radius. $\bar{\tau}_A$ is an averaged Alfv\'{e}n time
in units of $M_0$.  Model name NXHY denotes non-rotating models with
``X'' and ``Y'' being the values of $100\rho_c$ and the values of
$100H_{\rm mag}/|W|$, respectively.  RXHYTZ denotes rotating models
with ``Z'' being the values of $100T_{\rm rot}/|W|$, where meanings of
``X'' and ``Y'' are the same as the non-rotating models. 
The values shown in the brackets denote those in physical units, where 
the densities are normalized by $10^{15}{\rm g/cm^3}$, the masses by $M_{\odot}$, and 
the Alfv\'{e}n time are given in units of millisecond. Note that we set $\kappa$ as $1.6\times 10^5$ 
in cgs units.
}
\end{minipage}
\end{table*}

\begin{table*}
\centering
\begin{minipage}{140mm}
\caption{\label{tab:numset} Parameters for the grid structure employed
  in the numerical simulation. 
  }
\begin{tabular}{lccccc}
\hline\hline
Model                                    &
~~$N$~~                                  &
~~$N_0$~~                                &
~~$\Delta i$~~                           &
~~$\xi$~~                                & 
$L_{\rm NS}/\Delta x$                    \\
\hline
N22H5-l  & 161 & 120 & 30 & 22 &  80 \\
N22H5    & 200 & 150 & 30 & 20 & 100 \\
N22H5-h  & 253 & 195 & 30 & 20 & 130 \\
N22H1    & 200 & 150 & 30 & 20 & 100 \\
R22H2T8-l& 161 & 120 & 30 & 22 &  80 \\
R22H2T8  & 200 & 150 & 30 & 20 & 100 \\
R22H2T8-h& 253 & 195 & 30 & 21 & 130 \\
R22H08T8 & 226 & 180 & 30 & 22 & 100 \\
\hline\hline
\end{tabular}
\tablefoot{
The grid number for covering one
  positive direction $(N)$, that for the inner uniform grid zone
  $(N_0)$, the parameters for nonuniform-grid domain $(\Delta i,\xi)$,
  and the approximate grid number for covering the coordinate radius of 
  neutron star $(L_{\rm NS})$, respectively.
}
\end{minipage}
\end{table*}

\section{Local linear perturbation analysis}\label{Sec:Linear}

Before showing the results of nonlinear numerical simulation, it is
useful to remind the results of linear-stability analysis.  By using
the local linear perturbation analysis~\citep{Acheson:1978}, the
dispersion relations for the non-rotating and rotating models are
derived, and the (local) stability is determined. In the following,
the method of the analysis in the Newtonian framework is briefly
reviewed. 

In the local analysis, a perturbation quantity $\delta Q$
is assumed to be given by
\begin{eqnarray}
\delta Q=Q_0 \exp[i(l\varpi+m\varphi+nz-\sigma t)],
\end{eqnarray}
where $Q_0$ is a constant, $\sigma$ is the oscillation frequency, and
$(l,m,n)$ is the wave number vector.  To obtain the local dispersion
relations, we assume that $l \gg 1$, $n\gg 1$, and $m=O(1)$, which
implies that perturbations we consider have very short wavelength on
the meridional plane (for details, cf.~\citet{Acheson:1978}).

\subsection{Non-rotating case}

The dispersion relation for non-rotating models is given, 
irrespective of the density profile, by 
\begin{eqnarray}
&&\left(1+\frac{l^2}{n^2}\right)\left(1+\frac{v_A^2}{c_s^2}\right)
\sigma^4 \nonumber\\
&&+\left[
v_A^2\left(\frac{2}{\varpi}-\frac{\hat{G}}{c_s^2}\right)\partial_h
\ln\left(\frac{B_{(\varphi)}}{\rho\varpi}\right)
-\left(1+\frac{l^2}{n^2}\right)m^2\sigma_A^2\left(2+\frac{v_A^2}{c_s^2}\right)
\right]\sigma^2 \nonumber\\
&&+\left(1+\frac{l^2}{n^2}\right)m^4 \sigma_A^2 + m^2\sigma_A^4 \varpi^2 
\left(\frac{\hat{G}}{c_s^2}-\frac{2}{\varpi}\right)\partial_h 
\ln(B_{(\varphi)}\varpi)=0
\label{Eq:Disperse}
\end{eqnarray}
where $v_A=B_{(\varphi)}/\sqrt{4\pi\rho}$, $\sigma_A=v_A/\varpi$,
$\hat{G}=g_\varpi-(l/n) g_z$ with $g_\varpi$ and $g_z$ being the
gravitational acceleration,
$\partial_h=\partial_\varpi-(l/n)\partial_z,$ and $c_s$ is the sound
speed.  For the axisymmetric perturbation, i.e., $m=0$ perturbations,
the dispersion relation is reduced to a quadratic form of $\sigma$ as
\begin{eqnarray}
\left(1+\frac{l^2}{n^2}\right)\left(1+\frac{v_A^2}{c_s^2}\right)\sigma^2 
+\left[
v_A^2\left(\frac{2}{\varpi}-\frac{\hat{G}}{c_s^2}\right)\partial_h
\ln\left(\frac{B_{(\varphi)}}{\rho\varpi}\right)
\right]=0.
\end{eqnarray}

For the density profile which is necessary to solve Equation 
(\ref{Eq:Disperse}) for a specific model, 
we prepare the Newtonian spherical polytrope with $\Gamma=2$ as
\begin{eqnarray}
\rho = \rho_0 \frac{\sin(\pi r/r_s)}{\pi r/r_s} \label{Eq:NDEN}
\end{eqnarray}
where $\rho_0$ and $r_s$ are the central density and stellar 
radius. The spherical density profile is justified by the assumption
of a weak magnetic field, i.e., the magnetic field is too weak to
deform the star. In the Newtonian limit, Equation~(\ref{Eq:MAG}) becomes
\begin{eqnarray}
B_{(\varphi)}= B_0 \rho \varpi \, . \label{Eq:NMAG}
\end{eqnarray}

Because $B_{(\varphi)}/{\rho\varpi}=$const. in our models (see
Equations~((\ref{Eq:NDEN}) and (\ref{Eq:NMAG})), we have $\sigma=0$ for
$m=0$, i.e., the magnetic field is marginally stable against the
axisymmetric perturbation.  Physically, the magnetic field instability
caused by the axisymmetric perturbation is classified as the
interchange instability.  Our models are marginally stable against
such instability and we indeed found in the previous study 
that they were not unstable~\citep{Kiuchi:2008ss}. 
However, the marginally stable profile is not always kept in a stationary state 
in the presence of a perturbation. We return to this point in Section~\ref{subsec:rot}.

For non-axisymmetric perturbations, we
forecast that the Parker instability~\citep{Parker:1955,Parker:1966}
and the Tayler instability~\citep{Tayler:1973} set in. 
Following~\citet{Acheson:1978}, we define the critical radius as
\begin{eqnarray}
\varpi_c \equiv \frac{2c_s^2}{g_\varpi}.
\end{eqnarray}

Outside this critical radius, the magnetic buoyant force exceeds the
magnetic hoop stress, whereas inside it, the magnetic stress is
dominant.  The Tayler instability could set in near the axis of
$\varpi=0$~\citep{Spruit:1999}, which is inside the critical
radius. On the other hand, the Parker instability could play a role in
places where the magnetic buoyancy force surpasses the magnetic
tension.  Thus, we classify the instability which emerges inside and
outside the critical radius as the Tayler and Parker instabilities,
respectively.  Analyzing the dispersion relation near the axis of
$\varpi=0$, \citet{Acheson:1978} and \citet{Spruit:1999} concluded
that the dominant modes of the Tayler instability are $m=1$ and
$l/n\approx 0$ modes, which are associated with the horizontal motion
near the axis of $\varpi=0$.  However, to clarify the most relevant
unstable modes, we have to determine the growth time scales of the
instability for all the places inside the star.  Hence, we calculate
the maximum growth rates of the instability solving the dispersion
relation~(\ref{Eq:Disperse}) with varying $l/n$ and $m$.

Figure~\ref{Fig:fig1} displays the contours of the growth rate for the
model with $\rho_0=0.16$ and $B_0=0.2$, which give $H_{\rm
mag}/|W|=2.5\%$ and $M_b=0.2$ in the polytropic units.  These
parameters are chosen so as to mimic the initial models shown in
Table~\ref{Tab:model}.  The white colored region and the dotted curve
denote the stable region and critical radius, respectively. We note
that the most unstable mode at each point is the $l/n=0$ mode.
This figure shows that the fastest growing mode of the instability is
located near the stellar surface and is determined by the Parker
instability, because the location is outside the critical radius.  The
right panel of Figure~\ref{Fig:fig1} shows that not the $m=1$ mode but
a high-order $m$ mode is relevant for this fastest growing mode. The
local linear perturbation analysis predicts that the Parker
instability primarily emerges near the stellar surface in our
non-rotating models.

\subsection{Rotating-case}\label{sec:linear_rot}

For the rotating models, the dispersion relation is written as
\begin{eqnarray}
&&\left(1+\frac{l^2}{n^2}\right)\left(1+\frac{v_A^2}{c_s^2}\right)
\omega^4 +\Bigg[v_A^2\partial_h
\ln\left(\frac{B_{(\varphi)}}{\rho\varpi}\right)\left(\frac{2}{\varpi}
-\frac{\hat{G}}{c_s^2}\right)
\nonumber\\
&&-\left(1+\frac{l^2}{n^2}\right)m^2\sigma_A^2
\left(2+\frac{v_A^2}{c_s^2}\right)
-\frac{2\Omega}{\varpi}\partial_h(\Omega
\varpi^2)\left(1+\frac{v_A^2}{c_s^2}\right)
\Bigg] \omega^2 \nonumber\\
&&
-\frac{4\Omega m
 v_A^2}{\varpi}\left(\frac{2}{\varpi}-\frac{\hat{G}}{c_s^2}\right)
\omega 
+m^2\sigma_A^2 \Bigg[ 2\varpi\Omega\partial_h \Omega+
\left(1+\frac{l^2}{n^2}\right)m^2\sigma_A^2 \nonumber\\
&& - \left(\frac{2}{\varpi}-\frac{\hat{G}}{c_s^2}\right)
\partial_h \ln(B_{(\varphi)} \varpi) v_A^2
\Bigg]=0, \label{Eq:Disperse2}
\end{eqnarray}
where $\omega=\sigma-m\Omega$ with $\Omega$ being the angular
velocity. Following~\citet{Acheson:1978}, we focus on a low-frequency
and non-axisymmetric mode, for which the growth time scale is much
longer than the Alfv\'{e}n time scale as 
\begin{eqnarray}
|\omega|^2 \ll m^2\sigma_A^2. 
\end{eqnarray}
We also adopt a weak magnetic field approximation in which magnetic
energy is assumed to be everywhere much smaller than the
rotational kinetic energy as
\begin{eqnarray}
v_A^2 \ll \Omega^2 \varpi^2.\label{Eq:weak_mag}
\end{eqnarray}
 With these approximations, the bi-quadratic equation
(\ref{Eq:Disperse2}) is reduced to a quadratic equation and its
solution is
\begin{eqnarray}
&&\omega = \left(\frac{\hat{G}}{c_s^2}-\frac{2}{\varpi}\right)
\frac{mv_A^2}{2\Omega\varpi} \pm\frac{mv_A}{2\Omega\varpi}
\Bigg[
\left(\frac{\hat{G}}{c_s^2}-\frac{2}{\varpi}\right)v_A^2 \partial_h
\ln\left(\frac{B_{(\varphi)}}{\rho\varpi}\right)\nonumber\\
&&~~~~~~~~~+\Omega^2\varpi \partial_h (\ln\Omega^2) +
\frac{m^2v_A^2}{\varpi^2}\left(1+\frac{l^2}{n^2}\right)
\Bigg]^{1/2}. \label{Eq:growth}
\end{eqnarray}
This dispersion relation shows that the instability sets in 
if either of the following criteria is satisfied,
\begin{eqnarray}
&&-\Omega^2\varpi\partial_\varpi(\ln\Omega^2)-\left(\frac{g_\varpi}{c_s^2}
-\frac{2}{\varpi}\right)v_A^2
\partial_\varpi\ln\left(\frac{B_{(\varphi)}}{\rho\varpi}\right)>m^2\sigma_A^2
\label{Eq:cri2a}\\
&&-\frac{g_zv_A^2}{c_s^2}\partial_z
\ln\left(\frac{B_{(\varphi)}}{\rho\varpi}\right)>m^2\sigma_A^2\label{Eq:cri2b}
\end{eqnarray}
The inequalities (\ref{Eq:cri2a}) and (\ref{Eq:cri2b}) do not hold for
rigidly rotating stars with the magnetic field
profile~(\ref{Eq:NMAG}).  However, the angular velocity profile of any
star never remain constant in a strict sense, because the stars in
nature usually oscillate around their equilibria which makes a
gradient in the angular velocity profile. In the presence of negative
angular velocity gradient, the instability criterion~(\ref{Eq:cri2a})
may be satisfied because the first term in the left-hand side of
Equation~(\ref{Eq:cri2a}) can be the most dominant one due to the
weakness of the magnetic field as imposed in
Equation~(\ref{Eq:weak_mag}); for a region in which the Alfv\'{e}n
velocity is much smaller than the rotational velocity, a small
perturbation in the angular velocity is sufficient for satisfying the
inequality (\ref{Eq:cri2a}).  We will discuss this issue in
Section~\ref{subsec:rot} in more detail.

\section{Result}\label{Sec:Result}

\subsection{Non-rotating case}

We study the stability for two non-rotating models listed in
Table~\ref{Tab:model}.  The simulations are performed for a
sufficiently long time, more than ten times of the averaged Alfv\'{e}n
time scale or several thousands of time in units of $M_0$
(cf. Table~\ref{Tab:model}). This is necessary to clarify whether or
not any MHD instability, which grows approximately on an Alfv\'{e}n
time scale, sets in and to determine the final fate after the onset of
the instabilities. Figure~\ref{Fig:fig2} plots the evolution of the
central rest-mass density and the minimum value of the lapse function,
which characterize the compactness of the neutron star.  For the
non-rotating model N22H5, it is observed that the central density (the
minimum value of the lapse function) increases (decreases) for $t
\lesssim 200 M_0$ and subsequently decreases (increases) for $200M_0
\lesssim t \lesssim 600M_0$ and then stably oscillates for $t \gtrsim
600 M_0$.  This suggests that the star contracts first, then expands
slightly, and finally settles to another quasi-equilibrium state with
oscillations.  Note that the final central density is not beyond the
marginally stable point, $\rho_c \approx 0.318$. For model N22H1, the
qualitative features of the evolution are essentially the same as
those of N22H5, but the time scale is different from that of N22H5
because of the difference in the magnetic field strength.  The
behavior described above is well explained by the variation of
magnetic fields during the evolution as argued below. 

Figures~\ref{Fig:fig3} and \ref{Fig:fig4} plot the evolution of the
rest-mass density and magnetic energy density on the equatorial plane
and in one of meridian ($x$-$z$) planes, respectively, for model
N22H5.  The panels (a)--(c) in these figures show that the magnetic
field near the stellar surface is disturbed by the Parker instability, and
leaks out of the stellar surface. Because the plasma beta
is small and thus the matter inertia is small near the stellar surface
(as shown below), the matter is dragged by the magnetic force and
consequently the stellar surface is distorted.  Here, the plasma beta
is the ratio of the fluid pressure to the magnetic pressure, 
\begin{eqnarray}
\beta_{\rm plasma}\equiv\frac{8\pi P}{b^2} \propto 
\frac{\rho^2}{b_{(\varphi)}^2} \propto \frac{1}{\gamma_{\varphi\varphi}}
\end{eqnarray}
where we have used Equations (\ref{Eq:EOS}) and (\ref{Eq:MAG})
assuming $\Gamma=2$ and $k=1$.  The minimum value of the plasma beta
is initially $\approx 2 $ at the stellar surface, and after the onset
of the Parker instability, the leak-out magnetic field loop produces
even lower beta plasma near the stellar surface. As a result, a weak
wind expanding outward is driven. On the other hand, the ingoing
magnetic field loop enhances a turbulent motion in the neutron star.
During the transition from the state shown in panel (c) to (d) in
Figures~\ref{Fig:fig3} and~\ref{Fig:fig4}, the initial magnetic field
profile is completely destroyed and turbulent magnetic field is
produced. During the development of the turbulence, the Tayler
instability does not appear to play an important role. However, 
the region near the axis of $\varpi=0$ is not stable against 
this instability and thus no mechanism seems to help stabilizing 
there. 

The toroidal magnetic fields initially prepared behave like a rubber
belt, which fastens the ``waist'' of neutron stars. The disappearance
of the coherent toroidal magnetic fields, therefore, results in the
expansion of the star as shown in the panel (d) of Figures~\ref{Fig:fig3}
and \ref{Fig:fig4}.  After the magnetic field becomes turbulent, the
star stably oscillates around the hypothetical quasi-stationary
state. Although the density profile relaxes to a quasi-stationary
state, the turbulent motion is maintained.  We observe qualitatively
the same features for model N22H1, but the growth time scale of the
instability is longer than that of N22H5 as mentioned before. 

It is interesting to compare the simulation result with the linear analysis. 
We find the turbulent field develops in the region which is stable 
against the perturbation (see the stable region in Figure~\ref{Fig:fig1} 
and Figure~\ref{Fig:fig4} (d-f)). During the linear growth phase, i.e., Figure~\ref{Fig:fig4} 
(a-c), this region is likely to be stable. However, because the instability destroys the coherent initial 
magnetic field profile, i.e., the magnetic field is no longer pure toroidal, 
the stable region in Figure~\ref{Fig:fig1} is no longer stable after the onset of the instability. 
We also point out that the linear analysis in Figure~\ref{Fig:fig1} indicates that the higher $m$ mode is 
unstable near the stellar surface and we indeed find such a behavior (see Figure~\ref{Fig:fig3} (b)). 
For making the mode growth clear, we perform the mode analysis as follows. 
Because we are interested in the magnetic field behavior near the surface, we put a 
ring on the equator whose radius is nearly equal to the stellar radius. Then, the Fourier components are 
defined by
\begin{eqnarray}
C_m = \int^{2\pi}_0 b^2(r_{\rm ring},\pi/2,\varphi) {\rm e}^{im\varphi} d\varphi,
\end{eqnarray}
where $r_{\rm ring}$ is the ring radius which is chosen as $4M_0$ for model N22H5. 
Figure~\ref{Fig:fig5} (a) plots the evolution of $|C_m|$ with $1\le m \le 10$. It is found that 
all the mode (except for $m=4$ and $8$) start growing at $T \approx 150 M_0$, that shows that 
the Parker instabilities are in operation. For $m=4$ and $8$, $|C_m|$ starts growing 
at $T \approx 50 M_0$. This is the artifact due to our choice of the Cartesian coordinate grid.

The nonlinear growth of the Parker instability is well reflected in
the energy components as shown in Figure~\ref{Fig:fig5}. 
Here, the internal energy (\ref{eq:int}) is separated into the
adiabatic part $U_{\rm ad}$ and the heating part $U_{\rm heat}$, and
they are defined by replacing $\varepsilon$ in Equation~(\ref{eq:int})
to $\varepsilon_{\rm ad} \equiv \rho^{\Gamma-1}/(\Gamma-1)$ and to
$\varepsilon-\varepsilon_{\rm ad}$, respectively.
Figure~\ref{Fig:fig5} shows that the ADM mass is approximately
conserved, that implies that gravitational waves are not substantially
emitted during the evolution of the neutron stars.  The magnetic
energy gradually decreases and the kinetic energy increases during
$t\lesssim 500 M_0$ for model H22M5.  This illustrates that the
meridional circulation is excited due to the instability:
Figures~\ref{Fig:fig3} and~\ref{Fig:fig4} specifically show that this
meridional circular motion is induced by the displacement of the
toroidal magnetic field line. In other words, the magnetic field helps 
increasing the kinetic energy of the fluid element by liberating the gravitational potential energy. 
During this procedure, the kinetic energy
eventually reaches about ten percents of the magnetic energy at
$t\approx 200 M_0$.

For $t\gtrsim 500 M_0$, the decrease rate of the magnetic energy
becomes high and simultaneously the thermal energy ($U_{\rm heat}$)
quickly increases.  This indicates that the shock heating occurs
because of the turbulent motion induced by the distorted magnetic
fields.  The kinetic energy of the circulation gradually decreases,
and the magnetic and thermal energies settle approximately to constant
values at a late stage of the evolution. The adiabatic internal energy
also settles to a value lower than the initial one at $t\approx 500
M_0$. This implies that the density distribution changes, as shown in
Figures~\ref{Fig:fig3} and~\ref{Fig:fig4}.  As we pointed out above,
however, the magnetic field configuration never reaches to any
equilibrium state in contrast to the axisymmetric
case~\citep{Kiuchi:2008ss}.

All the features found in the evolution of the energy components for
model N22H1 are essentially the same as for model N22H5 except for the
relevant time scale, as found from Figures~\ref{Fig:fig2} and
\ref{Fig:fig5}.  The variation time scales of the various components
of the energy for model N22H5 are systematically shorter than those
for model N22H1.  This is natural because the instability develops
approximately in the Alfv\'{e}n time scale.

As found from Figure~\ref{Fig:fig4}, the meridian circulation is
excited by the magnetic field instability. Thus, the growth time scale
of the kinetic energy in the rising-up phase ($t \lesssim 200M_0$)
should be of order of the Alfv\'{e}n time scale. To check if this is
indeed the case, we evaluate the growth time of the kinetic
energy. For this evaluation, the data of $t/M_0 \in [50,100]$ and of
$t/M_0 \in [50,200]$ are used for models N22H5 and N22H1, and are
fitted assuming the function form of $\propto e^{t/\tau_k}$ where
$\tau_k$ is the growth time. We find that $\tau_k \approx 14M_0$
and $\approx 22M_0$ for models N22H5 and N22H1, respectively. Because
these values are substantially smaller than the averaged Alfv\'{e}n
time scale given in Table~\ref{Tab:model}, $\bar{\tau}_A$ is less
appropriate for characterizing the growth time scale of the magnetic
field instability. 
Rather, we find it appropriate to employ Alfv\'{e}n time scale in the vicinity of 
the stellar surface because both the linear analysis and 
our simulations indicate that the instability primarily grows there. 
The Alfv\'{e}n time scale $\tau_A$ estimated at $r/r_s=0.95$ is 
$60M_0$ for model N22H5 and $120M_0$ for model N22H1. 
Furthermore, the linear analysis suggests that the mode of $m \sim O(10)$ is 
most unstable and we find such a feature in our simulations as discussed 
above. Because the growth rate is proportional to $m$ as we see in Equation~(\ref{Eq:Disperse}), 
the growth time scale should be given by $\tau_A/m$, $\sim 6M_0$ for model N22H5 
and $12M_0$ for model N22H1, which agrees with $\tau_k$ within a factor of 
two. In any case, the growth time scale of the instability is
approximately proportional to the Alfv\'{e}n time scale normalized by $m$. 
Therefore, we can conclude that the primary instability is the Parker instability as
expected in the linear analysis. 

\subsection{Rotating case}\label{subsec:rot}

We study the stability for two rigidly rotating models listed in
Table~\ref{Tab:model}.  Again, long-term simulations are performed as
in the non-rotating models.  In Figure~\ref{Fig:fig2}, the evolution
of the central density and the minimum value of the lapse function for
models R22H2T8 and R22H08T8 are plotted together. We note again that
for both models, the angular velocity is approximately as large as the
Kepler limit at the equatorial stellar surface.  The central density
for model R22H2T8 (R22H08T8) remains approximately constant until
$t\approx 800 M_0 (1000M_0)$, and then, gradually increases for $t
\lesssim 2000M_0 (2400M_0)$. Eventually, it settles to a new 
value for $t \gtrsim 2000 M_0 (2400M_0)$.  The reason that 
the final central density is larger than the initial one will 
be described below. 

Although the magnetic field strength and central density for the
rotating model R22H08T8 are comparable with those for the non-rotating
model N22H1, the evolution process in the central region is
significantly different. For model R22H08T8, the central density
remains approximately constant for a time much longer that for model
N22H1. This is due to the fact that the rotation stabilizes the Tayler
instability which may set in for the non-rotating models, as expected
in the local linear perturbation analyses (see
Section~\ref{sec:linear_rot}): The Tayler instability is a primary
magnetic field instability associated with the toroidal field near the
axis of $\varpi=0$ for the non-rotating models, but this is not the
case for the rapidly rotating models. 

However, the instability is not suppressed by the presence of rapid
rotation for all the places inside the neutron star, as
Figure~\ref{Fig:fig2} illustrates that the central density varies for
a longer-term evolution with $t \approx 2000M_0$. This result is
totally different from that in the axisymmetric
case~\citep{Kiuchi:2008ss}, in which the rapid rotation suppresses the
onset of the interchange instability.  Note that one of the models
studied in~\citet{Kiuchi:2008ss} has similar model parameters to those
of R22H08T8, which is stable for the axisymmetric perturbations.  This
implies that a non-axisymmetric instability is excited for the rapidly
rotating models this time.  To clarify the relevant instability, we
generate Figures~\ref{Fig:fig6} and~\ref{Fig:fig7}, in which the
rest-mass density and the magnetic energy density for model R22H2T8
are plotted on the equatorial and meridian planes, respectively.
Until $t \approx 200 M_0$, we observe that 
the coherent magnetic field and density profiles remain. 
However, at $t
\approx 400 M_0$, the magnetic field profile {\em near the stellar surface} is
deformed in the same manner as found for the non-rotating
models. During the subsequent evolution, the coherent magnetic field
structure is totally destroyed and a turbulent motion is excited.  The
surface expands because the plasma beta near the surface is below
unity due to the leak-out of the magnetic field. Because the
coherent toroidal magnetic field, which fastens the waist of the neutron
star, disappears, the radius of the neutron star increases.  All these
features are essentially the same as for the non-rotating models.

The evolution of the magnetic field near the rotation axis in the
rotating models is slightly different from that in the non-rotating
models. Figures~\ref{Fig:fig8} (a) and (b) plot the snapshots of the
magnetic pressure defined by $b^2/8\pi$ along the $x$ and $y$-axes on the
equator for models R22H2T8 and N22H5, respectively. 
In the rotating 
model R22H2T8, the profile near the center does not drastically change.
 However, the magnetic pressure increases gradually and systematically, and by the
increased magnetic pressure, the matter near the axis is pinched. 
This increase proceeds in the axisymmetric way because 
the profiles along $y$-axis is approximately identical to those along $x$-axis 
in Figure~\ref{Fig:fig8} (a). This might seem to be incompatible with our previous results~\citep{Kiuchi:2008ss}. 
To clarify this point, 
we reconsider the criterion for the axisymmetric perturbation :
\begin{align}
&4\Omega^2 \left(1+\frac{v_A^2}{c_s^2}\right)
+\left(\frac{g_\varpi}{c_s^2}-\frac{2}{\varpi}\right)v_A^2 \partial_\varpi 
\ln\left(\frac{B_{(\varphi)}}{\varpi \rho}\right) <0  \label{Eq:stab-a1}\\
& g_z \frac{v_A^2}{c_s^2}\partial_z \ln\left(\frac{B_{(\varphi)}}{\varpi \rho}\right) < 0\label{Eq:stab-a2}\\
& 2 g_z \varpi \Omega^2 \frac{v_A^2}{c_s^2}\partial_z \ln\left(\frac{B_{(\varphi)}}{\varpi \rho}\right) < 0
\label{Eq:stab-a3}.
\end{align}
The axisymmetric instability ignites if any of these inequalities holds. 
Equations ~(\ref{Eq:stab-a2}) and (\ref{Eq:stab-a3}) implies 
the magnetic field profile given by Equation (\ref{Eq:NMAG}) is marginally stable. 
The magnetic field line strength evolves as
\begin{align}
(\partial_t + {\rm e}^{i(a)} v_{(a)} \partial_i)\left(\frac{B_{(\varphi)}}{\varpi \rho}\right)
=\frac{{\rm e}^{i(a)} B_{(a)}}{\rho}\partial_i \Omega,
\end{align}
where ${\rm e}^{i(a)}$ is a tetrad basis. In axisymmetric simulations, 
the field line strength is completely preserved because 
the poloidal magnetic field, i.e. $B_{(\varpi)}$ and $B_{(z)}$, is never generated 
if the initial magnetic field is purely toroidal. Then, the inequalities (\ref{Eq:stab-a2}) and (\ref{Eq:stab-a3}) 
are never satisfied during the evolution of the magnetized star. 
On the other hand, the field line strength deviates from its initial value in three-dimensional simulations 
because the poloidal field is generated by a perturbation induced by a truncation error in general (in addition to the change of 
the angular velocity profile as discussed below). As the result, the inequalities (\ref{Eq:stab-a2}) 
and (\ref{Eq:stab-a3}) may hold during the evolution and the axisymmetric instability could set in. 
This is the reason why the magnetic pressure increases and angular velocity is distorted near the center 
as seen in Figures~\ref{Fig:fig8} (a) and (c). Note that the things are the same in model R22T08T8.
 It is interesting to note that the stellar surface in the rotating model expands whereas
the central part contracts due to the redistribution of the magnetic
field profile. Contrary to the rotating model, the coherent profile of
the magnetic fields near the stellar center in the non-rotating model
N22H5 disappears after the onset of the instability, which results in
the systematic expansion of the star as mentioned in the previous
subsection. 
The non-rotating models are also marginally stable against the interchange instability 
as shown in Equation~(\ref{Eq:stab-a2}). However, these models are unstable against the non-axisymmetric 
mode (see Figure~\ref{Fig:fig1} (b)) and this mode overcomes the the interchange mode. 
It is also interesting to note that the non-axisymmetric
instability develops in an earlier time than that expected from the
central density's evolution, e.g., at $t\approx 800M_0(1000M_0)$ for
R22H2T8 (R22T08T8) (see Figure~\ref{Fig:fig2} (a)). 
However, the instability develops before $t \sim 600 M_0$ for 
these rotating models as shown in Figure~\ref{Fig:fig9}. 
This also reflects the fact that the instability,
which affects the global structure of the neutron star, sets in 
both near the stellar surface and in the central region. 
Namely, the Parker (interchange) instability plays an important role  
in the outer (central) region for the rotating models. 

The local linear perturbation analysis predicts that our rigidly
rotating models are stable against the non-axisymmetric perturbation 
as described in
Section~\ref{sec:linear_rot}.  Thus, our numerical result does not
seem to agree with that in the linear analysis.  However, this is not
the case, because our rotating model is close to a marginally stable
state and is destabilized by a slight nonlinear perturbation to the
rotational velocity. To explain this fact, we plot the snapshots of
the angular velocity in Figure~\ref{Fig:fig8}, in which the angular
velocity profiles along the $x$-axis on the equator are plotted for
models R22H2T8 and R22H08T8. This shows that the angular velocity
deviates slightly from the constant profile and the negative
gradients, in particular, near the stellar surface are developed
during the evolution.  This time variation is due to the oscillation
of the neutron star which is initially triggered by a perturbation of
numerical origin. Note that the initial conditions we gave are in
equilibria, but a small numerical error associated with the finite
grid resolution induces a perturbation and then the neutron stars
start oscillating around their equilibrium states. Although the
perturbation is induced by a numerical error in this case, it is quite
natural to expect that any star in nature oscillates and thus the
precisely rigid rotation is not realized. Once the angular velocity
profile has the negative gradient, the instability criterion
(\ref{Eq:cri2a}) could be satisfied because the first term in the
left-hand side has a substantially large positive value in the rapidly
rotating models with weak magnetic fields, i.e., the criterion is
satisfied even for a small angular velocity gradient. 

The growth process of the instability is well reflected in several
energy components. Figure~\ref{Fig:fig9} plots their evolution.  This
shows that the ADM mass, adiabatic internal energy, and rotational
kinetic energy remain approximately constant during the evolution. The
decrease in the magnetic energy is prominent at $t\approx 1500M_0$ for
model R22H2T8 and $t\approx 2000 M_0$ for R22H08T8. 
Note that the magnetic energy increases only by $\sim 2\%$ during the development 
of the interchange instability, e.g, for $t\lesssim 400 M_0$ for 
model R22H2T8. The kinetic
energy, in which contribution of the rotational kinetic energy is
excluded, increases with time. This shows that the meridian motion is
developed. We find again that the rapid rotation is not enough to
stabilize the instability associated with the presence of toroidal
magnetic fields and the magnetic field never reaches an equilibrium
state as in the non-rotating models. 

It is well known that negative gradient of angular velocity 
in the presence of magnetic field leads to the MRI~\citep{Balbus:1991}. 
As discussed above, the negative angular velocity appears in the vicinity 
of the stellar surface. Hence, we hypothetically estimate the MRI growth rate 
as $\sigma_{\rm MRI} \sim \Omega \varpi |\partial_\varpi \ln \Omega|$. 
We fit the angular velocity profile as a function of the radius in Figure~\ref{Fig:fig8} (c-d) 
and obtain the gradient $\partial_\varpi \ln \Omega$. Putting the value of the 
gradient, angular velocity, and the stellar radius into the equation, we estimate the growth rate 
as $\sigma_{\rm MRI} M_0 \approx 0.09$ $(0.1)$ for model R22H2T8 (R22H08T8). 
On the other hand, we infer the instability growth rate from the increase 
of the kinetic energy shown in Figure~\ref{Fig:fig9} with the same strategy for 
the non-rotating case. The resulting growth rate $\sigma M_0$ is $0.008$ $(0.007)$ for 
model R22H2T8 (R22H08T8), where we use the data of $t/M_0 \in [0,600]$ in 
both the models. Therefore, the growth rates do not match. 
However, Equations~(\ref{Eq:Disperse}) and (\ref{Eq:growth}) tell us that the growth rates of the 
interchange and Parker instability depend on the magnetic field strength. Hence, we conclude 
that the instability we find in the rotating model is not the MRI.

Before closing this section, we show that the convergence is achieved
in our numerical results with an accuracy high enough to draw a
quantitatively reliable conclusion. Figures~\ref{Fig:fig10} and
\ref{Fig:fig11} plot the evolution of (a) the central density, (b) the
minimum value of the lapse function, (c) the ADM mass, (d) the
internal energy, (e) the magnetic energy, (f) the kinetic energy, and
(g) the Hamiltonian constraint violation for models N22H5 and R22H2T8.
The Hamiltonian constraint violation is defined by
\begin{eqnarray}
{\rm ERROR}=\frac{1}{M_b}\int \rho w \sqrt{\gamma} |V| d^3 x
\end{eqnarray}
with
\begin{eqnarray}
V \equiv \frac{\tilde{\Delta}\psi-\frac{\psi}{8}{\tilde{R}^k}_k
+2\pi \rho_{\rm H}\psi^5 +\frac{\psi^5}{8}
\left(\tilde{A}_{ij}\tilde{A}^{ij}-\frac{2}{3}K^2\right)}
{|\tilde{\Delta}\psi|+|\frac{\psi}{8}{\tilde{R}^k}_k|
+|2\pi \rho_{\rm H}\psi^5| +\frac{\psi^5}{8}
\left(\tilde{A}_{ij}\tilde{A}^{ij}+\frac{2}{3}K^2\right)}
\end{eqnarray}
where $\tilde{\Delta}$ is the Laplacian associated with
$\tilde{\gamma}_{ij}$, $\psi$ is the conformal factor, and
$\rho_H=(\rho h + b^2/4\pi)w^2-(P+b^2/8\pi)-(\alpha b^t)^2/4\pi$.  In all
the panels, we plot the results of three different grid
resolutions. In Figure~\ref{Fig:fig10} (a), (b), (d), and (e), we find
a transition time at $t \sim 500 M_0$ depends slightly on the grid
resolution, but on the whole, the results appear to be convergent.
The ADM mass, which is approximately conserved in the present
situation, is conserved within 99.8 percent accuracy (see
Figure~\ref{Fig:fig10} (c)) and the error in the Hamiltonian
constraint violation is in an acceptable level (less than $\sim 1$\%
error; see Figure~\ref{Fig:fig10} (g)).  In the panel (e), we observe
that the convergence is lost for $t\gtrsim 1000M_0$. 
This is likely to be caused by the turbulent field developing 
both outside and inside the star as shown in Figures~\ref{Fig:fig3} (f) 
and \ref{Fig:fig4} (f). To handle such a region and/or turbulent field, 
we would need more sophisticated numerical scheme.

From these figures, we conclude that all the qualitative features of
the magnetic field instability found in this work are independent of
the grid resolution.  In the rotating model R22H2T8, we find
essentially the same features in the convergence study. 

\section{Summary \& Discussion}\label{Sec:Summary}
\subsection{Summary}\label{subsec:summary}

We explored the non-axisymmetric instability of neutron stars
with purely toroidal magnetic fields.  Preparing the non-rotating and
rotating neutron stars in equilibrium as the initial conditions, the
three-dimensional GRMHD simulations were performed.  For the
non-rotating models, the local linear perturbation analysis predicts
that the Parker instability would be the primary instability and we
confirmed this.  Due to the Parker instability, a turbulent state is
developed and the initially coherent magnetic field profile is totally
varied. The magnetic field profile never reaches an equilibrium
state. This fact is in sharp contrast with that in the axisymmetric
instability of~\citet{Kiuchi:2008ss}. The growth time scale of the
Parker instability depends on the magnetic field strength, i.e., the
Alfv\'{e}n time scale, and this result also agrees with the local
linear perturbation analysis.  The present result strongly suggests
that three-dimensional treatment is crucial to clarify the instability
of a neutron star with toroidal magnetic fields.  In other words, any
a priori assumption of the spacetime symmetry (e.g., axisymmetric
symmetry) could prevent from deriving the correct conclusion. 

We also explored the instability of rigidly and rapidly rotating
neutron stars. The linear analyses have suggested that rapid rotation
could play a role as a stabilizing agent. We confirm that the rapid
rotation stabilizes the Tayler instability, which may occur near the
axis of $\varpi=0$ in the non-rotating case. 
However, the interchange instability could play a minor role 
because the neutron stars are marginally stable against it. 
We note that the interchange mode never develops in the axisymmetric simulations 
due to the symmetry imposed. We also find that the
Parker instability which is relevant near the stellar surface may not
be stabilized by the rapid rotation.  The reason is that by a
perturbative oscillation, neutron stars may have a region in which the
gradient in the angular velocity profile is negative ($\partial
\Omega/\partial\varpi < 0$). This negative gradient can induce the 
Parker instability in the case that the neutron star has rapid
rotation and weak magnetic fields. As in the non-rotating model, a
turbulent state is subsequently developed in the outer region of the
neutron star.  This result also gives us a message that the
three-dimensional simulation is essential for investigating a magnetic
field instability.

\subsection{Discussion}\label{subsec:discussion}

As mentioned in Introduction, a large number of core-collapse
supernova simulations based on the magneto-rotational mechanism have
shown that toroidal magnetic fields are dominantly amplified in the
proto-neutron stars via the winding mechanism. In the assumption of
axial symmetry, these fields may be in quasi-equilibrium after the
saturation is reached. However, this could disagree with the result in
the linear perturbation analysis, i.e., the neutron stars with purely
toroidal magnetic fields are often unstable. The present study indeed
suggests that such neutron stars are unstable and thus the assumptions
of axial symmetry and rapid rotation, which are imposed for most of
the magneto-hydrodynamical supernova simulations, would be 
inappropriate. (Note that in axial symmetry, the rapid rotation
stabilizes neutron stars with toroidal magnetic fields, as illustrated
in~\citet{Kiuchi:2008ss}.)  In a non-axisymmetric simulation, we may 
find that a proto-neutron star with strongly toroidal magnetic fields
is unstable and a turbulent motion inside it is excited.
Magneto-hydrodynamic simulations have to be performed in fully three
spatial dimensions.

Stability of more generic magnetic field configurations, i.e., mixed
poloidal-toroidal fields, is quite important because such fields would
be in general realized. Recently, 
\citet{Braithwaite:2005xi,Braithwaite:2005ps,Duez:2010} studied 
the stability of Newtonian stars with such fields, and reported that
the mixed field may play an important role in stabilization; they
showed that the equilibrium profile is maintained over several
Alfv\'{e}n time scales. We plan to study this issue with the weak
magnetic field solution obtained by~\citet{Ioka:2003nh} in the fully
general relativistic framework.

\begin{figure*}
  \begin{center}
  \vspace*{40pt}
    \begin{tabular}{cc}
      \begin{minipage}{0.5\hsize}
      \includegraphics[width=10.0cm]{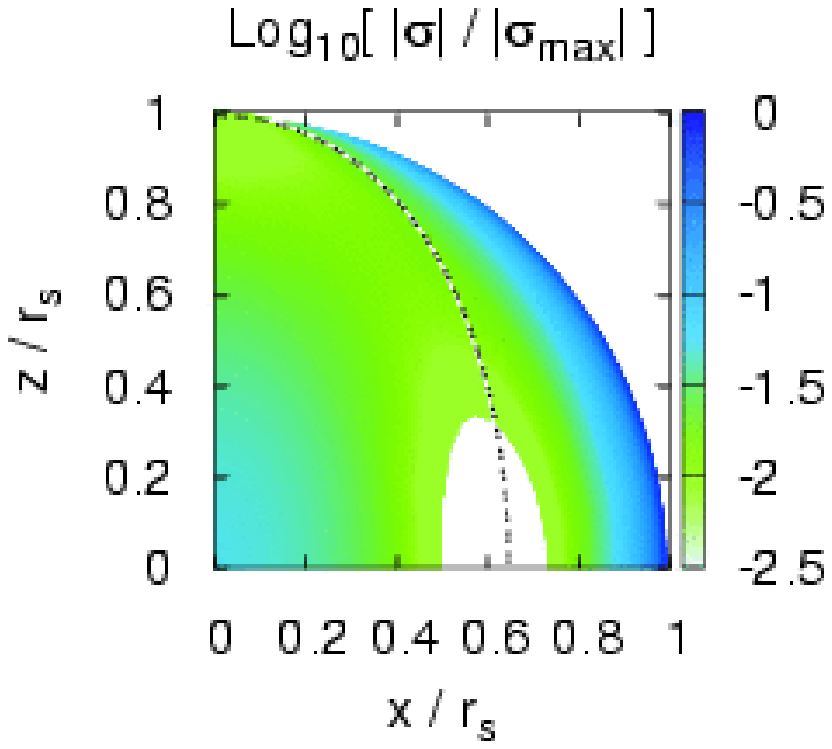}
      \end{minipage}
      \hspace{-1.0cm}
      \begin{minipage}{0.5\hsize}
      \includegraphics[width=10.0cm]{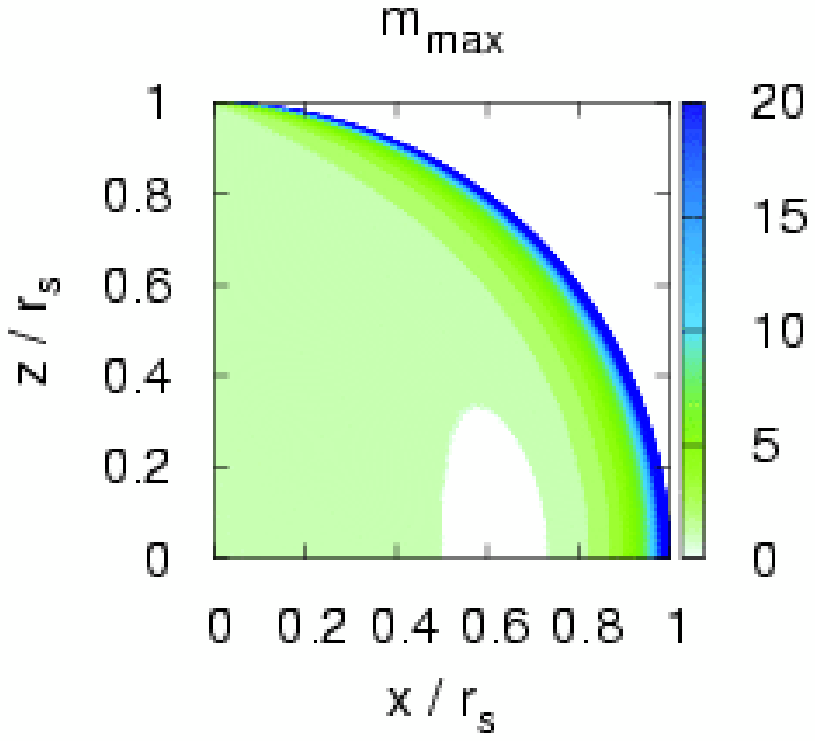}
      \end{minipage}
    \end{tabular}
    \caption{\label{Fig:fig1} The growth rate of the instability of a
    neutron star with toroidal magnetic fields normalized by the
    maximum growth rate $|\sigma_{\rm max}|$ (left) and corresponding
    $m$ mode (right) on the meridian plane for non-rotating Newtonian
    polytrope with $\rho_0=0.16$ and $B_0=0.2$.  The white colored
    region and the dotted curve denote the stable region and critical
    radius, respectively. $r_s$ denotes the stellar radius.}
    \end{center}
\end{figure*}

\begin{figure*}
  \begin{center}
  \vspace*{40pt}
    \begin{tabular}{cc}
      \begin{minipage}{0.5\hsize}
      \includegraphics[width=8.0cm]{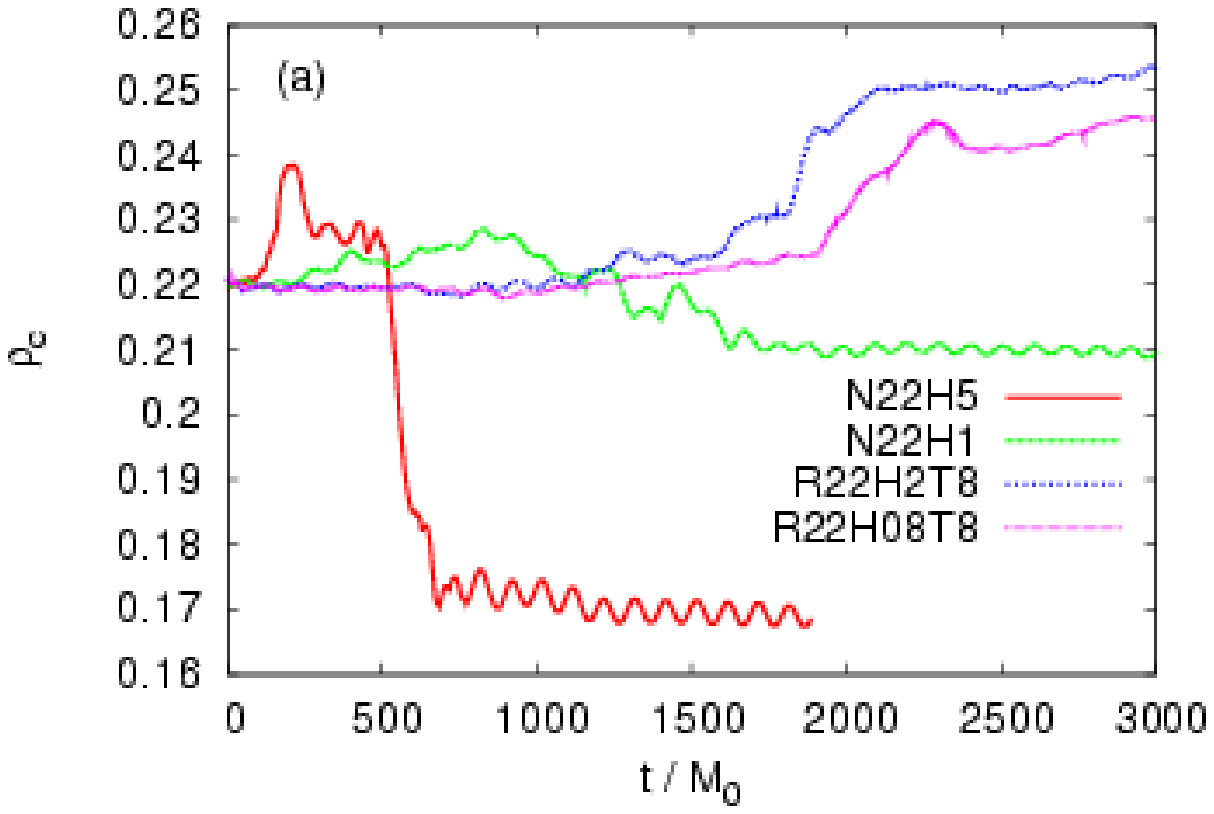}
      \end{minipage}
      \hspace{-1.0cm}
      \begin{minipage}{0.5\hsize}
      \includegraphics[width=8.0cm]{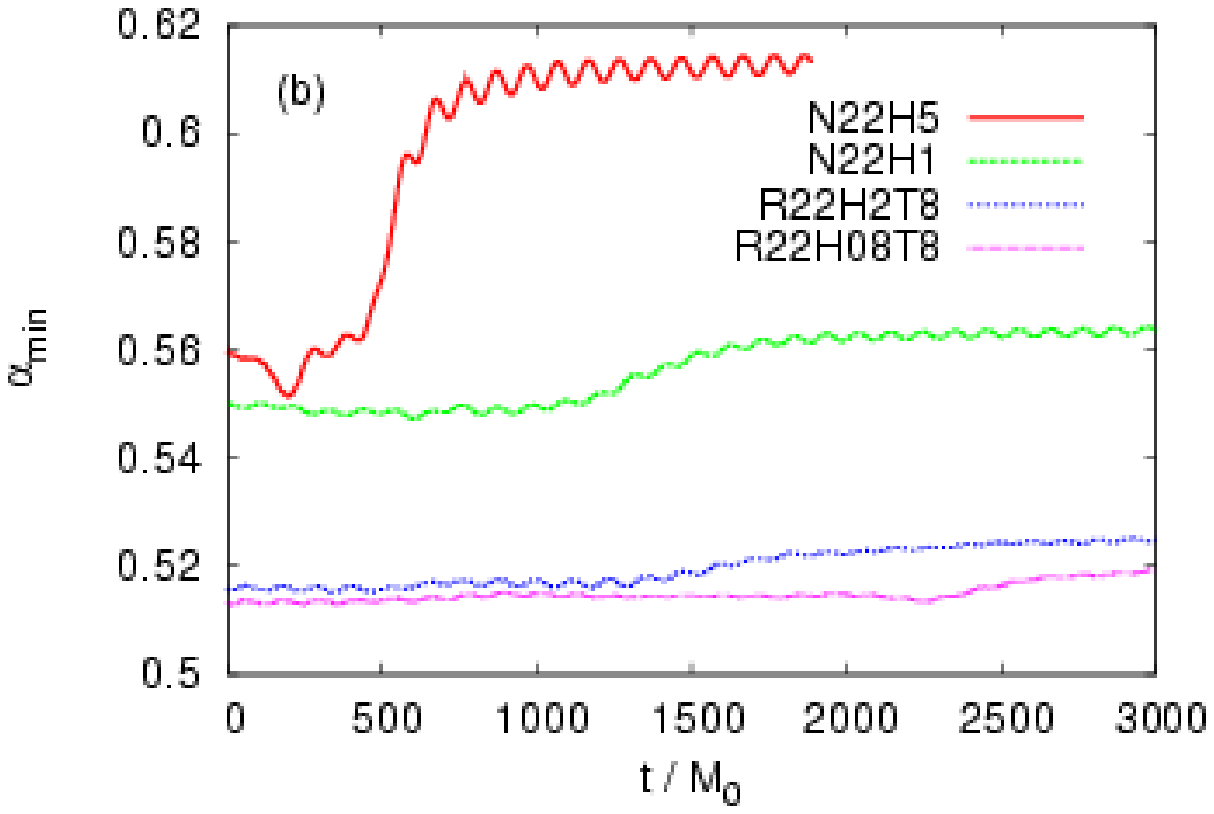}
      \end{minipage}
    \end{tabular}
    \caption{\label{Fig:fig2}
    The evolution of (a) central rest-mass density $\rho_c$ 
    and (b) minimum value of lapse function $\alpha_{\rm min}$ 
    for non-rotating models N22H1 and N22H5.
    }
  \end{center}
\end{figure*}

\begin{figure*}
  \begin{center}
  \vspace*{40pt}
    \begin{tabular}{cc}
      \begin{minipage}{0.5\hsize}
      \includegraphics[width=9.0cm]{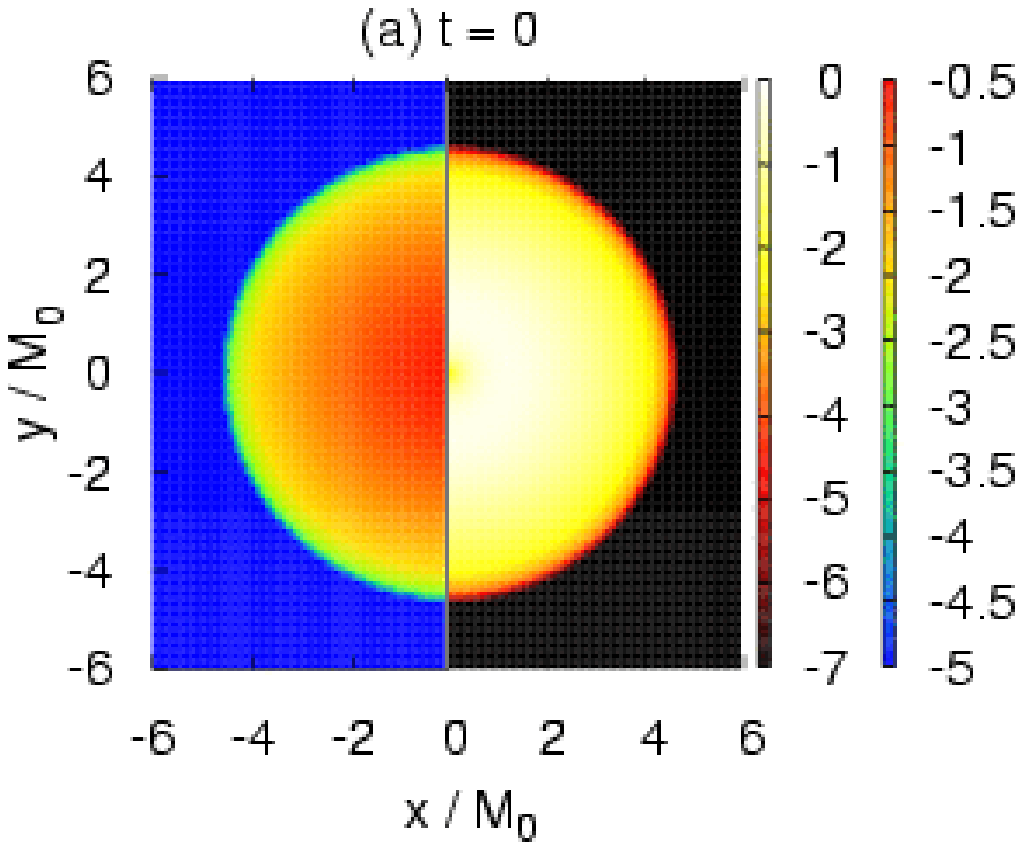}
      \end{minipage}
      \hspace{-1.0cm}
      \begin{minipage}{0.5\hsize}
      \includegraphics[width=9.0cm]{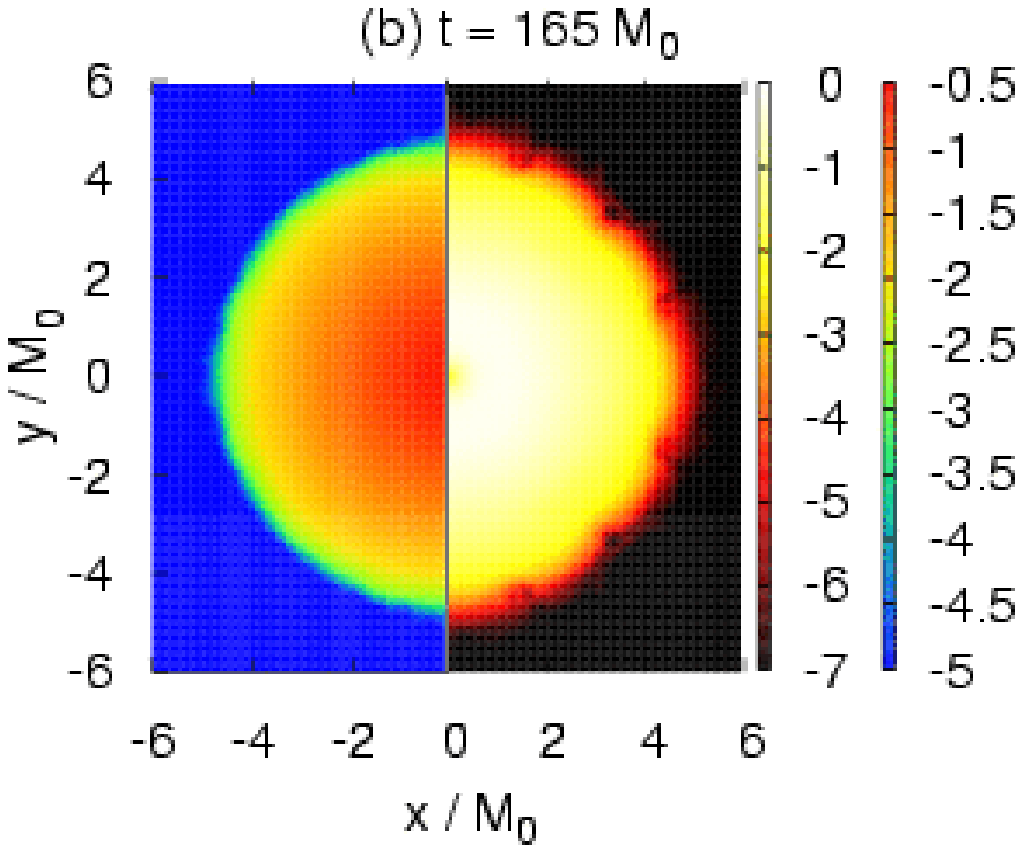}
      \end{minipage}\\
      \begin{minipage}{0.5\hsize}
      \includegraphics[width=9.0cm]{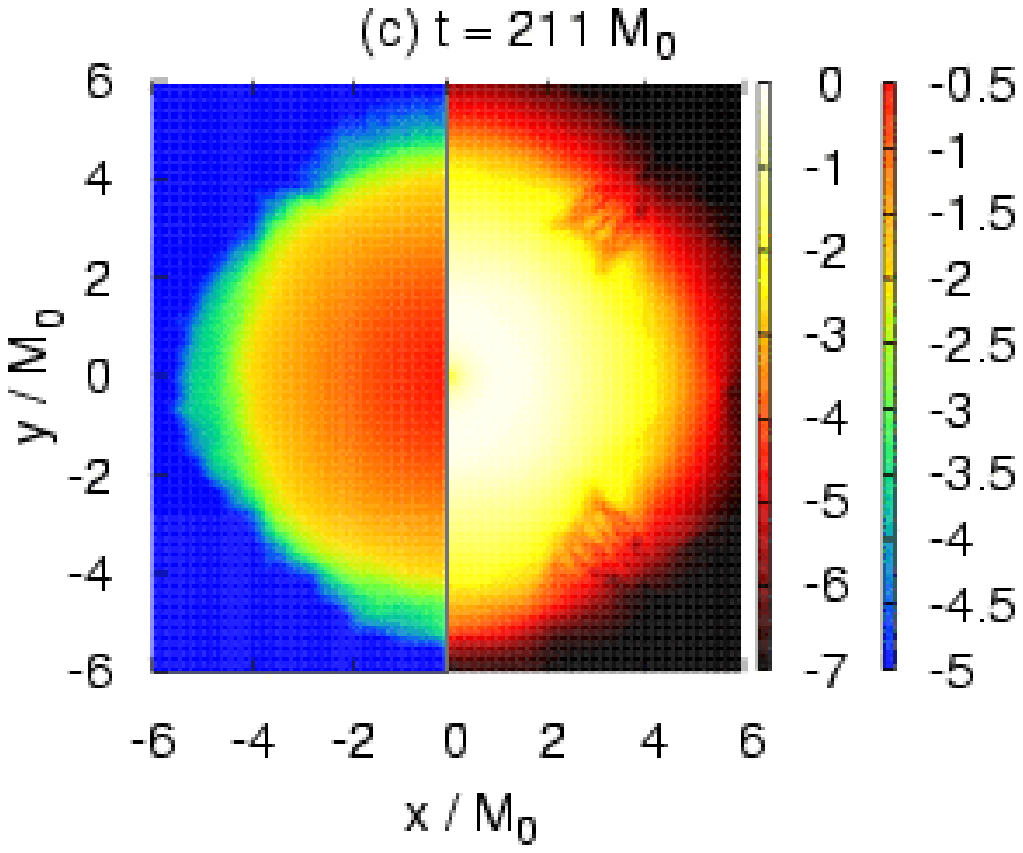}
      \end{minipage}
      \hspace{-1.0cm}
      \begin{minipage}{0.5\hsize}
      \includegraphics[width=9.0cm]{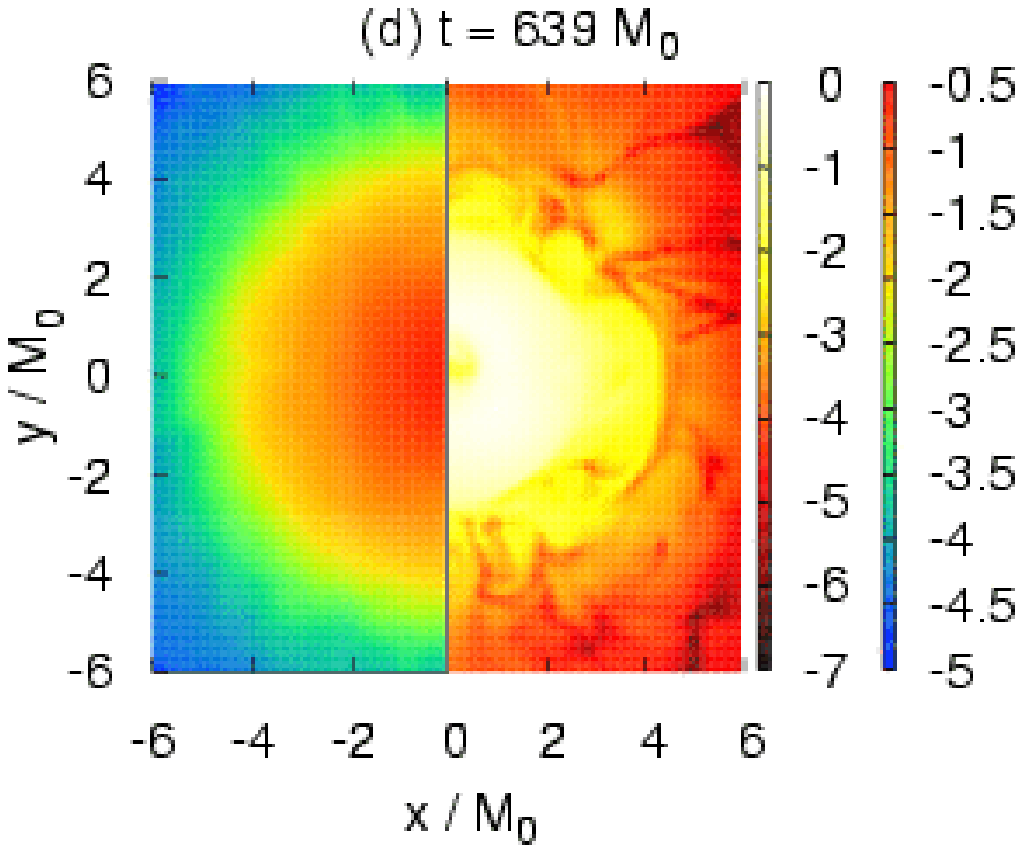}
      \end{minipage}\\
      \begin{minipage}{0.5\hsize}
      \includegraphics[width=9.0cm]{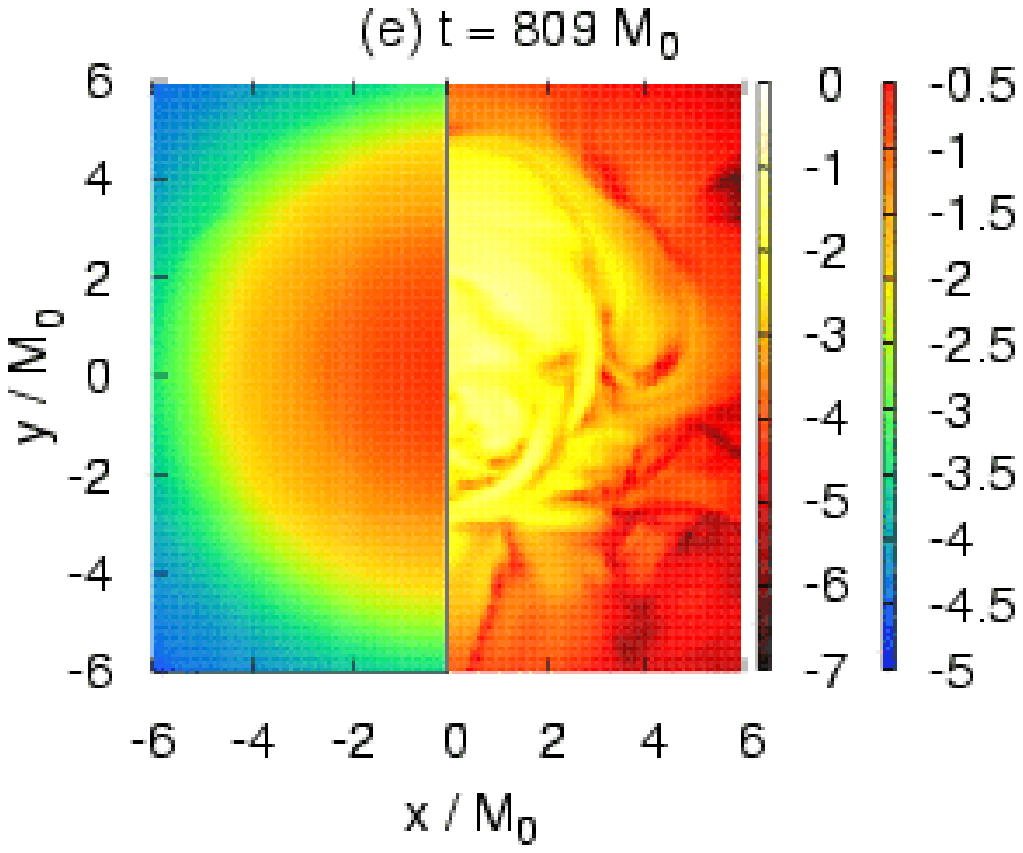}
      \end{minipage}
      \hspace{-1.0cm}
      \begin{minipage}{0.5\hsize}
      \includegraphics[width=9.0cm]{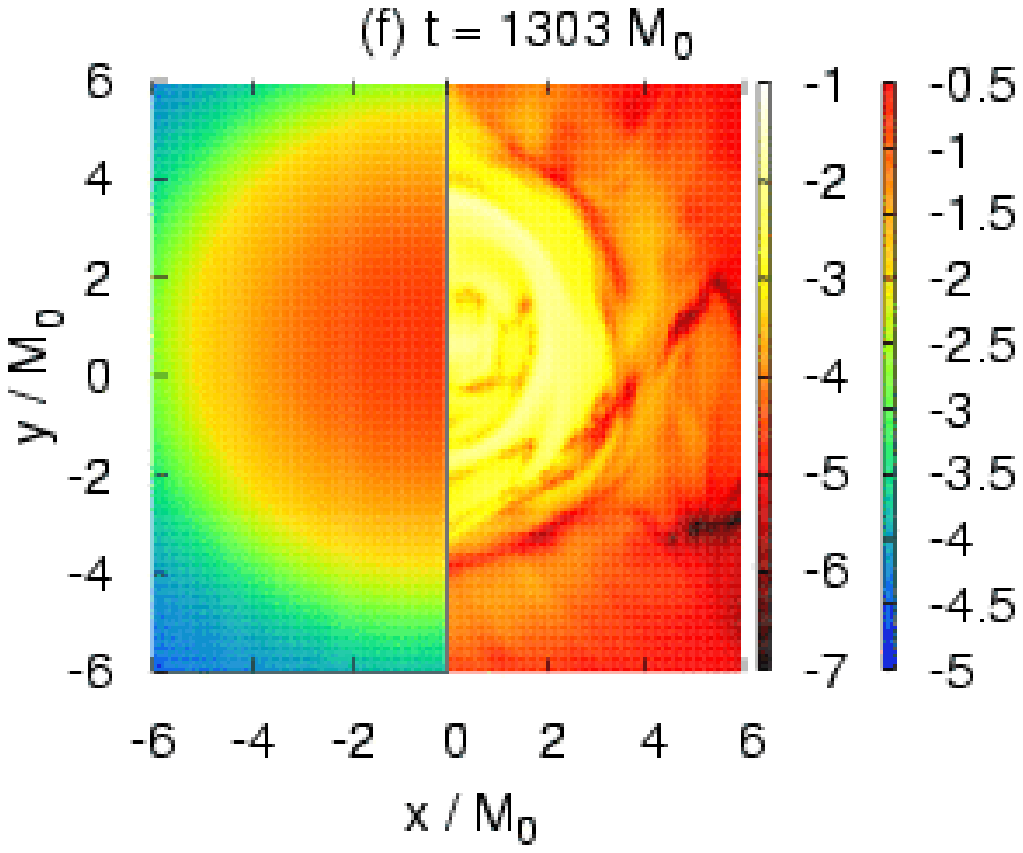}
      \end{minipage}\\
    \end{tabular}
    \caption{\label{Fig:fig3} The evolution of rest-mass density
    (left) and magnetic energy density (right) on the equator for
    N22H5. Both of them are plotted in the logarithmic scale. 
    The coordinate time at each slice is shown in each panel.  } \end{center}
\end{figure*}

\begin{figure*}
  \begin{center}
  \vspace*{40pt}
    \begin{tabular}{cc}
      \begin{minipage}{0.5\hsize}
      \includegraphics[width=9.0cm]{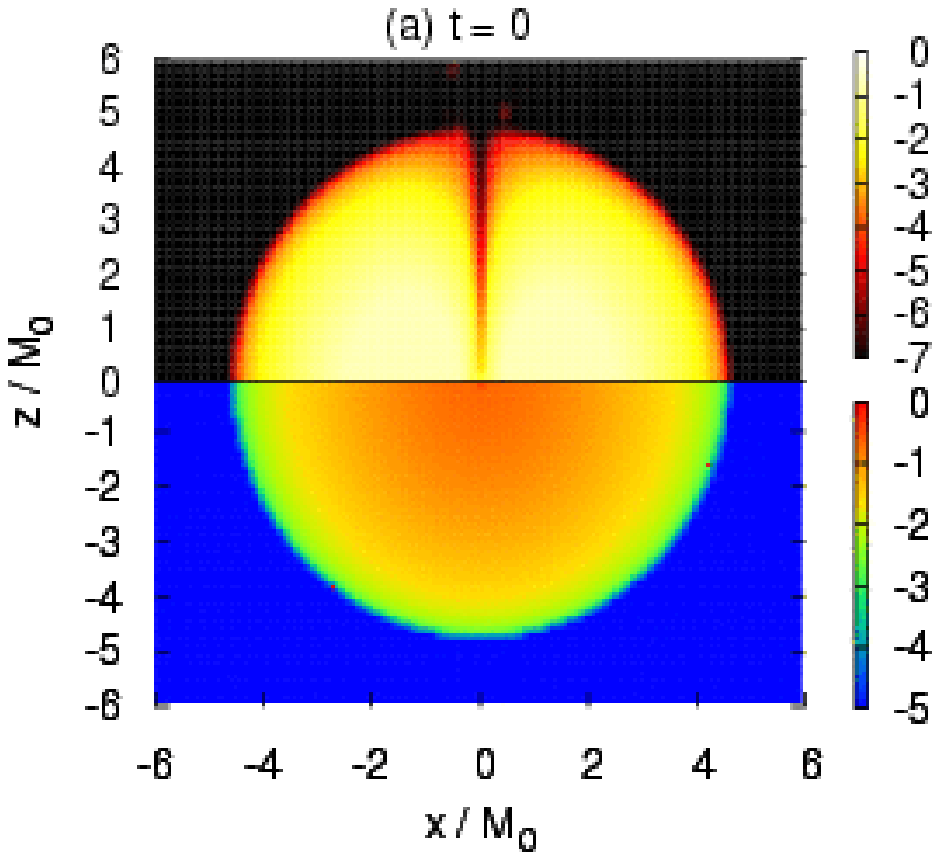}
      \end{minipage}
      \hspace{-1.0cm}
      \begin{minipage}{0.5\hsize}
      \includegraphics[width=9.0cm]{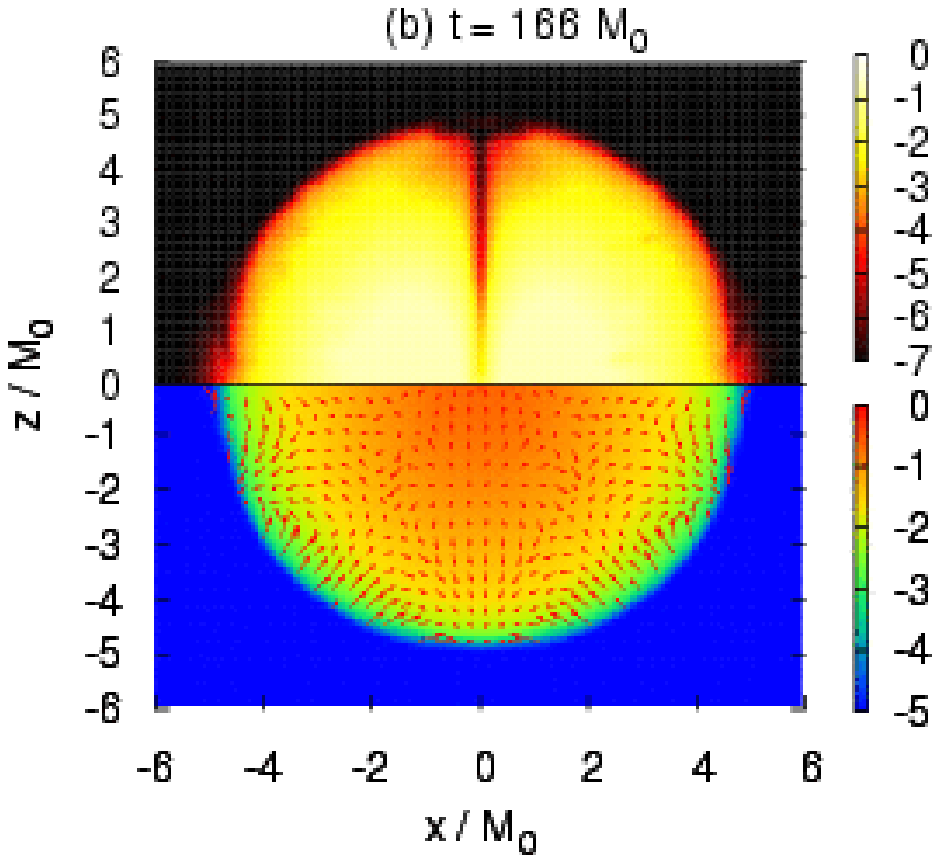}
      \end{minipage}\\
      \begin{minipage}{0.5\hsize}
      \includegraphics[width=9.0cm]{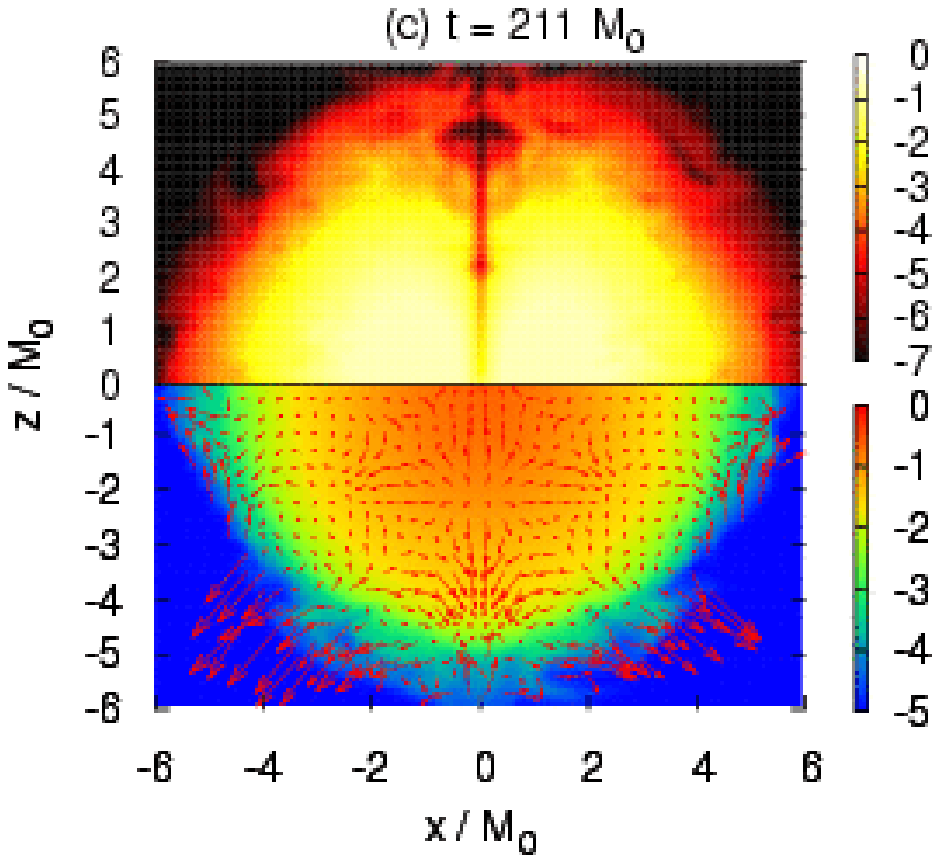}
      \end{minipage}
      \hspace{-1.0cm}
      \begin{minipage}{0.5\hsize}
      \includegraphics[width=9.0cm]{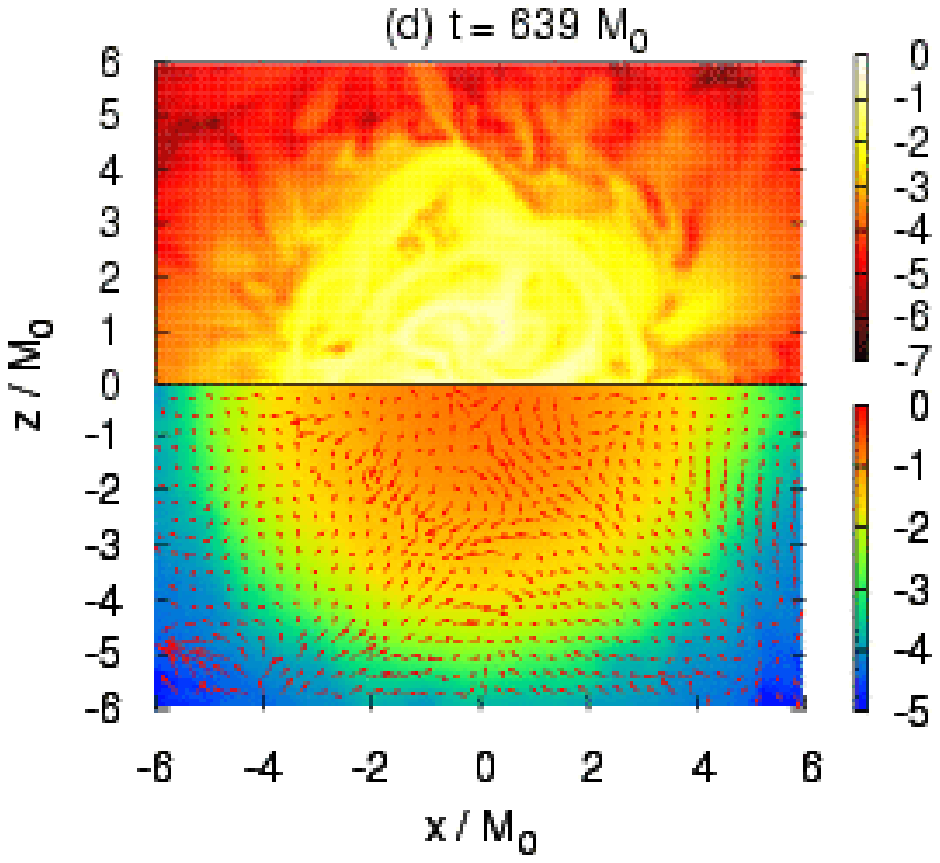}
      \end{minipage}\\
      \begin{minipage}{0.5\hsize}
      \includegraphics[width=9.0cm]{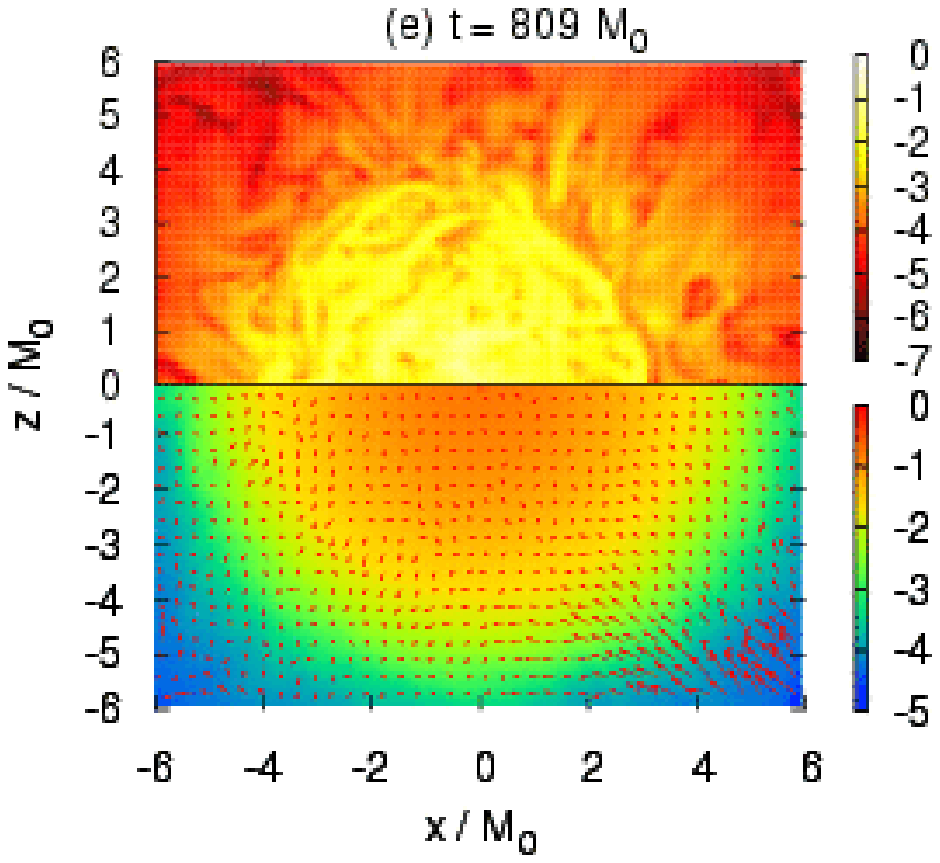}
      \end{minipage}
      \hspace{-1.0cm}
      \begin{minipage}{0.5\hsize}
      \includegraphics[width=9.0cm]{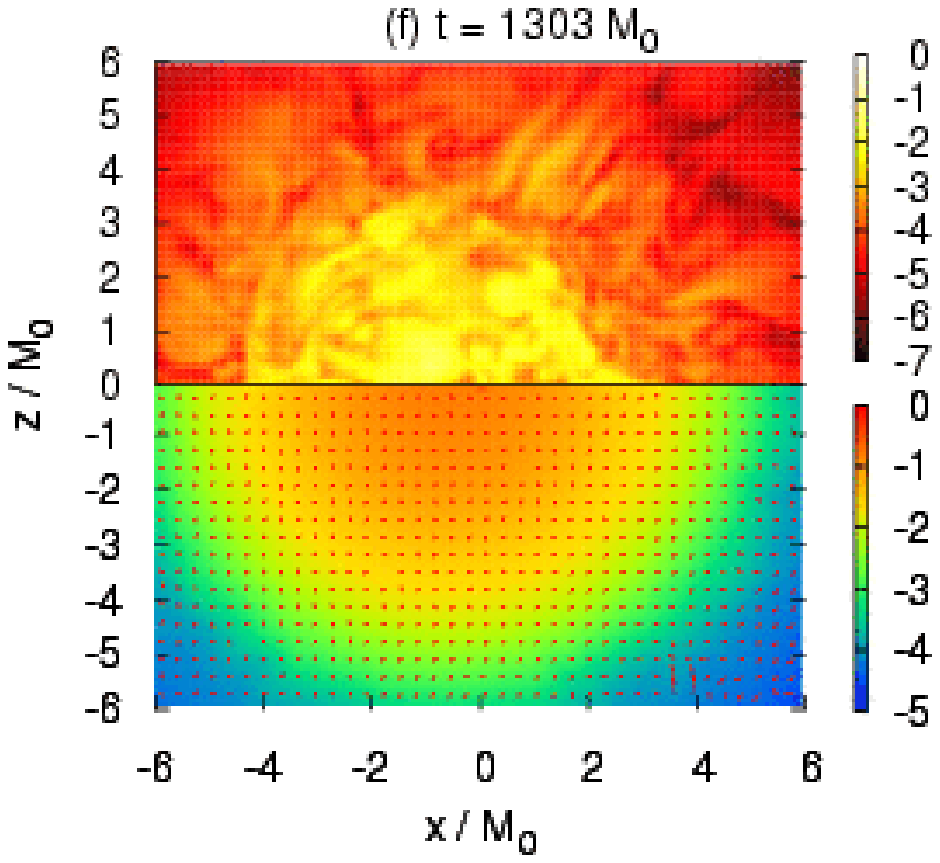}
      \end{minipage}\\
    \end{tabular}
    \caption{\label{Fig:fig4} The evolution of the rest-mass density
    (bottom) and magnetic energy density (top) on a meridian ($x$-$z$) 
     plane for N22H5 in the logarithmic scale. The arrows indicate the
     velocity fields.  } \end{center}
\end{figure*}

\begin{figure*}
  \begin{center}
  \vspace*{40pt}
    \begin{tabular}{cc}
      \begin{minipage}{0.5\hsize}
      \includegraphics[width=8.0cm]{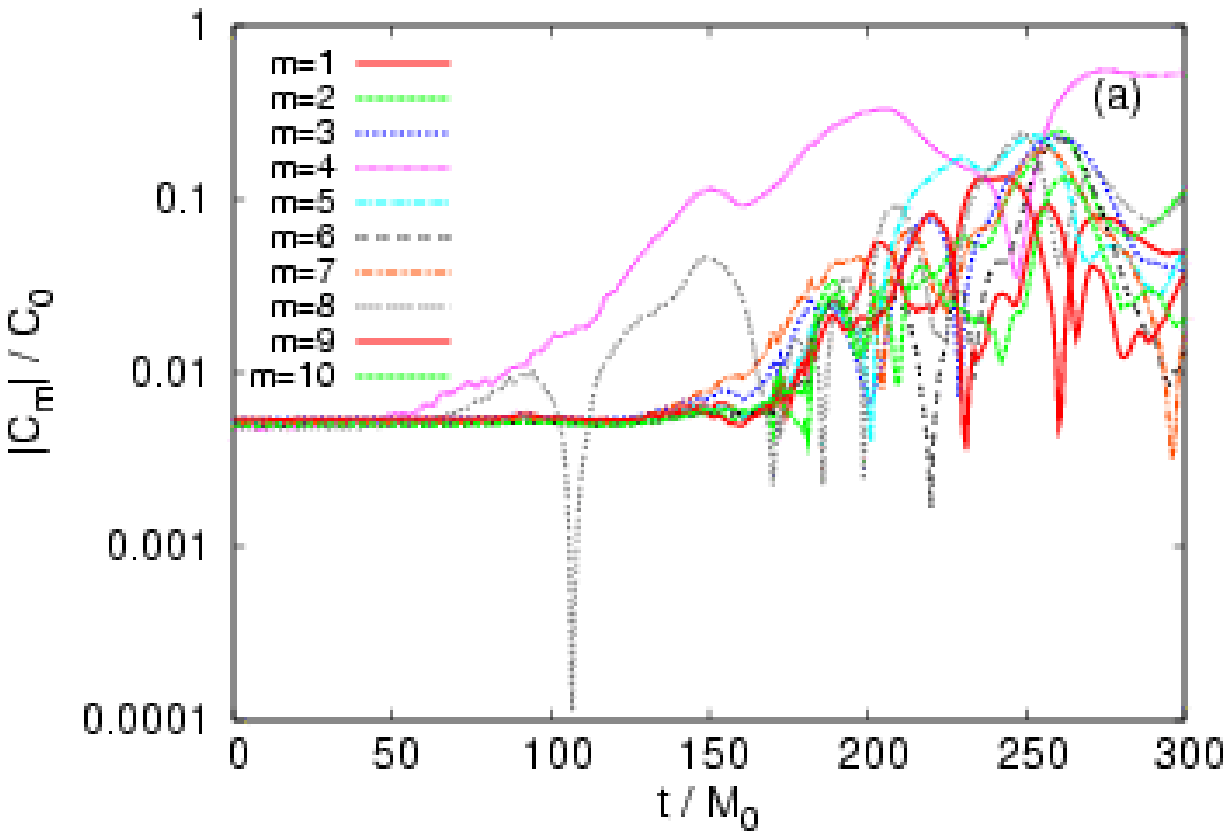}
      \end{minipage}
      \hspace{-1.0cm}
      \begin{minipage}{0.5\hsize}
      \includegraphics[width=8.0cm]{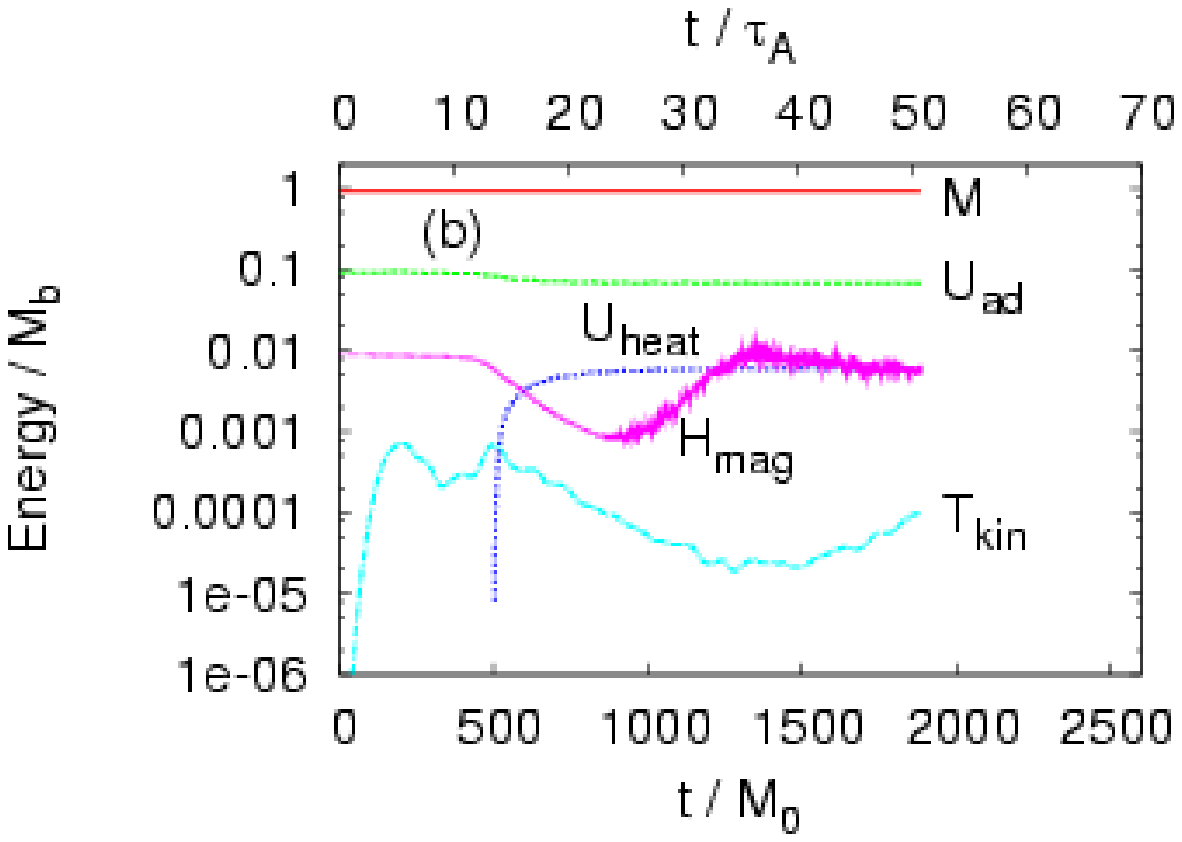}
      \end{minipage}\\
      \begin{minipage}{0.5\hsize}
      \includegraphics[width=8.0cm]{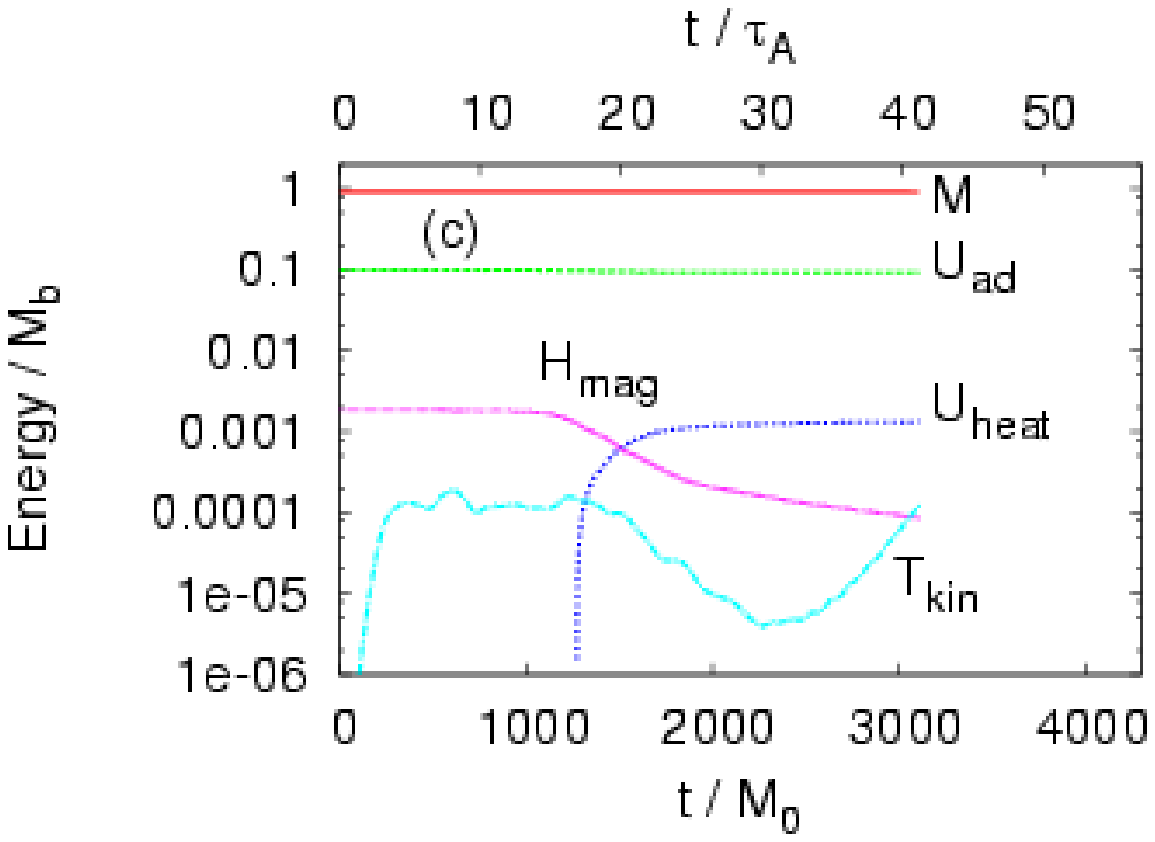}
      \end{minipage}
    \end{tabular}
    \caption{\label{Fig:fig5} 
    (a) Evolution of the Fourier mode for the magnetic energy density for model N22H5, 
    (b) evolution of several energy components for model N22H5, and (c) the same as (b) but for model N22H1. 
    All the energy
    components are normalized by the rest mass, and the time axes are
    shown in units of $M_0$ (bottom) and $\bar{\tau}_A$ (top).  Note
    that the internal energy is split into the adiabatic and heating
    parts (see the text for more details).  
    } \end{center}
\end{figure*}

\begin{figure*}
  \begin{center}
  \vspace*{40pt}
    \begin{tabular}{cc}
      \begin{minipage}{0.5\hsize}
      \includegraphics[width=9.0cm]{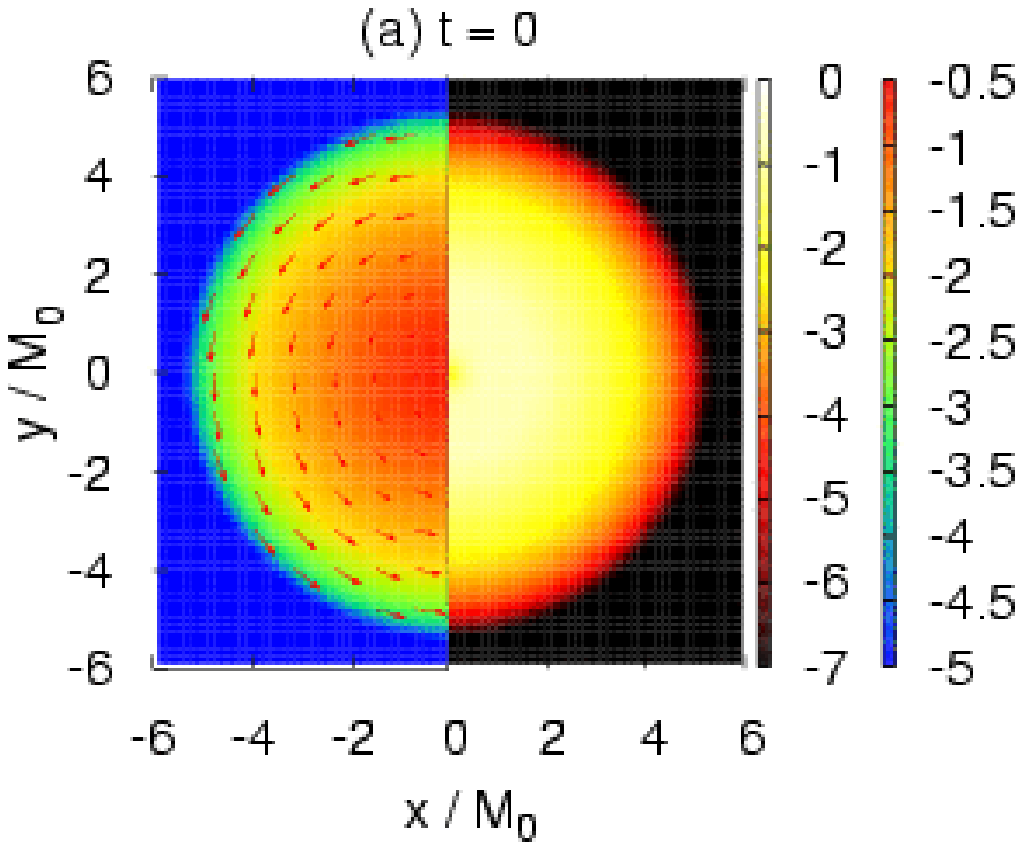}
      \end{minipage}
      \hspace{-1.0cm}
      \begin{minipage}{0.5\hsize}
      \includegraphics[width=9.0cm]{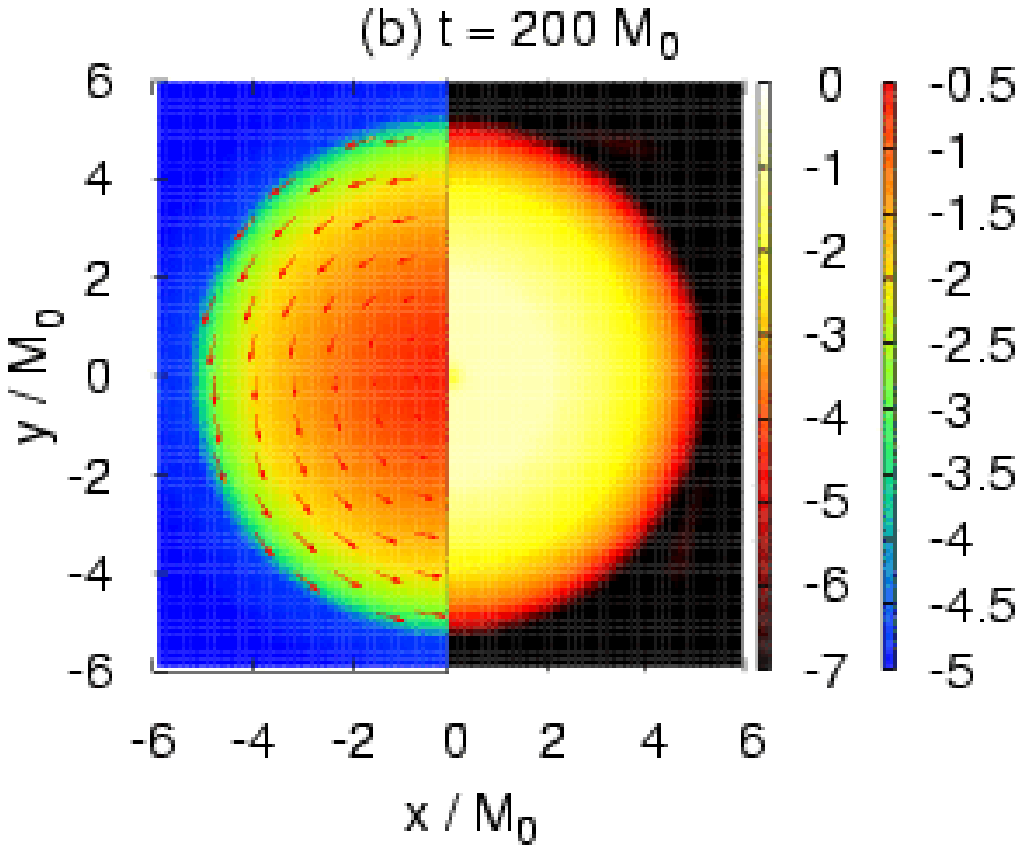}
      \end{minipage}\\
      \begin{minipage}{0.5\hsize}
      \includegraphics[width=9.0cm]{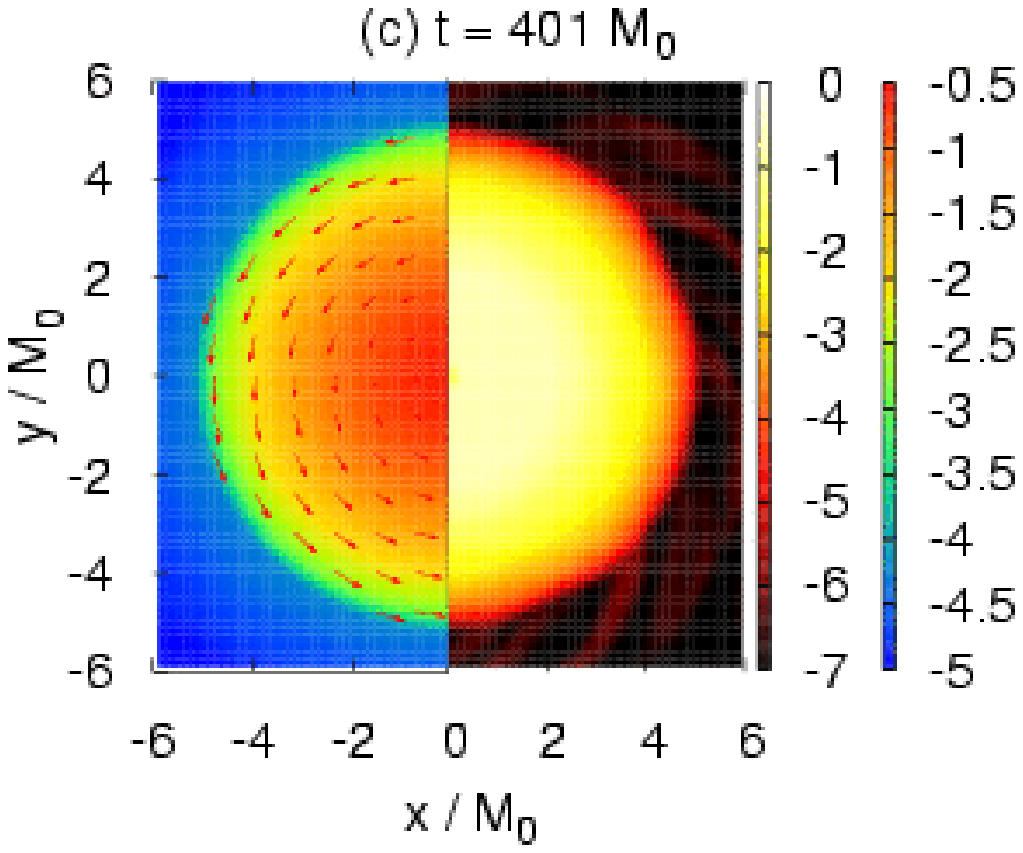}
      \end{minipage}
      \hspace{-1.0cm}
      \begin{minipage}{0.5\hsize}
      \includegraphics[width=9.0cm]{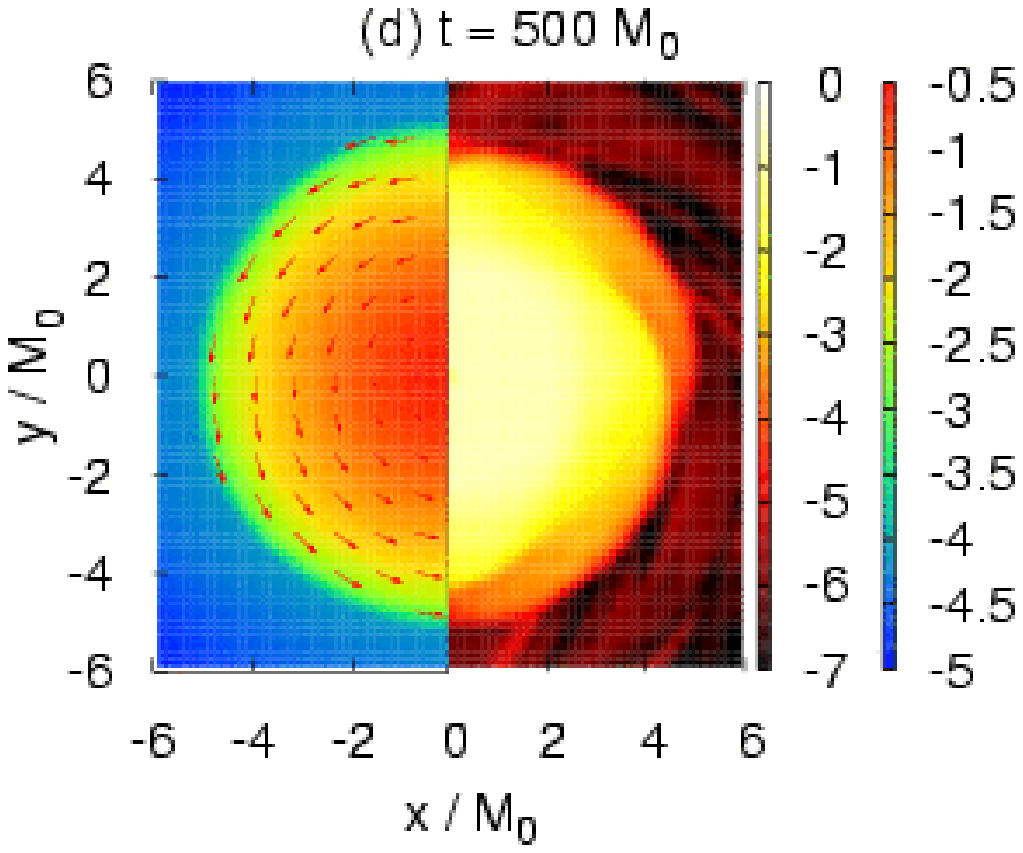}
      \end{minipage}\\
      \begin{minipage}{0.5\hsize}
      \includegraphics[width=9.0cm]{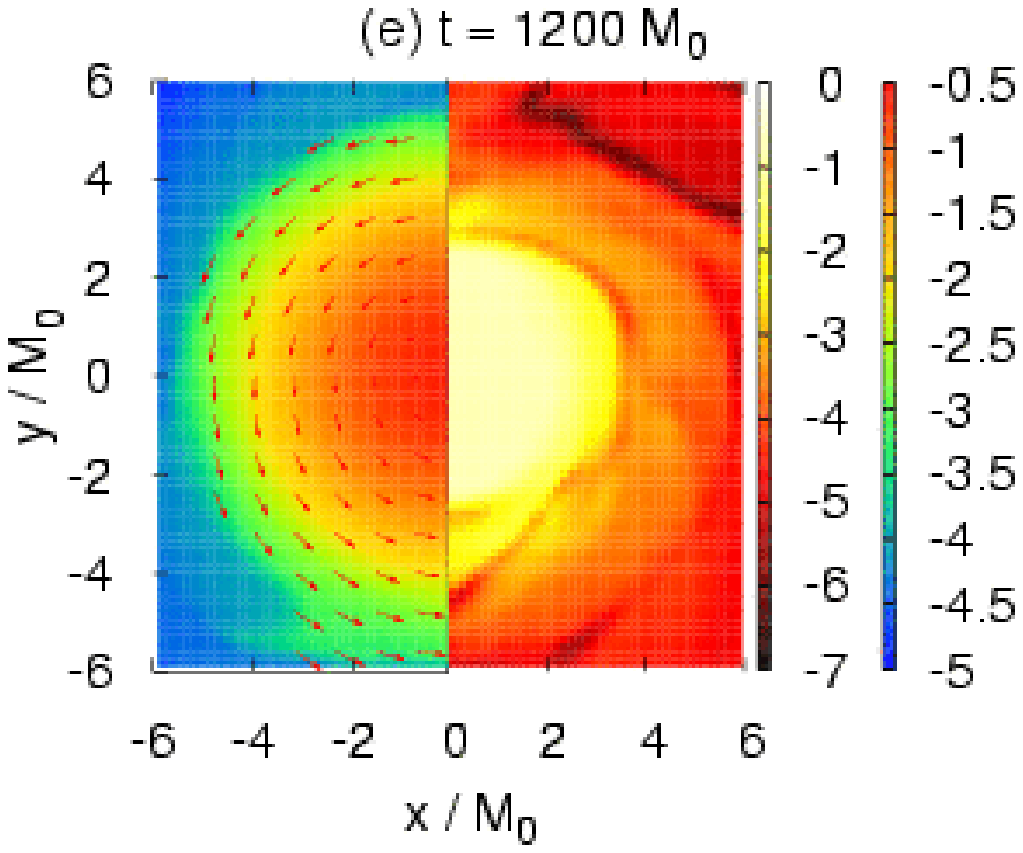}
      \end{minipage}
      \hspace{-1.0cm}
      \begin{minipage}{0.5\hsize}
      \includegraphics[width=9.0cm]{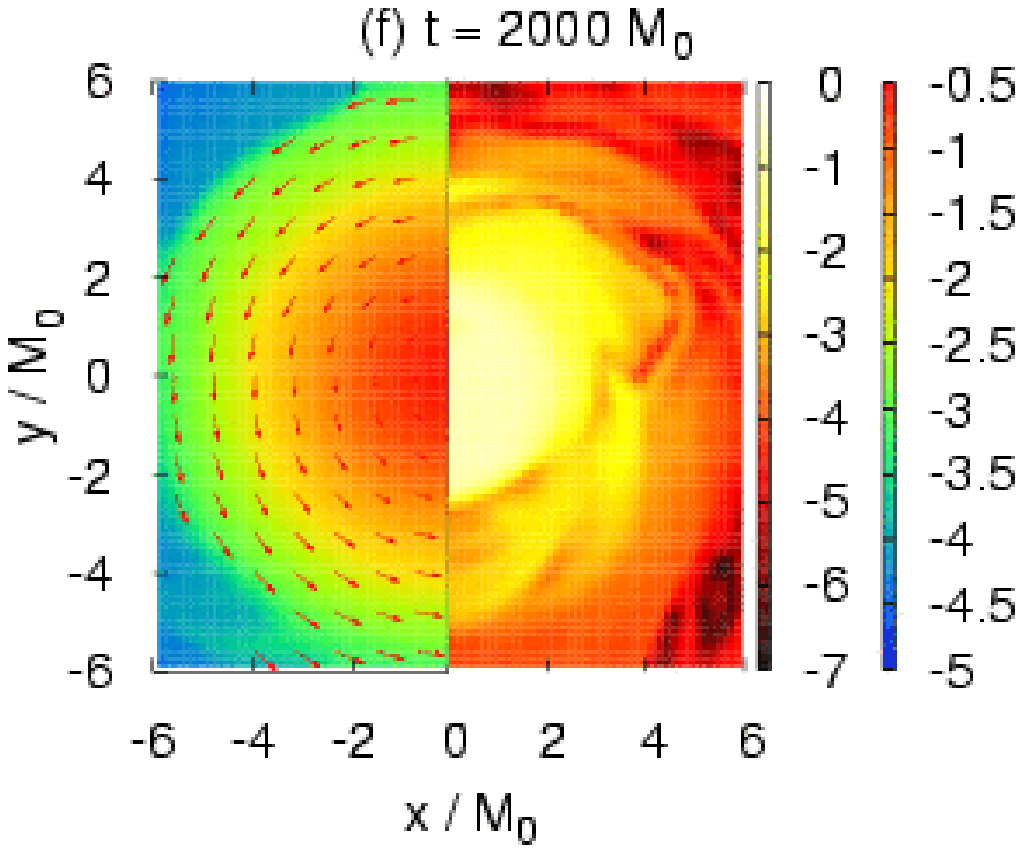}
      \end{minipage}\\
    \end{tabular}
    \caption{\label{Fig:fig6} The same as Figure~\ref{Fig:fig3}, but for 
    a rotating model R22H2T8. The arrows indicate the velocity fields. 
    }
  \end{center}
\end{figure*}

\begin{figure*}
  \begin{center}
  \vspace*{40pt}
    \begin{tabular}{cc}
      \begin{minipage}{0.5\hsize}
      \includegraphics[width=9.0cm]{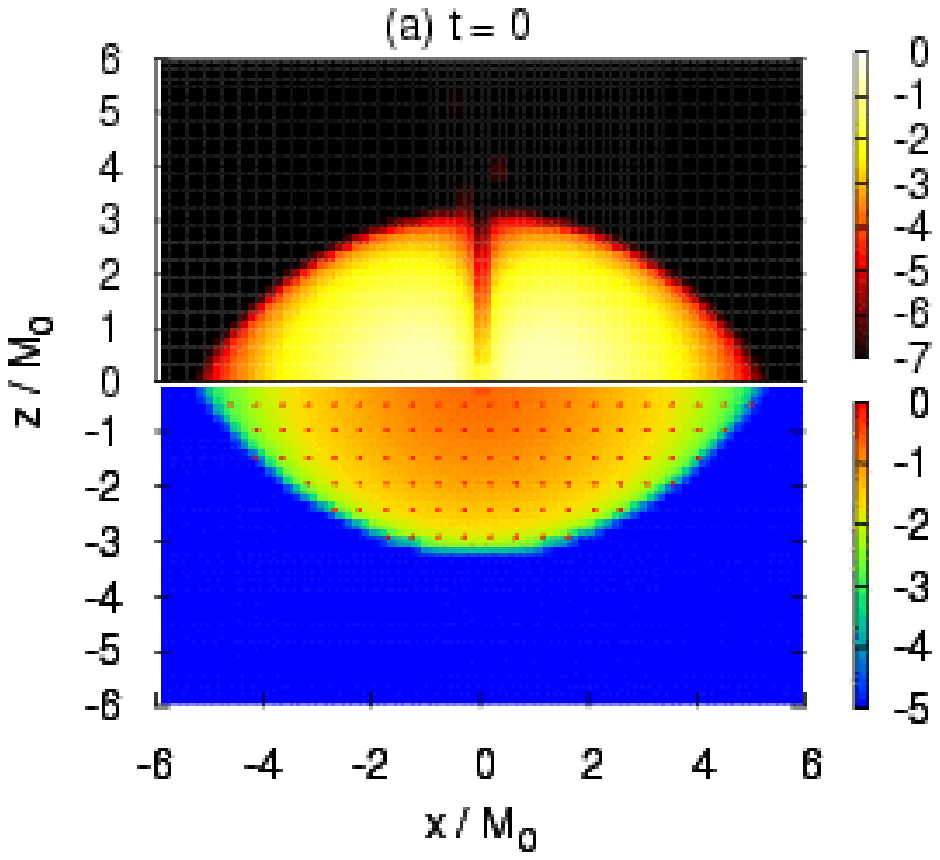}
      \end{minipage}
      \hspace{-1.0cm}
      \begin{minipage}{0.5\hsize}
      \includegraphics[width=9.0cm]{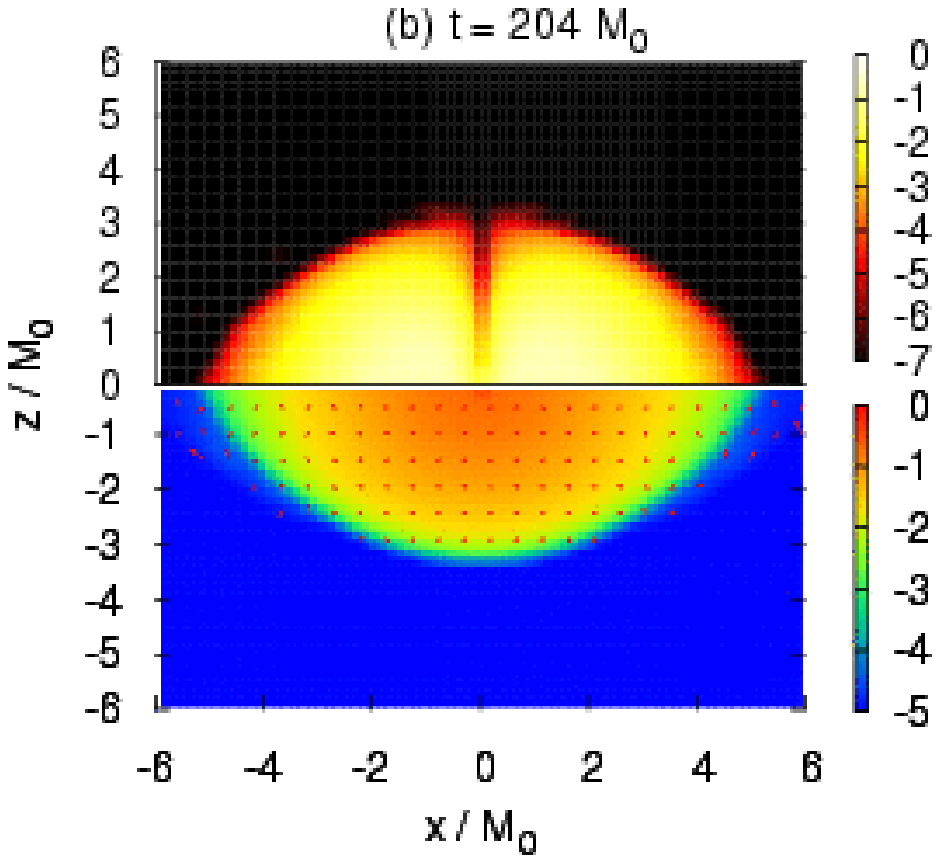}
      \end{minipage}\\
      \begin{minipage}{0.5\hsize}
      \includegraphics[width=9.0cm]{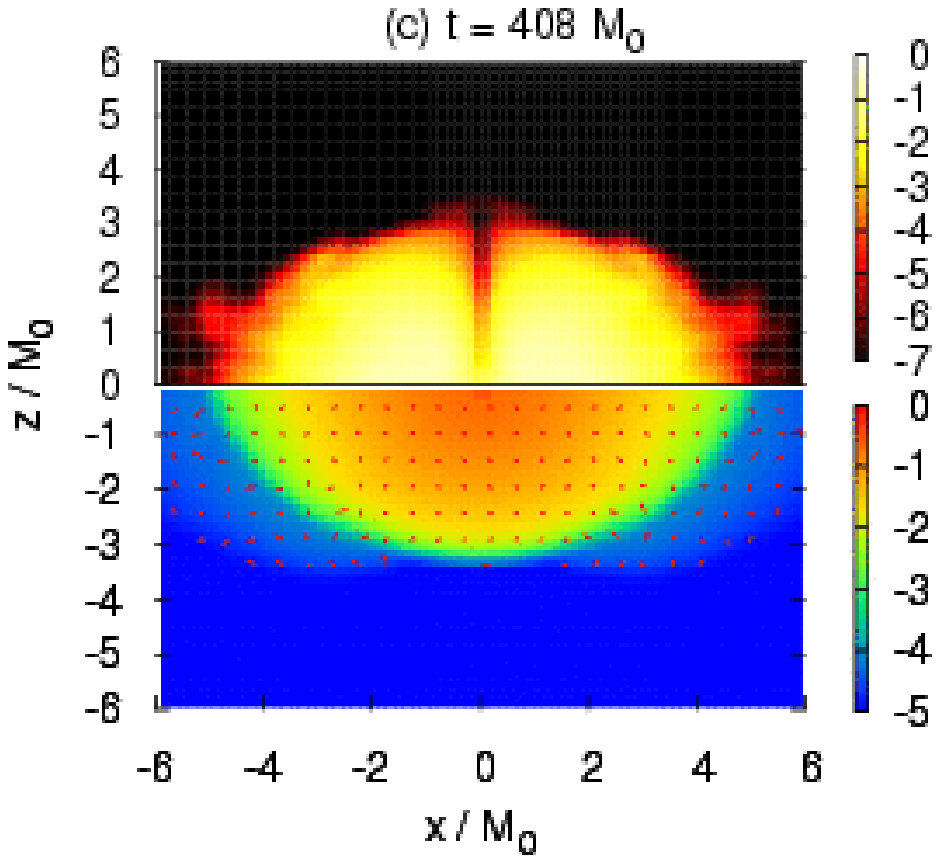}
      \end{minipage}
      \hspace{-1.0cm}
      \begin{minipage}{0.5\hsize}
      \includegraphics[width=9.0cm]{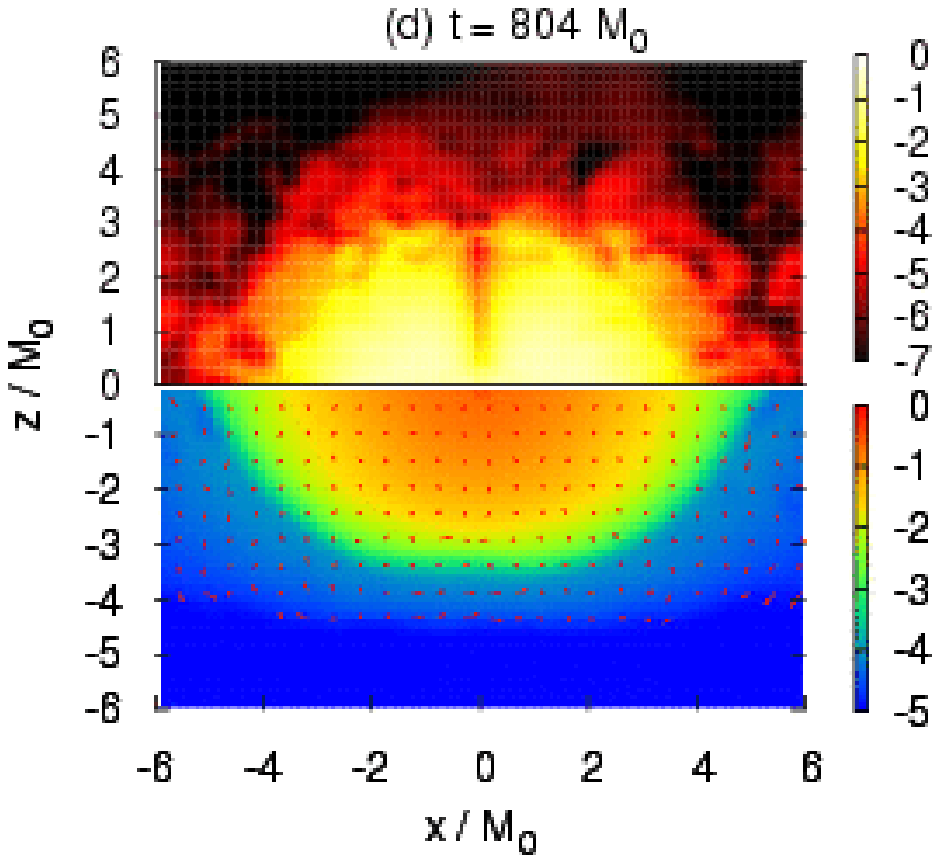}
      \end{minipage}\\
      \begin{minipage}{0.5\hsize}
      \includegraphics[width=9.0cm]{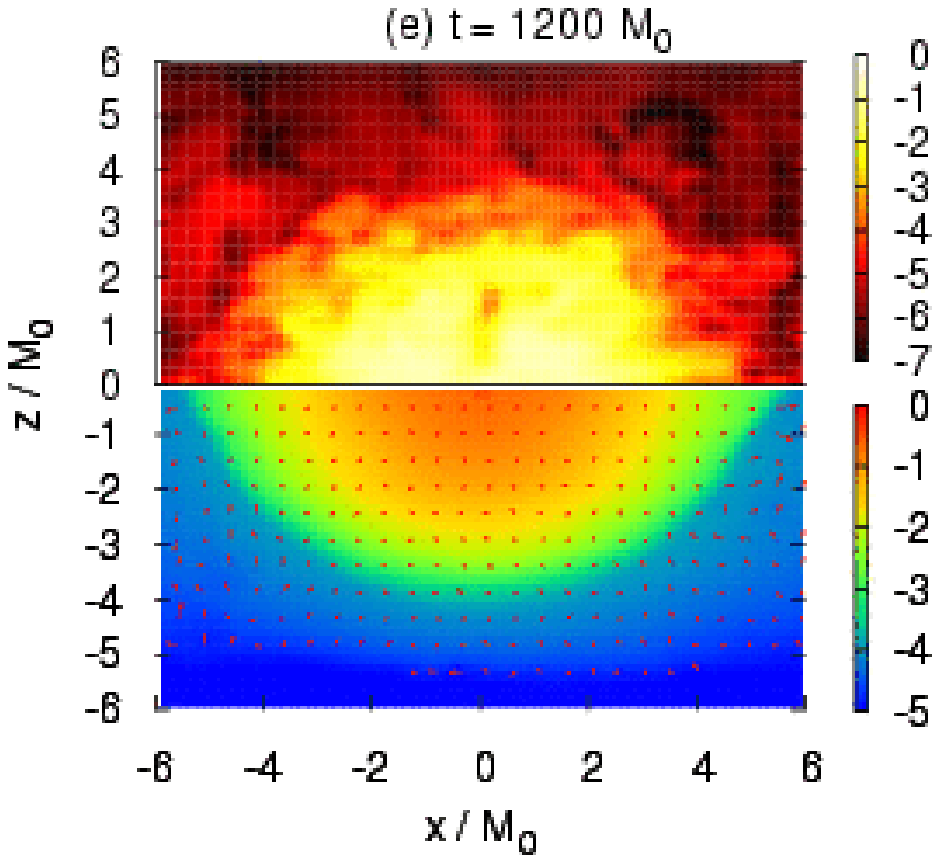}
      \end{minipage}
      \hspace{-1.0cm}
      \begin{minipage}{0.5\hsize}
      \includegraphics[width=9.0cm]{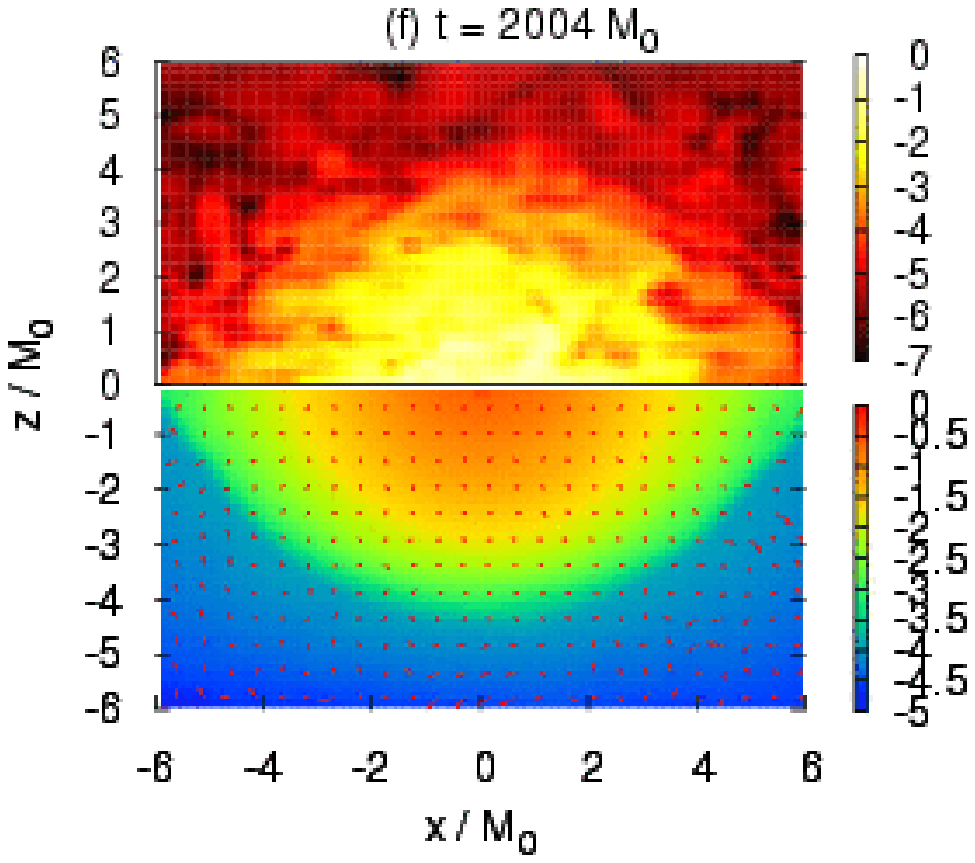}
      \end{minipage}\\
    \end{tabular}
    \caption{\label{Fig:fig7} The same as Figure~\ref{Fig:fig4}, but for 
    a rotating model R22H2T8. 
    }
  \end{center}
\end{figure*}

\begin{figure*}
  \begin{center}
  \vspace*{40pt}
    \begin{tabular}{cc}
      \begin{minipage}{0.5\hsize}
      \includegraphics[width=8.0cm]{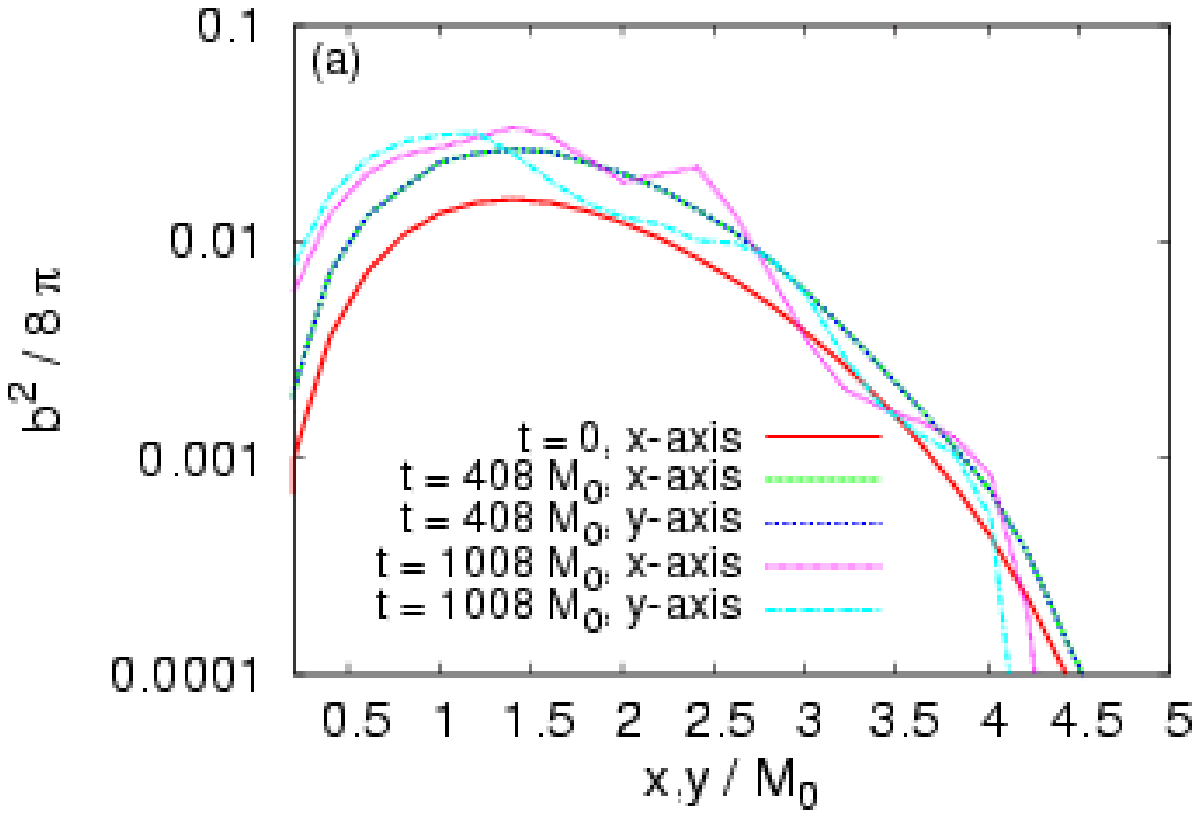}
      \end{minipage}
      \hspace{-1.0cm}
      \begin{minipage}{0.5\hsize}
      \includegraphics[width=8.0cm]{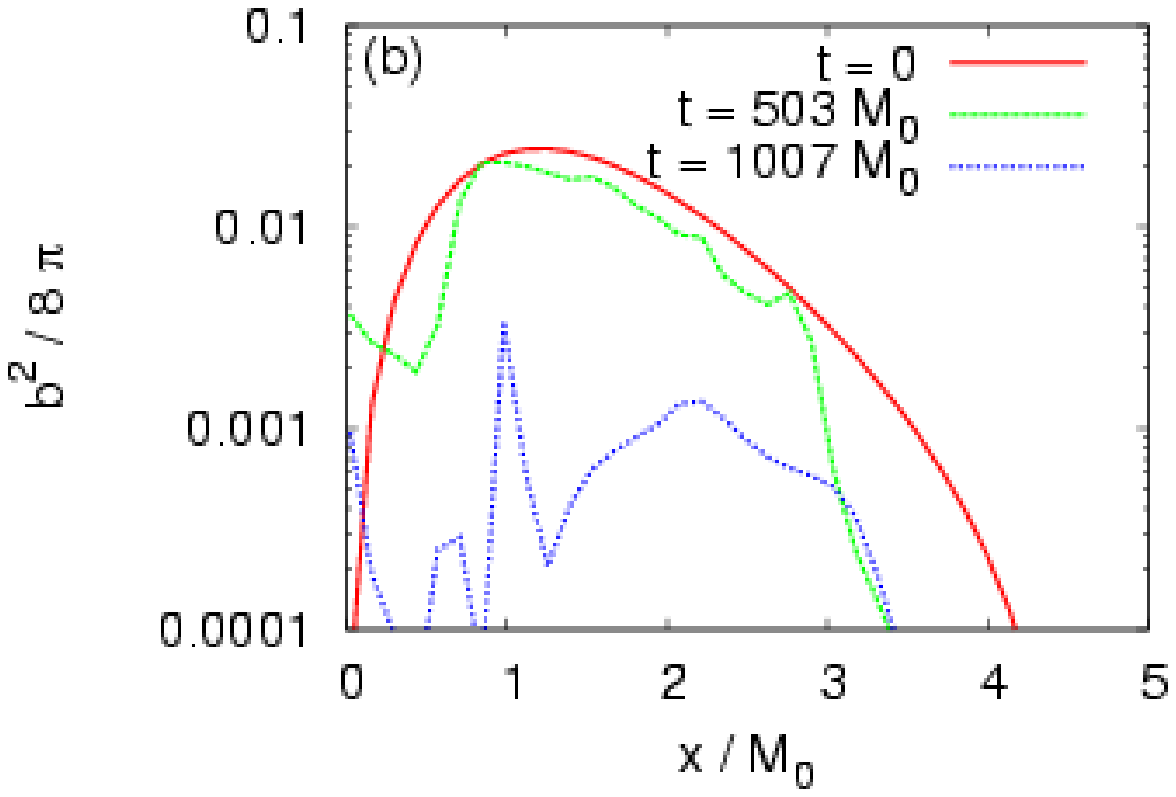}
      \end{minipage}\\
      \begin{minipage}{0.5\hsize}
      \includegraphics[width=8.0cm]{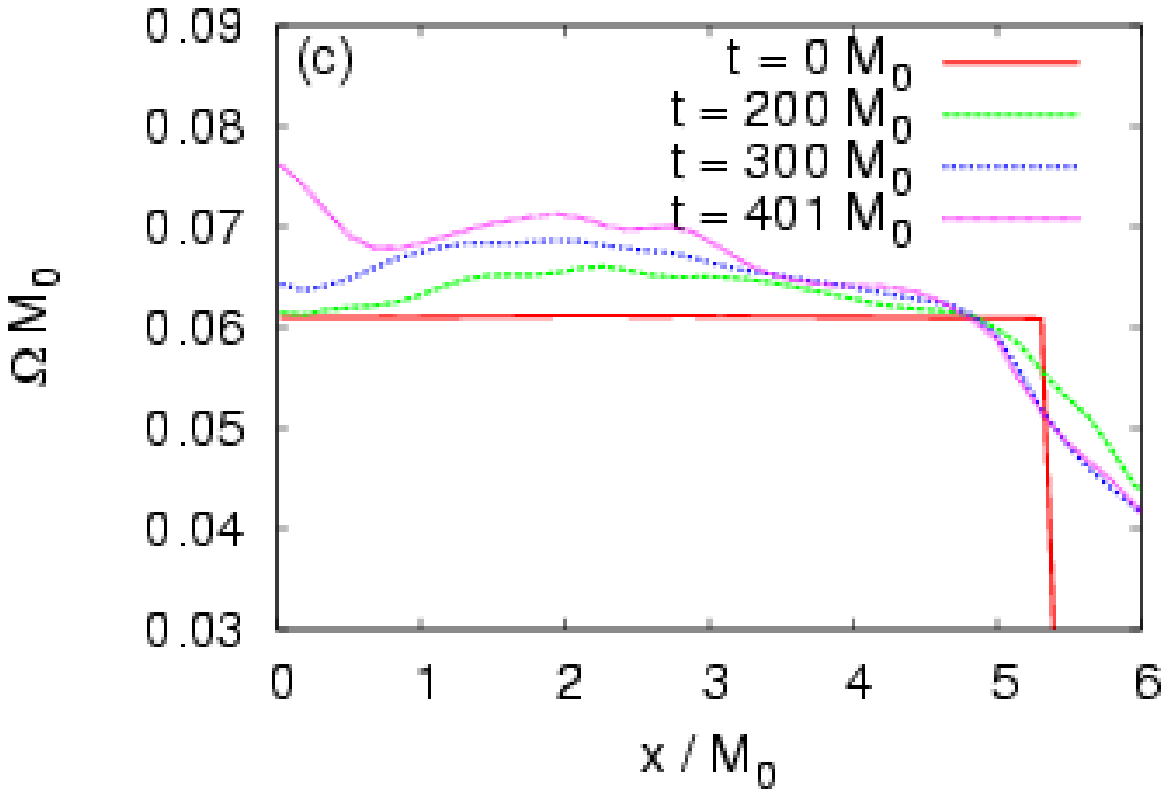}
      \end{minipage}
      \hspace{-1.0cm}
      \begin{minipage}{0.5\hsize}
      \includegraphics[width=8.0cm]{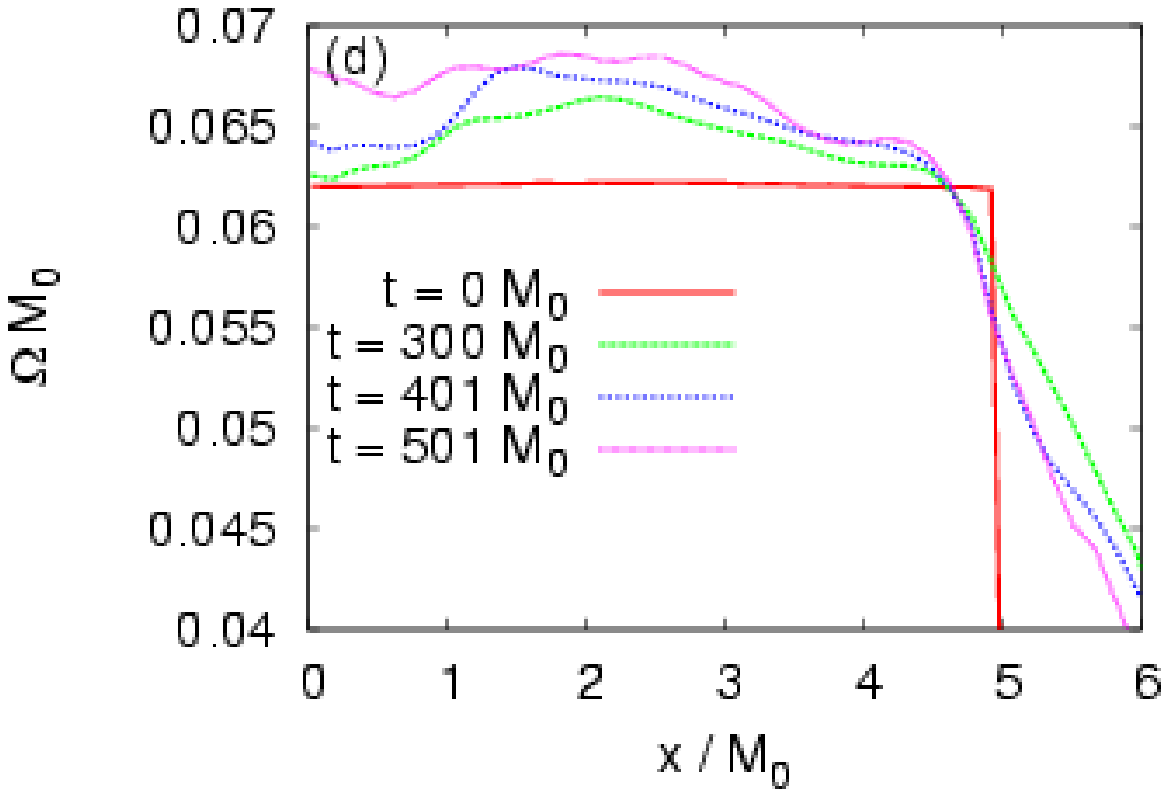}
      \end{minipage}
    \end{tabular}
    \caption{\label{Fig:fig8} Snapshots of magnetic pressure (a) for
    model R22H2T8 and (b) for mode N22H5, and those of angular
    velocity profiles (c) for model R22H2T8 and (d) for model
    R22H08T8. Both profiles are plotted along the $x$-axis on the
    equator. In the panel (a), the profile along $y$-axis is also plotted.
    In the panels (c) and (d), it is seen that the negative
    gradients of the angular velocity profiles are developed 
    near the stellar surface.  } \end{center}
\end{figure*}

\begin{figure*}
  \begin{center}
  \vspace*{40pt}
    \begin{tabular}{cc}
      \begin{minipage}{0.5\hsize}
      \includegraphics[width=8.0cm]{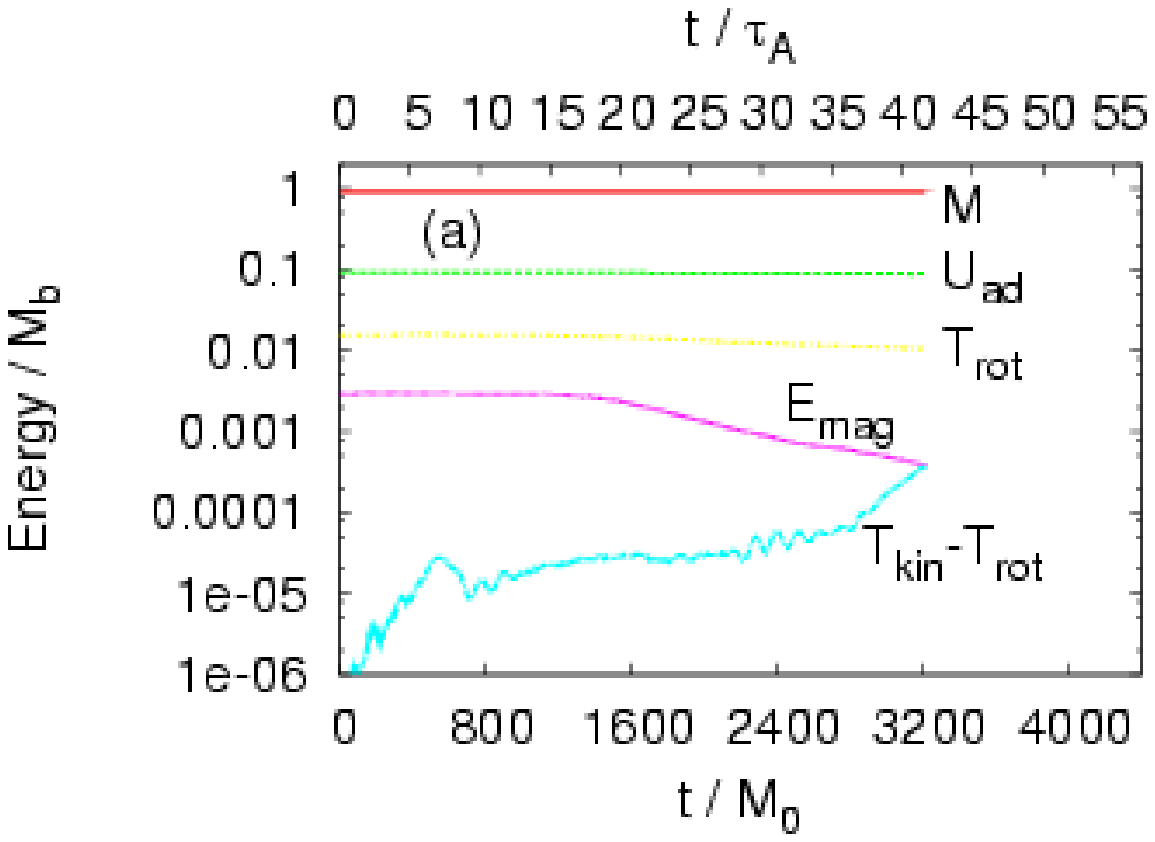}
      \end{minipage}
      \hspace{-1.0cm}
      \begin{minipage}{0.5\hsize}
      \includegraphics[width=8.0cm]{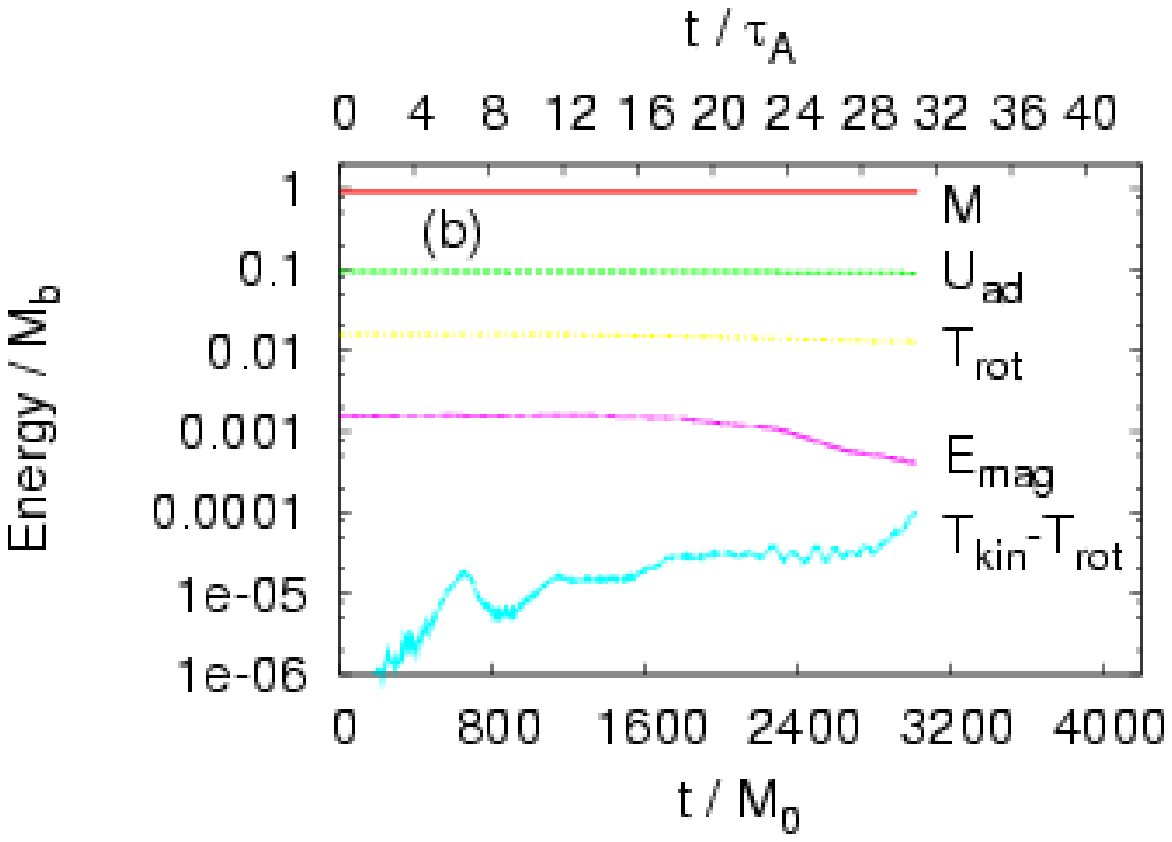}
      \end{minipage}
    \end{tabular}
    \caption{\label{Fig:fig9}
    Time evolution of several energy components 
    (a) for model R22H2T8 and (b) for model R22H08T8.
    }
  \end{center}
\end{figure*}

\begin{figure*}
  \begin{center}
  \vspace*{40pt}
    \begin{tabular}{cc}
      \begin{minipage}{0.5\hsize}
      \includegraphics[width=8.0cm]{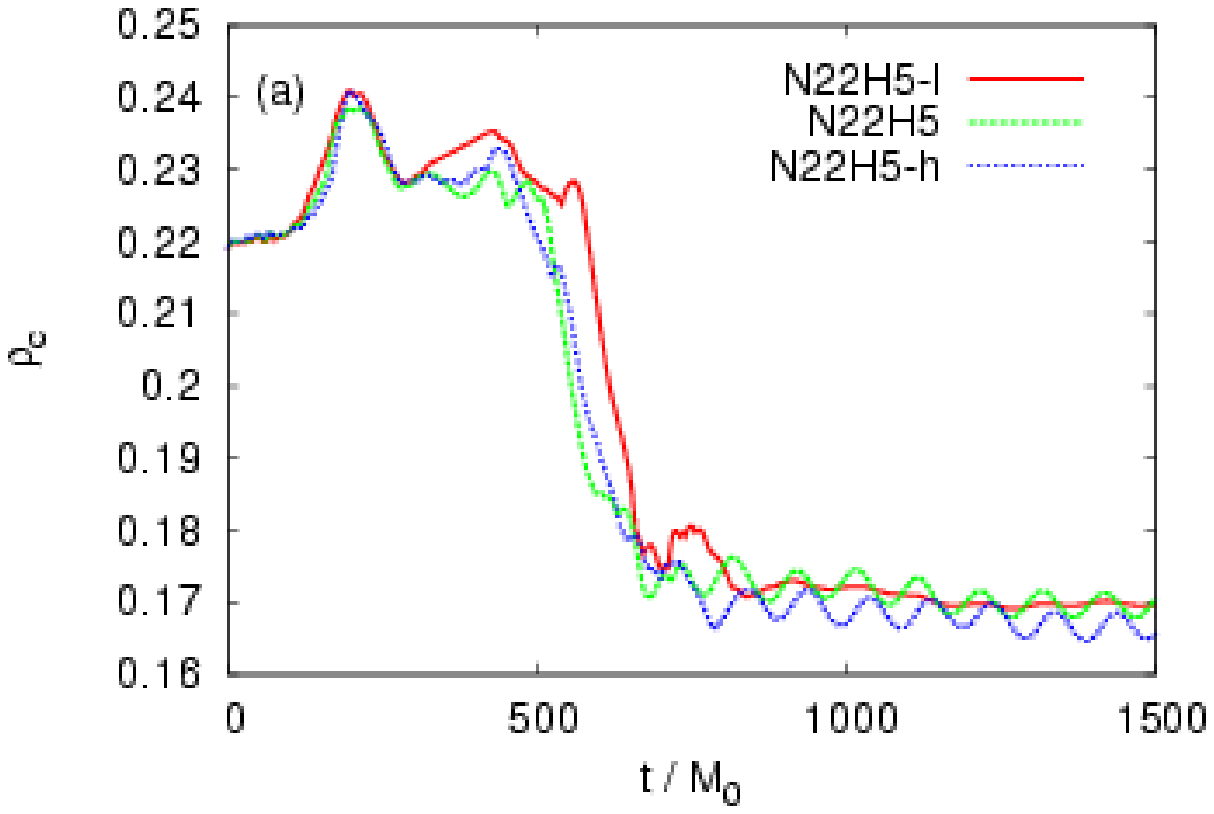}
      \end{minipage}
      \hspace{-1.0cm}
      \begin{minipage}{0.5\hsize}
      \includegraphics[width=8.0cm]{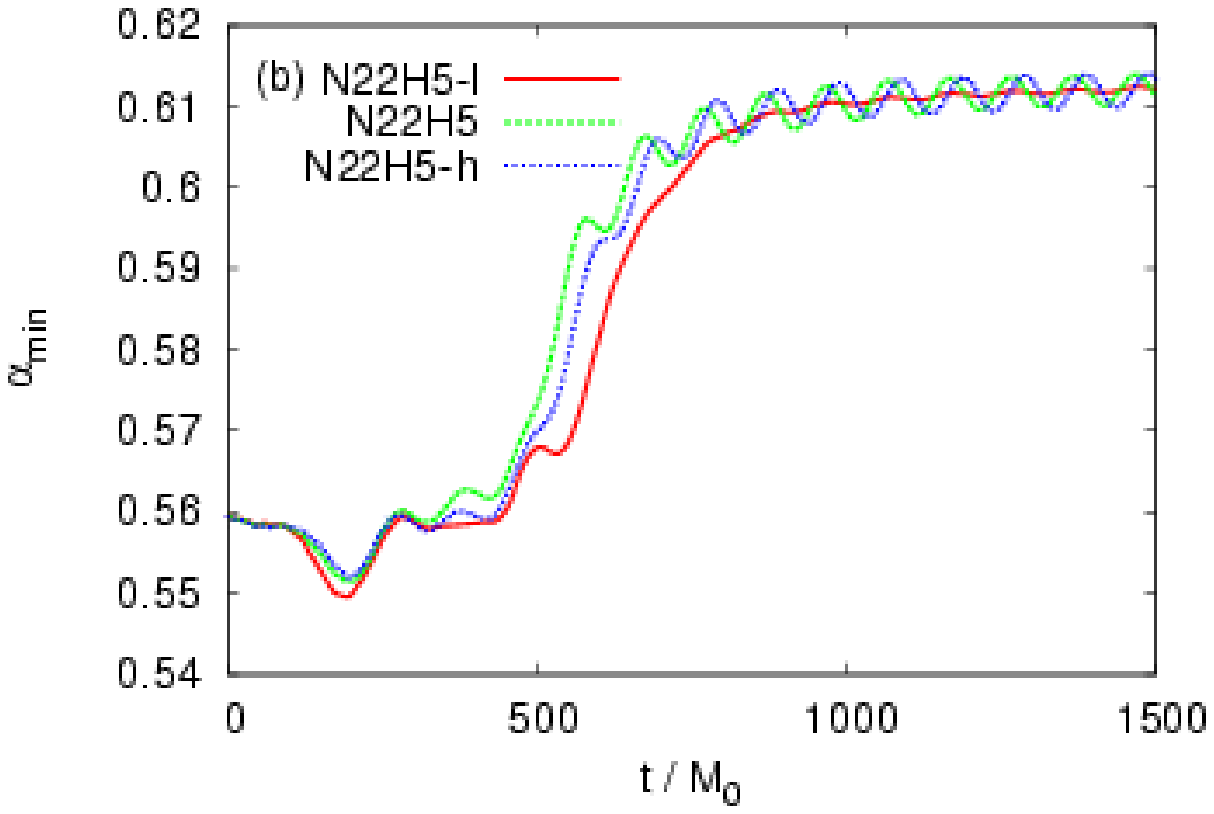}
      \end{minipage}\\
      \begin{minipage}{0.5\hsize}
      \includegraphics[width=8.0cm]{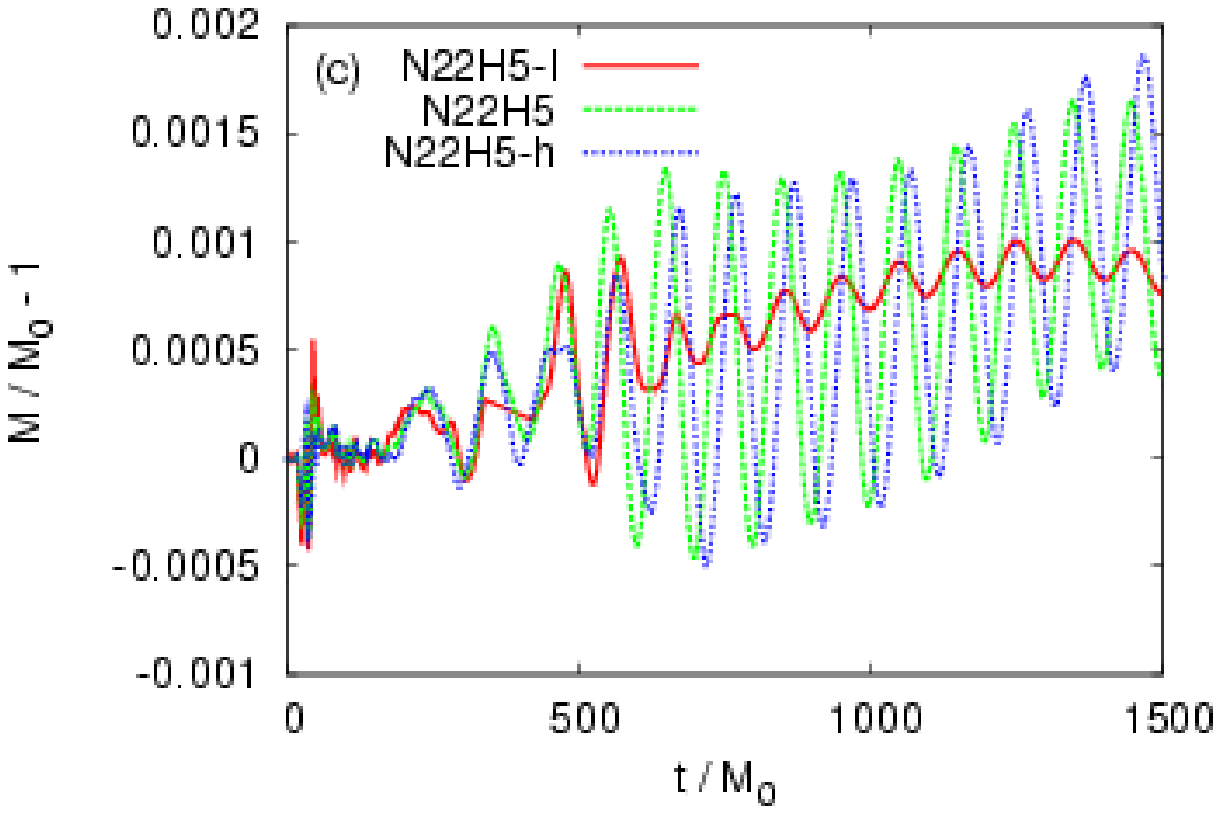}
      \end{minipage}
      \hspace{-1.0cm}
      \begin{minipage}{0.5\hsize}
      \includegraphics[width=8.0cm]{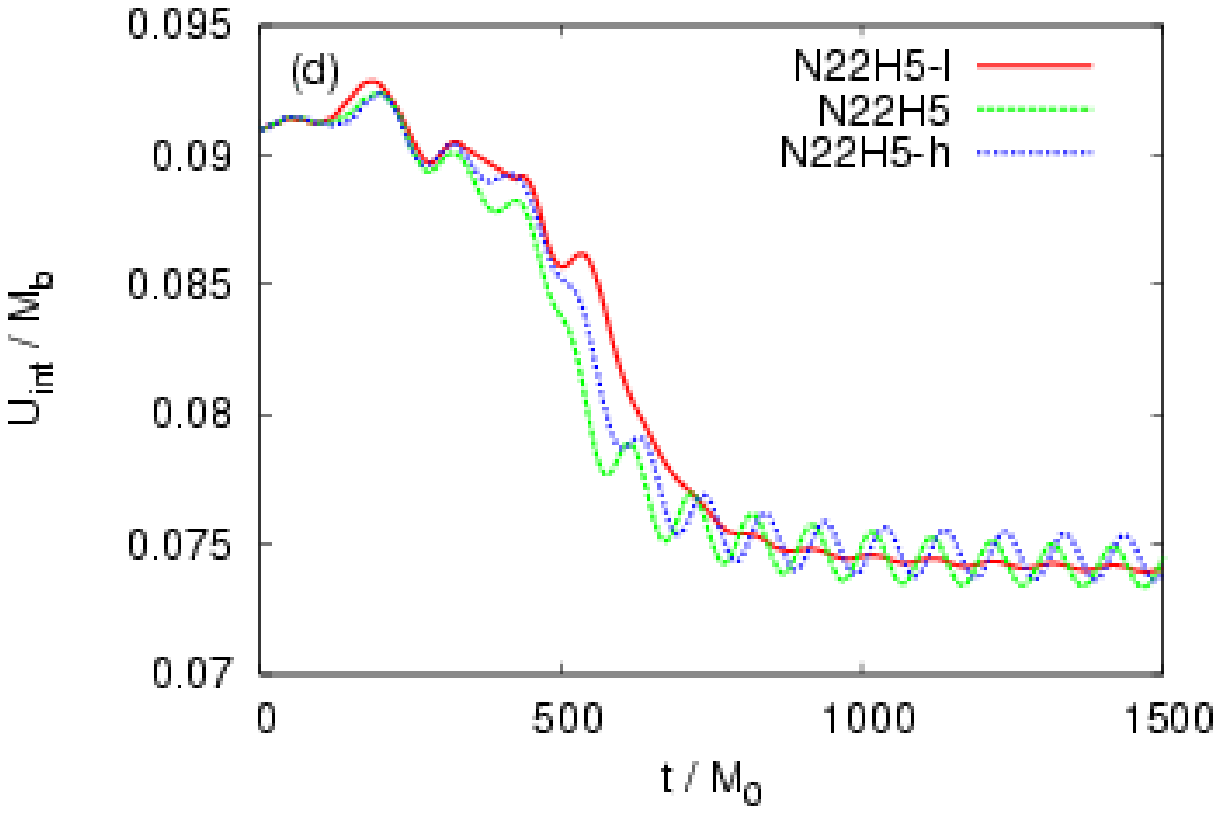}
      \end{minipage}\\
      \begin{minipage}{0.5\hsize}
      \includegraphics[width=8.0cm]{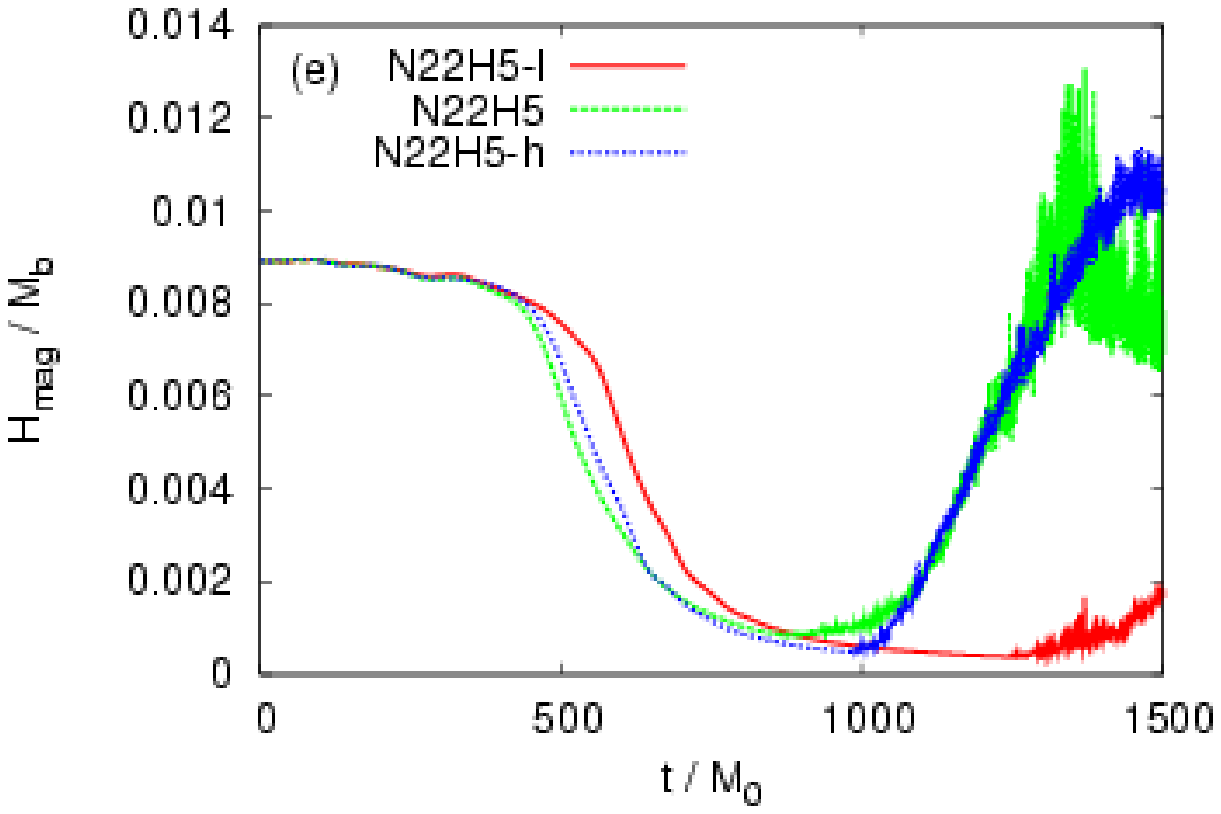}
      \end{minipage}
      \hspace{-1.0cm}
      \begin{minipage}{0.5\hsize}
      \includegraphics[width=8.0cm]{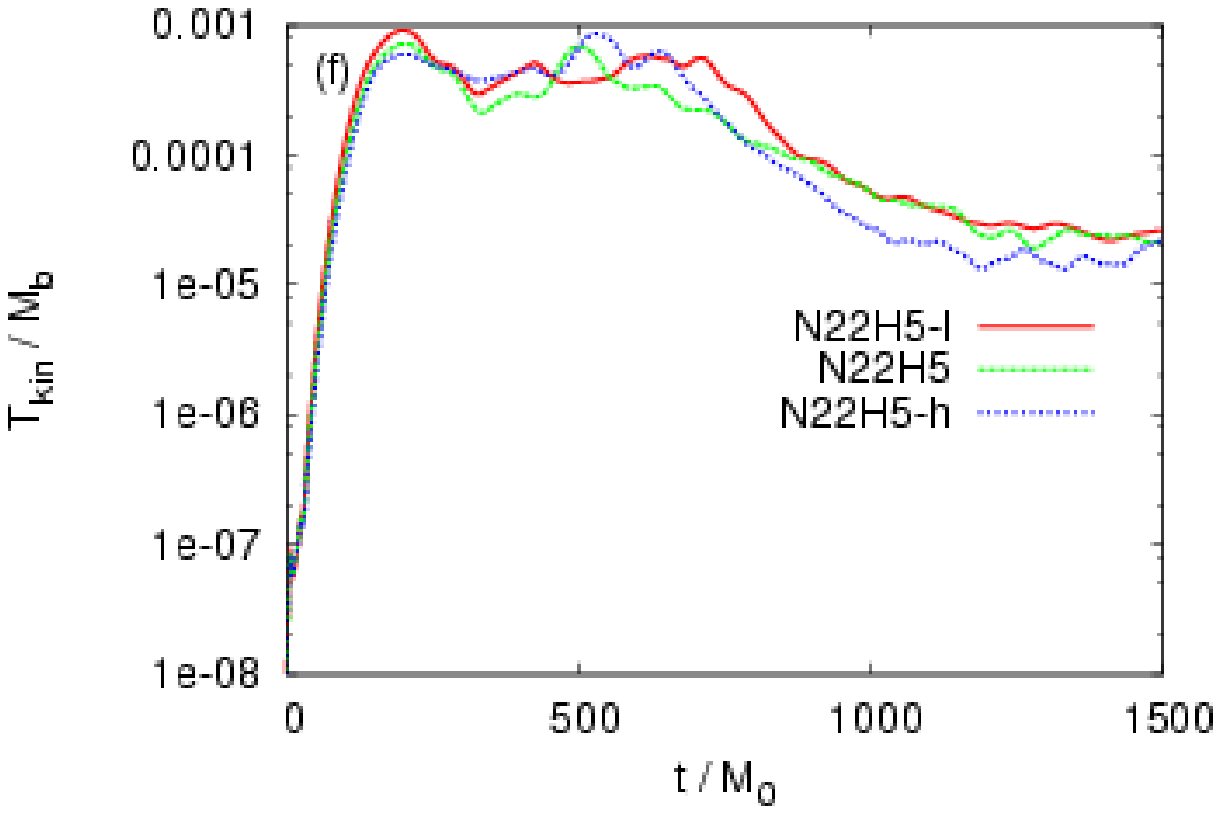}
      \end{minipage}\\
      \begin{minipage}{0.5\hsize}
      \includegraphics[width=8.0cm]{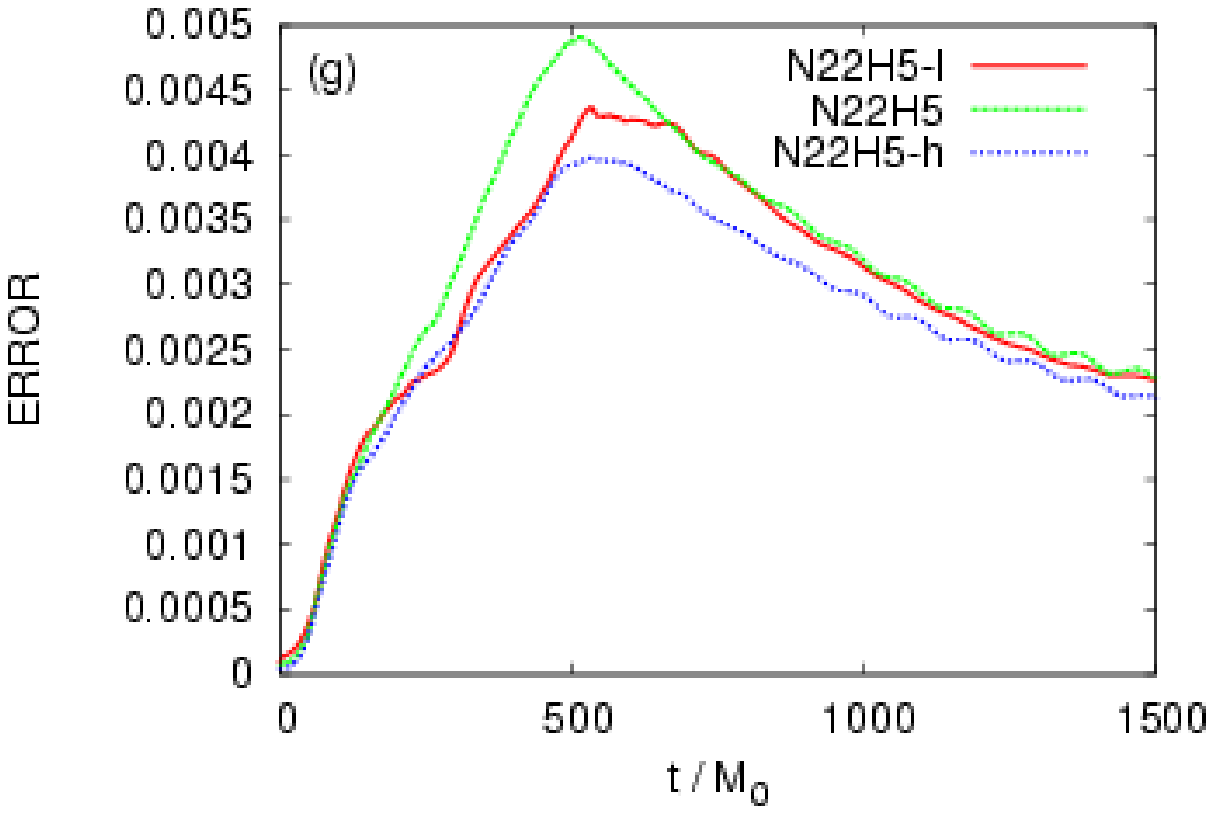}
      \end{minipage}
    \end{tabular}
    \caption{\label{Fig:fig10} The results of convergence tests for
    model N22H5; (a) the central density, (b) the minimum value of
    lapse function, (c) the ADM mass normalized by its initial value,
    (d) the internal energy, (e) the magnetic energy, (f) the kinetic
    energy, and (g) the normalized Hamiltonian constraint.  }
    \end{center}
\end{figure*}

\begin{figure*}
  \begin{center}
  \vspace*{40pt}
    \begin{tabular}{cc}
      \begin{minipage}{0.5\hsize}
      \includegraphics[width=8.0cm]{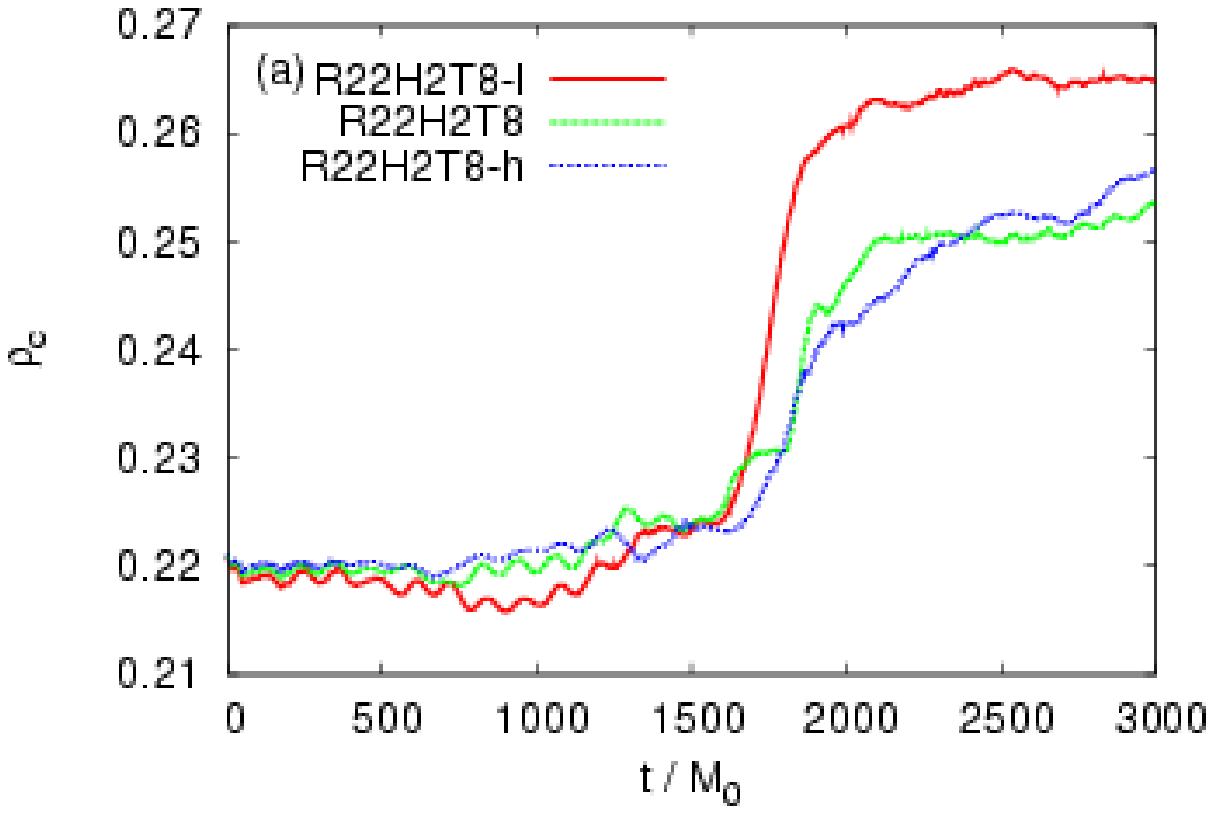}
      \end{minipage}
      \hspace{-1.0cm}
      \begin{minipage}{0.5\hsize}
      \includegraphics[width=8.0cm]{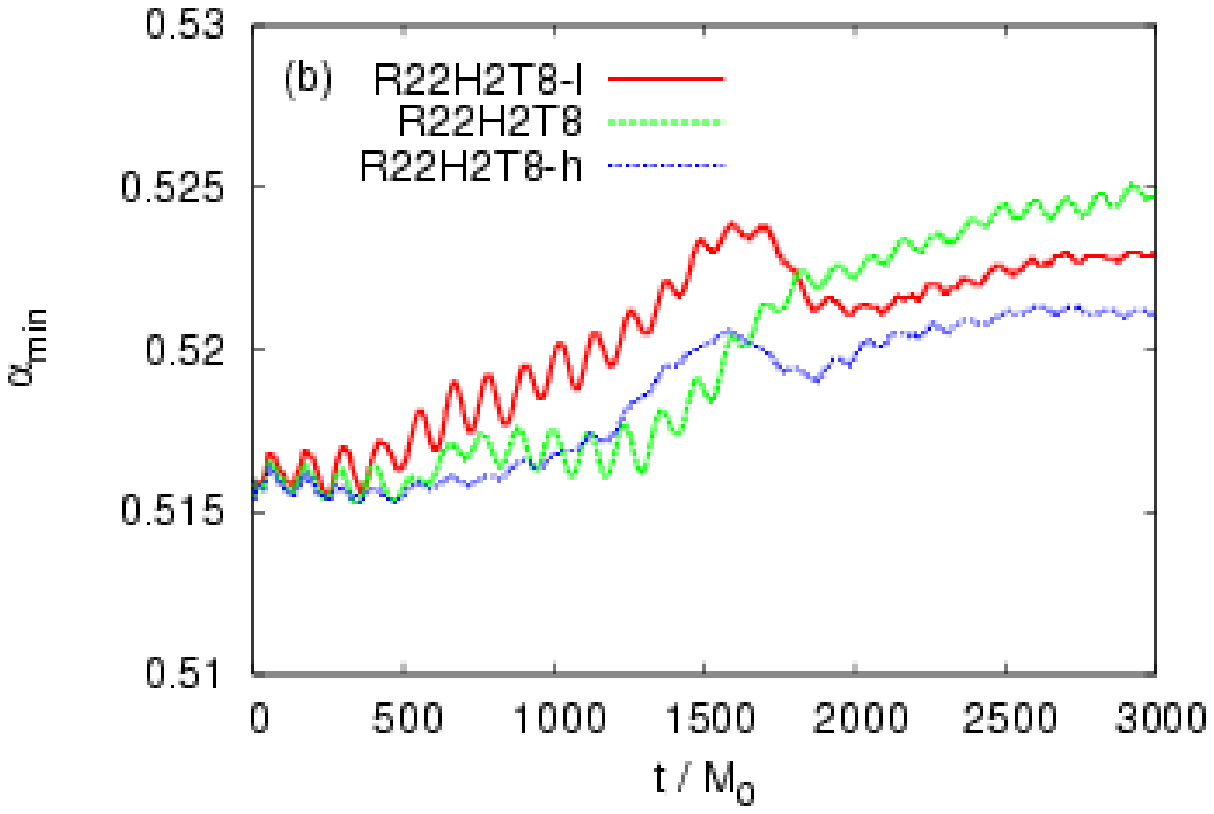}
      \end{minipage}\\
      \begin{minipage}{0.5\hsize}
      \includegraphics[width=8.0cm]{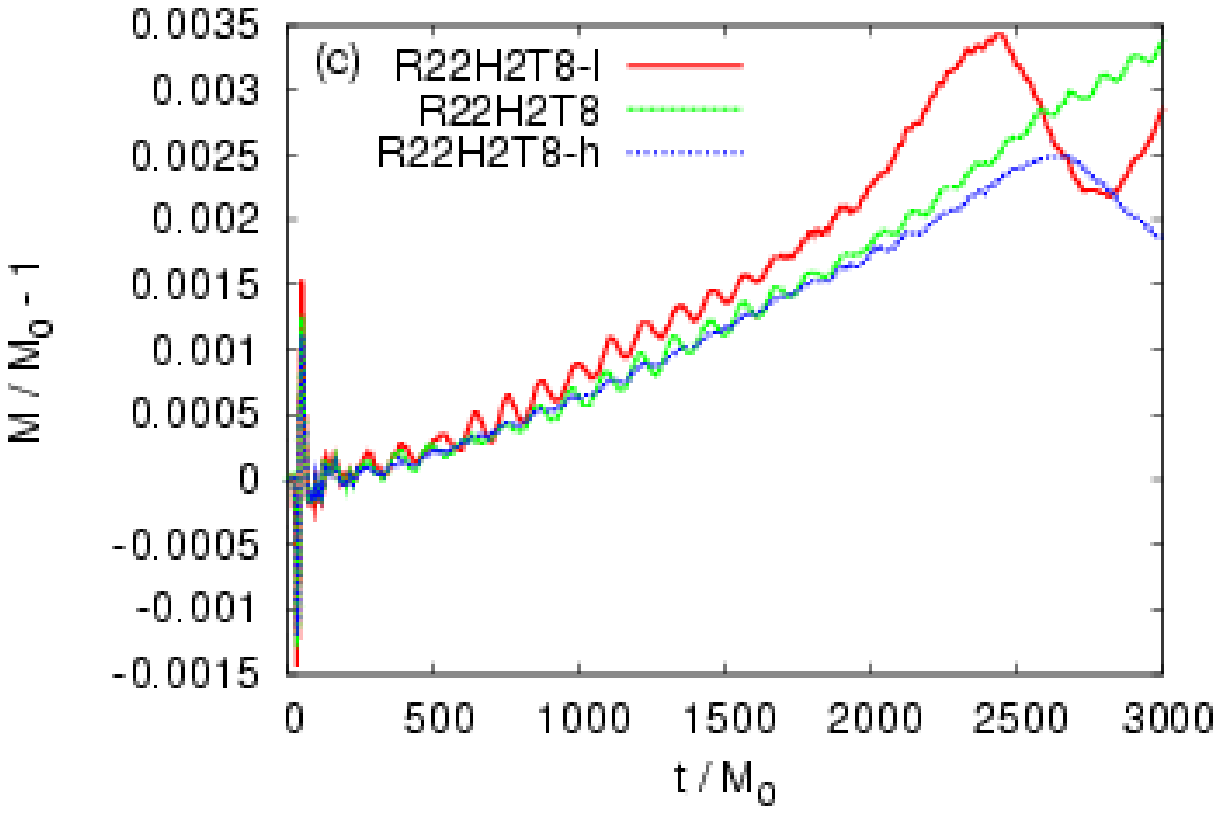}
      \end{minipage}
      \hspace{-1.0cm}
      \begin{minipage}{0.5\hsize}
      \includegraphics[width=8.0cm]{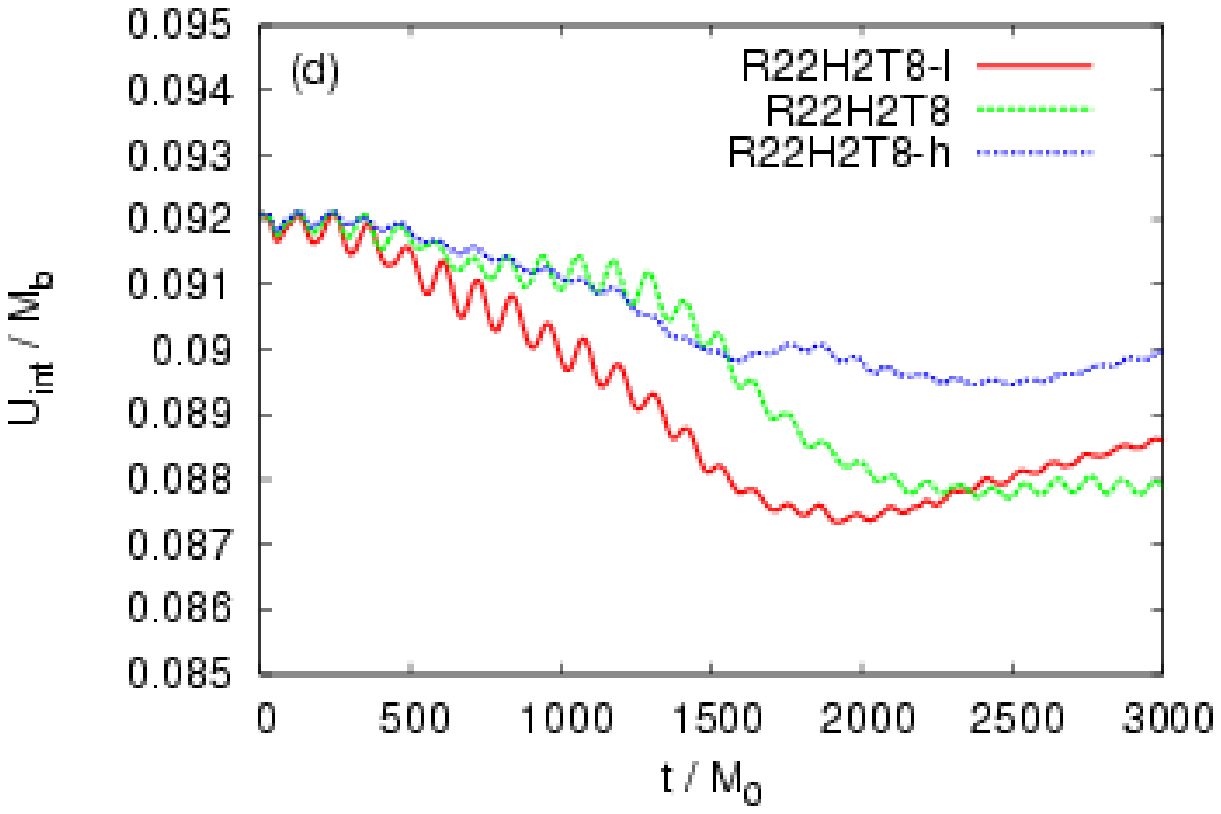}
      \end{minipage}\\
      \begin{minipage}{0.5\hsize}
      \includegraphics[width=8.0cm]{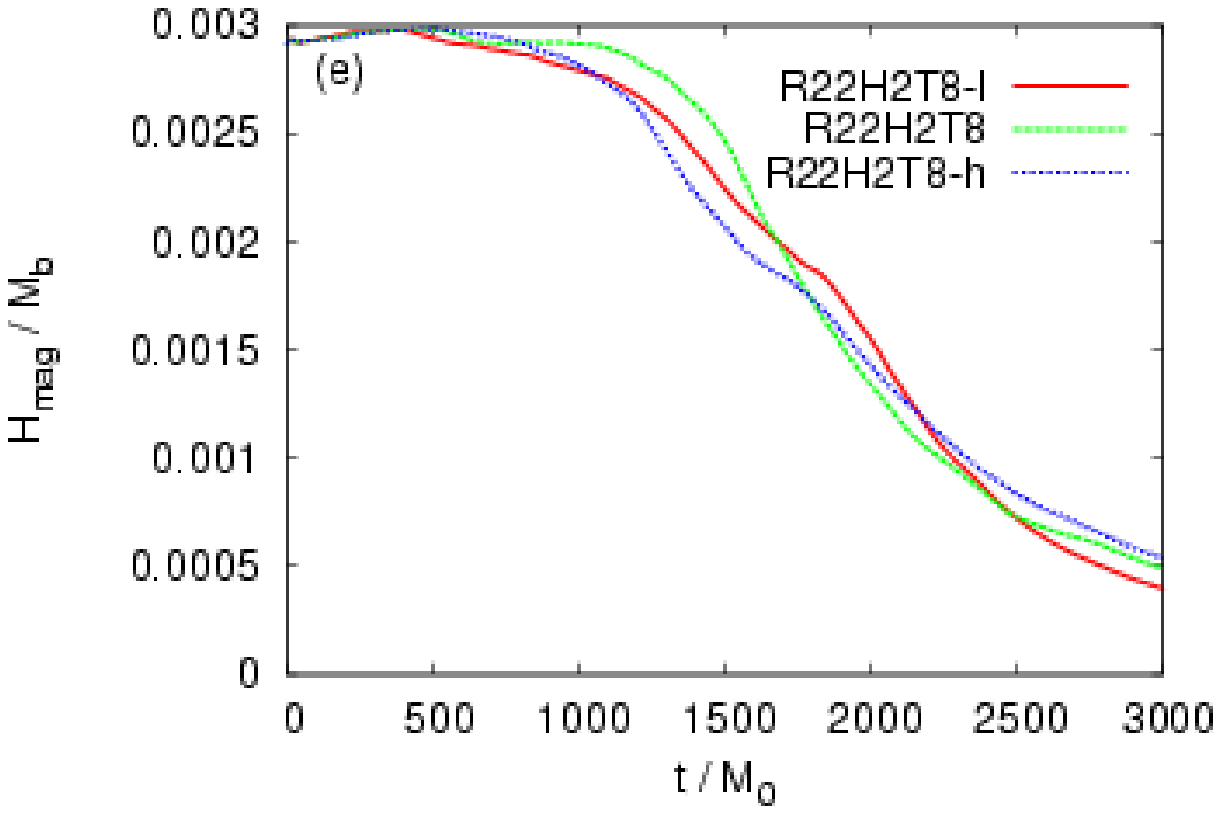}
      \end{minipage}
      \hspace{-1.0cm}
      \begin{minipage}{0.5\hsize}
      \includegraphics[width=8.0cm]{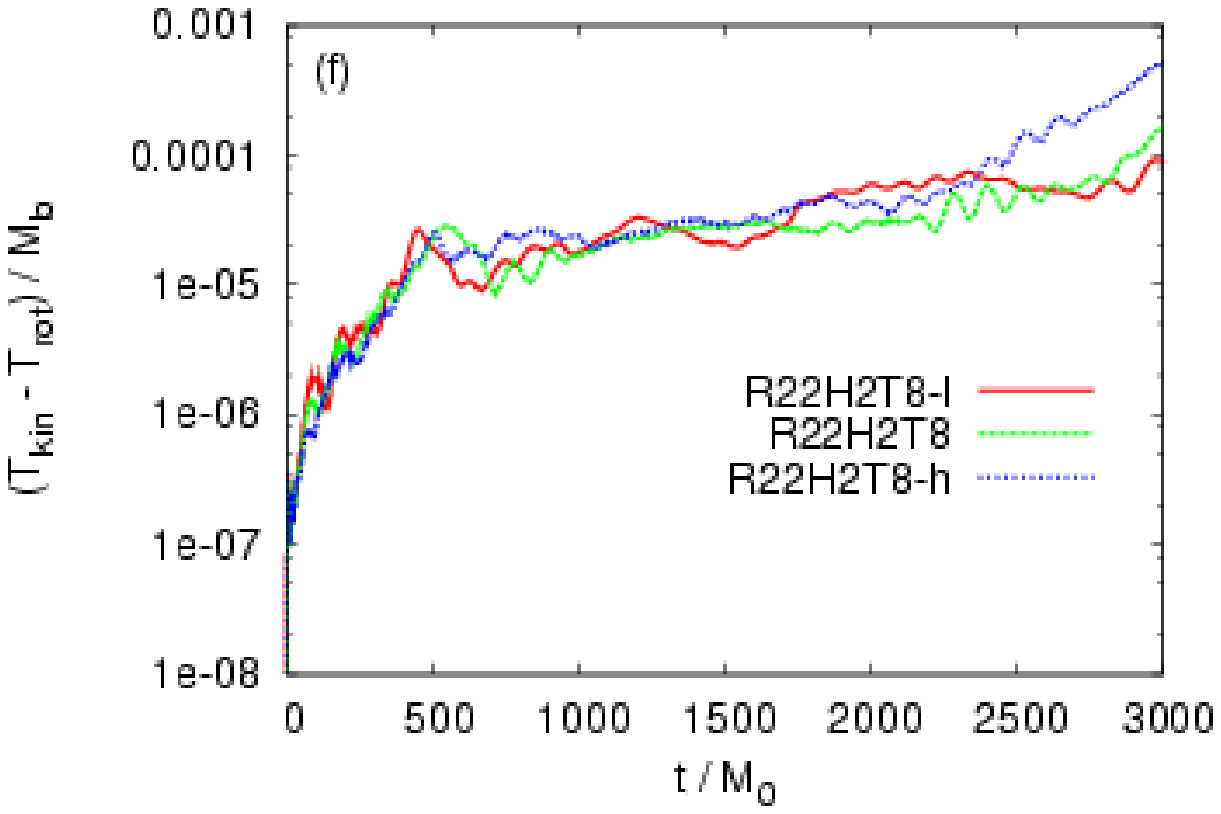}
      \end{minipage}\\
      \begin{minipage}{0.5\hsize}
      \includegraphics[width=8.0cm]{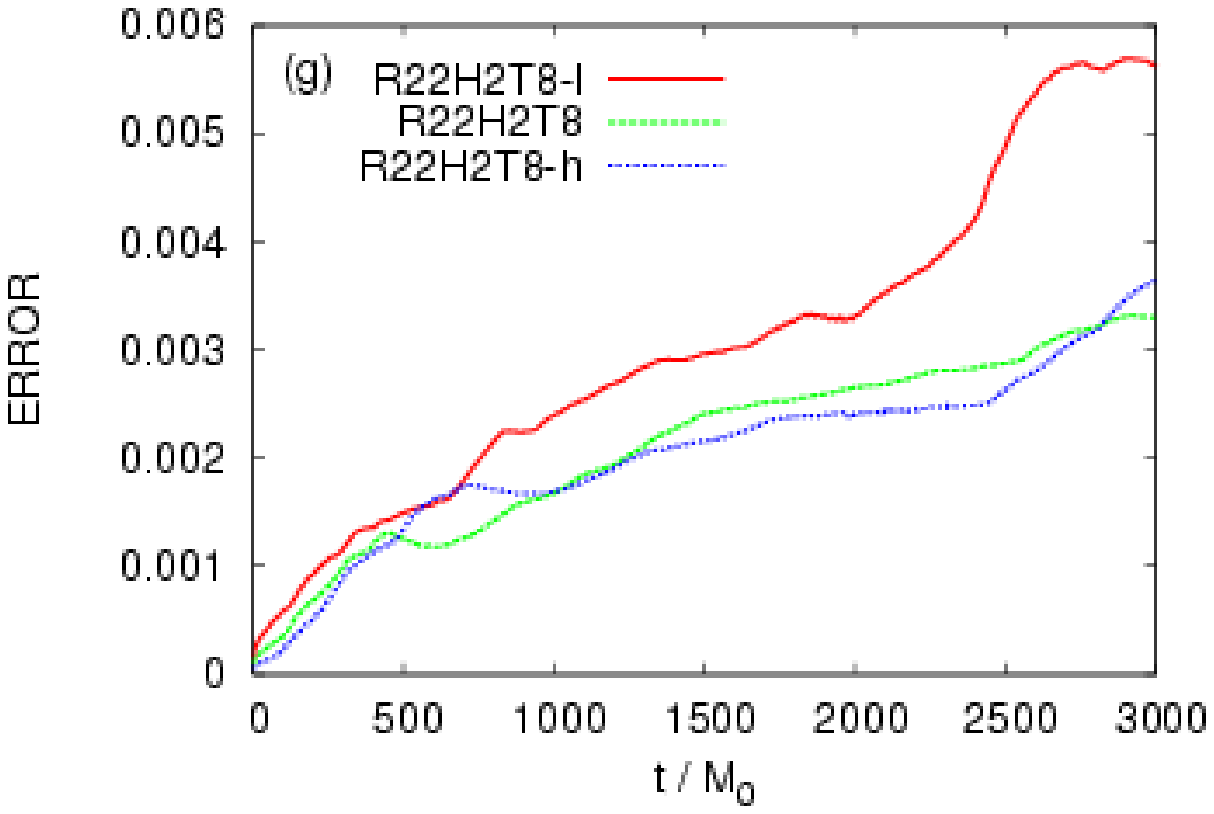}
      \end{minipage}
    \end{tabular}
    \caption{\label{Fig:fig11}
    The same as Figure~\ref{Fig:fig10} but for model R22H2T8.
    }
  \end{center}

\end{figure*}

Recently, \citet{Lander:2009ws} and \citet{Lander:2010br} studied the
instability associated with the presence of toroidal magnetic fields
by solving the linearized Newtonian MHD equations with their
time-domain code. They showed that the Tayler instability
characterized by the azimuthal mode number $m=1$ primarily occurs
near the axis of $\varpi=0$ and that no pronounced Parker instability
sets in near the surface of the star .  Similar results were obtained
by \cite{Duez:2010}, in which Newtonian resistive MHD simulations are
performed.  These results seem to be incompatible with our present
results. The reason for this discrepancy seems to be the following: 
\citet{Lander:2009ws} and~\citet{Lander:2010br} restrict their studies 
to low-order azimuthal modes ($m \le 6$) and \cite{Duez:2010} employs
some kinds of (artificial) viscosity and resistivity for the evolution
to remove numerical instability caused by a short-wavelength
oscillation.  Note that \citet{Lander:2009ws} and
\citet{Lander:2010br} also use artificial viscosity to stabilize their
computations.  This suggests that in their simulations,
short-wavelength modes might be suppressed.  As can be seen from
Figures~\ref{Fig:fig1} and~\ref{Fig:fig3}, the Parker instability
found in this study is characterized by a high-order azimuthal mode
number, which implies that the unstable modes of the Parker
instability near the surface have a short wavelength. 
In \citet{Duez:2010}, they used a stably stratified model with 
the polytrope index $n=3$ and forecasted the instability as we did 
in Figure~\ref{Fig:fig1}. They found the primarily instability is 
not the Parker instability, but the Taylor instability in their model. 
This is likely that the restoring force due to the entropy gradient near 
the surface stabilizes the magnetic buoyant force.


Finally, we comment on a possible potential effect for stabilizing the
magnetic field instabilities which are not taken into account in our
present work.  In a stably stratified region of the star, the buoyant
force inhibits the interchange, Parker, and Tayler instabilities from
growing~\citep{Acheson:1978}.  For the cold neutron star, the
composition gradient induced by chemical inhomogeneities can stably
stratify the neutron star matter and supports the gravity mode
(g-mode) oscillations.
\citet{Finn:1987} considered crustal g-modes due to the composition
discontinuities in the outer envelopes, whose typical oscillation
period is about 5~ms for a canonical neutron star
model. \citet{Reisenegger:1992} studied the effect of g-modes
associated with buoyancy induced by proton-neuron composition
gradients in the core, whose typical oscillation period is about 2~ms
for a canonical neutron star model.  If the growth time scale of the
Parker instability found in our study is longer than these periods of
the g-modes, it is possible that the buoyant forces inside the neutron
star suppress the growth of the Parker instability. In the present
study, we find that a typical growth time scale of the Parker
instability is about $20M_0$, which gives a typical growth time scale
of 0.1~ms for a canonical neutron star model with a strong magnetic
field of $10^{16}$~G.  For the present models, thus, it seems that the
buoyancy inside the neutron star could not suppress the onset of the
Parker instability.  However, for a weaker magnetic field strength
$\approx 10^{15}$~G, the Parker instability could be suppressed by the
buoyancy because a typical growth time scale of the Parker instability
becomes 1~ms.  To derive a definite answer to this problem, whether or
not the buoyancy can stabilize the Parker instability inside the
neutron star, we have to know the local growth rate of the Parker
instability.  Unfortunately, in our simulations, it is difficult to
estimate it in the vicinity of the stellar surface because the local
magnetic structure highly depends on the numerical resolution.  To
investigate this point more precisely, we have to take into account
the chemical inhomogeneities; this implies that it is necessary to
implement equations of state which depend on the chemical compositions
and to obtain the evolution of the chemical compositions.  This is
beyond the scope of this paper. 

\begin{acknowledgements}
  KK thanks to Y. Sekiguchi and K. Kyutoku for a fruitful discussion.  Numerical
  computations were performed on XT4 at the Center for Computational 
  Astrophysics in NAOJ and on NEC-SX8 at YITP in Kyoto University.
  This research has been supported by Grant-in-Aids for Young
  Scientists (B) 22740178, for Scientific Research (21340051),
  and for Scientific Research on Innovative Area (20105004) of 
  Japanese MEXT. 
\end{acknowledgements}

\bibliographystyle{aa}

\begin{thebibliography}{99}

\bibitem[Acheson (1978)]{Acheson:1978}
  Acheson,~D.~J., 1978, 
  Phil.~Trans.~Roy.~Soc.~Lond.~A, 289, 459

\bibitem[Baker et al. (2006)]{Baker:2006}
  Baker,~J.~G., Centrella,~J., Choi,~D.-I., Koppitz,~M., and
  van Meter,~J., 2006, Phys. Rev. Lett. 96, 111102

\bibitem[Balbus \& Hawley (1991)]{Balbus:1991}
  Balbus,~S.~A., and Hawley,~J.~F.,
  1991, Astrophys.\ J.\ ,376, 214

\bibitem[Baumgarte \& Shapiro (1999)]{Baumgarte:1998te}
  Baumgarte,~T.~W., and Shapiro,~S.~L., 1999, 
  Phys.\ Rev.\  D 59, 024007 

\bibitem[Bildsten et al. (1997)]{Bildsten:1997}
  Bildsten,~L. et al.,
  1997, Astrophys.\ J.\ Suppl., 113, 367

\bibitem[Braithwaite \& Nordlund (2005)]{Braithwaite:2005xi}
  Braithwaite,~J. and Nordlund,~A.,
  2005, Astron. Astrophys. 450, 1077

\bibitem[Braithwaite \& Spruit (2004)]{Braithwaite:2005ps}
  Braithwaite,~J., and Spruit,~H.~C., 
  2004, Nature, 431, 819

\bibitem[Burrows et al. (2007)]{Burrows:2007yx}
  Burrows,~A., Dessart,~L., Livne,~E., Ott,~C.~D., and Murphy,~J., 
  2007, Astrophys.\ J.\  664, 416

\bibitem[Brugmann et al. (2008)]{Brugmann:2008zz}
  Brugmann,~B., Gonzalez,~J.~A., Hannam,~M., Husa,~S., 
  Sperhake,~U., and Tichy,~W., 2008, 
  Phys.\ Rev.\  D 77, 024027

\bibitem[Campanelli et al. (2006)]{Campanelli:2005dd} 
  Campanelli,~M., Lousto,~C.~O., Marronetti,~P., 
  and Zlochower,~Y., 2006, 
  Phys.\ Rev.\ Lett.\ 96, 111101

\bibitem[Cook et al. (1994)]{Cook:1994}
  Cook,~G.~B., Shapiro,~S.~L., and Teukolsky,~S.~A.
  1994,  Astrophys.\ J.\  422, 227
  
\bibitem[Duez et al. (2010)]{Duez:2010}
  Duez,~V., Braithwaite,~J., Mathis,~S., 
 arXiv:1009.5384 [astro-ph.SR]

\bibitem[Finn (1987)]{Finn:1987}
  Finn,~L.~S., 
  1987, Mon.~Not.~R.~Astro.~Soc., 227, 265

\bibitem[Goossens (1980)]{Goossens:1980}
  Goossens,~M., 1980, 
  Geo.~Astro.~Fluid.~Dyn., 15, 123

\bibitem[Gavriil et al. (2002)]{Gavriil:2002mc}
  Gavriil,~F.~P., Kaspi,~V.,~M., and Woods,~P.~M., 
  2002, Nature, 419, 142

\bibitem[Harding \& Lai (2006)]{Harding:2006qn}
  Harding,~A.~K., and Lai,~D., 
  2006, Rept.\ Prog.\ Phys.\  69, 2631

\bibitem[Ibrahim et al. (2003)]{Ibrahim:2003}
  Ibrahim,~A.~I., Swank,~J.~H., Parke,~W., 
  2003, Astrophys.~J.~, 584, L17

\bibitem[Ioka \& Sasaki (2004)]{Ioka:2003nh}
  Ioka,~K. and Sasaki,~M.
  2004, Astrophys.\ J.\  600, 296

\bibitem[Kotake et al. (2004)]{Kotake:2004}
  Kotake,~K., Sawai,~H., Yamada,~S., and Sato,~K., 
  2004, Astrophys.~J.~, 608, 391

\bibitem[Kiuchi et al. (2009)]{Kiuchi:2009jt}
  Kiuchi,~K., Sekiguchi,~Y., Shibata,~M., and Taniguchi,~K., 2009,
  Phys.\ Rev.\  D 80, 064037

\bibitem[Kiuchi et al.(2008)]{Kiuchi:2008ss}
  Kiuchi,~K., Shibata,~M., and Yoshida,~S., 2008, 
  Phys.\ Rev.\  D 78, 024029

\bibitem[Kiuchi \& Yoshida (2008)]{Kiuchi:2008ch}
  Kiuchi,~K., and Yoshida,~S., 2008,
  Phys.\ Rev.\  D 78, 044045

\bibitem[Kurganov \& Tadmor (2000)]{KT} 
  Kurganov,~A., and Tadmor,~E., 2000, 
  J.~Comput.\ Phys. 160, 214

\bibitem[Lai (2001)]{Lai:2001}
  Lai,~ D., 2001
  Rev.~Mod.~Phys., 73, 629

\bibitem[Lander et al. (2010)]{Lander:2009ws}
  Lander,~S.~K., Jones~D.~I., and Passamonti,~A.,
  2010, Mon.~Not.~R.~Astro.~Soc., 405, 318

\bibitem[Lander \& Jones (2010)]{Lander:2010br}
  Lander,~S.~K., and Jones~D.~I.,  
  arXiv:1009.2453 

\bibitem[Lucas-Serrano et al. (2004)]{lucas} 
  Lucas-Serrano,~A., Font,~J.~A., Ib\'anez,~J.~M. and 
  Mart\'i,~J.~M., 2004, 
  Astron. Astrophys. 428, 703

\bibitem[Manchester et al.(2005)]{Manchester:2005}
Manchester, R. N., Hobbs, G. B., Teoh, A., and Hobbs, M., 2005, 
AJ, 129, 1993: 
see http://www.atnf.csiro.au/research/pulsar/psrcat/ for 
the latest number of pulsars

\bibitem[Obergaulinger et al. (2006)]{OAM2006}
 Obergaulinger,~M., Aloy, M.A., and M\"uller, 2006, 
Astron. Astrophys. 450, 1107

\bibitem[Orlandini \& Fiume (2001)]{Orlandini:2001ue}
  Orlandini,~M. and Fiume,~D.~D., 
  2001, AIP Conference Proceedings, 599, 283, 

\bibitem[Cerda-Duran et al. (2007)]{CerdaDuran:2007cr}
  Cerda-Duran,~P., Font,~J.~A., and Dimmelmeier,~H., 
  2007, Astron. Astrophys. 474, 169

\bibitem[Parker (1955)]{Parker:1955}
  Parker,~E.~N., 1955, ApJ, 121, 49

\bibitem[Parker (1966)]{Parker:1966}
  Parker,~E.~N., 1966, ApJ, 145, 811


\bibitem[Reisenegger \& Goldreich (1992)]{Reisenegger:1992}
  Reisenegger,~A., and Goldreich,~P.,
  1992, Astrophys.\ J.\  395, 240

\bibitem[Rea et al. (2003)]{Rea:2003mx}
  Rea,~N. {\it et al.},
  2003, Astrophys.\ J.\  586, L65

\bibitem[Scheidegger et al. (2008)]{Scheidegger:2007nk}
  Scheidegger,~S., Fischer,~T., and Liebendoerfer,~M., 
  2008, Astron. Astrophys., 490, 231

\bibitem[Shibata (2003)]{S03} 
  Shibata,~M., 2003, Phys. Rev. D 67, 024033

\bibitem[Shibata \& Nakamura (1995)]{Shibata:1995we}
  Shibata,~M. and Nakamura,~T., 1995,
  Phys.\ Rev.\  D 52, 5428

\bibitem[Shibata \& Sekiguchi (2005)]{Shibata:2005gp}
  Shibata,~M. and Sekiguchi,~Y.~i., 2005,
  Phys.\ Rev.\  D 72, 044014

\bibitem[Shibata et al. (2006)]{Shibata:2006hr}
  Shibata,~M., Liu,~Y.,~T., Shapiro,~S.~L., and Stephens,~B.~C., 
  2006, Phys.\ Rev.\  D 74, 104026

\bibitem[Spruit (1999)]{Spruit:1999}
 Spruit,~H.~C., 1999, 
 Astron. Astrophys., 349, 189

\bibitem[Takiwaki et al. (2009)]{Takiwaki:2007sf}
  Takiwaki,~T., Kotake,~K., and Sato,~K.,
  2009, Astrophys.\ J.\  691, 1360

\bibitem[Tayler (1973)]{Tayler:1973}
  Tayler,~R.~J., 1973, 
  Mon.~Not.~R.~Astro.~Soc., 161, 365

\bibitem[Thompson \& Duncan (1993)]{Thompson:1993hn}
  Thompson,~C., and Duncan,~R.~C., 
  1993, Astrophys.\ J.\  408, 194

\bibitem[Thompson \& Duncan (1995)]{Thompson:1995gw}
  Thompson,~C., and Duncan,~R.~C., 
  1995, Mon.~Not.~R.~Astro.~Soc., 275, 255

\bibitem[Thompson \& Duncan (1996)]{Thompson:1996pe}
  Thompson,~C., and Duncan,~R.~C., 
  1996, Astrophys.\ J.\  473, 322

\bibitem[Thompson \& Duncan (2001)]{Thompson:2001}
  Thompson,~C., and Duncan,~R.~C., 
  2001, Astrophys.\ J.\ , 561, 980

\bibitem[Wickramasinghe \& Ferrario (2005)]{Wickramasinghe:2005}
  Wickramasinghe,~D., and Ferrario,~L.
  2005, Mon.\ Not.\ Roy.\ Astron.\ Soc.\  356, 1576

\bibitem[Woods \& Thompson (2004)]{Woods:2004kb}
  Woods,~P.~M., and Thompson,~C., 
  arXiv:astro-ph/0406133.

\bibitem[Yamada \& Sawai (2004)]{Yamada:2004}
  Yamada,~S., \& Sawai,~H., 
  2004, Astrophys.\ J.\ , 608, 907

\end{thebibliography}

\end{document}